\documentclass[longauth]{aa}

\usepackage[utf8]{inputenc}
\usepackage[T1]{fontenc}

\usepackage[dvipsnames]{xcolor}
\definecolor{ochre}{rgb}{0.8, 0.47, 0.13}

\definecolor{BlueF}{rgb}{0.03, 0.27, 0.49}

\usepackage{float}
\usepackage{placeins} 

\usepackage{subcaption}
\usepackage{graphicx}
\usepackage{txfonts}
\usepackage[export]{adjustbox}

\usepackage[bottom]{footmisc}

\PassOptionsToPackage{hyphens}{url}\usepackage{hyperref}
\usepackage{mathtools}
\usepackage{amsmath}

\newcommand*\diff{\mathop{}\!\mathrm{d}}

\usepackage{multirow}
\usepackage{makecell}

\usepackage{stackengine}

\usepackage{float}

\newcommand{\Pth}{\ensuremath{P_\text{th}}}
\newcommand{\Guv}{\ensuremath{G_0}} 
\newcommand{\AV}{\ensuremath{A_V^{\text{tot}}}}
\newcommand{\paramAngle}{\ensuremath{\alpha}}
\newcommand{\scalingParam}{\kappa}

\newcommand{\Av}{\AV{}}
\newcommand{\Gnaught}{\Guv{}}

\newcommand{\ch}[1]{\ensuremath{\mathrm{#1}}}
\newcommand{\HH}{\ch{H_2}}

\usepackage{xifthen}
\newcommand{\ifequals}[3]{\ifthenelse{\equal{#1}{#2}}{#3}{}}
\newcommand{\case}[2]{#1 #2} 
\newenvironment{switch}[1]{\renewcommand{\case}{\ifequals{#1}}}{}

\newcommand{\latexmol}[1]{%
    \begin{switch}{#1}%
        \case{12cn}{\ch{^{12}CN}}%
        \case{12co}{\ch{^{12}CO}}%
        \case{12cs}{\ch{^{12}CS}}%
        \case{13co}{\ch{^{13}CO}}%
        \case{32so}{\ch{^{32}SO}}%
        \case{34so}{\ch{^{34}SO}}%
        \case{c3h2}{\ch{C_3H_2}}%
        \case{c17o}{\ch{C^{17}O}}%
        \case{c18o}{\ch{C^{18}O}}%
        \case{cch}{\ch{CCH}}%
        \case{ch3oh}{\ch{CH_3OH}}%
        \case{hcn}{\ch{HCN}}%
        \case{hcop}{\ch{HCO^+}}%
        \case{hnc}{\ch{HNC}}%
        \case{n2hp}{\ch{N_2H^+}}%
        \case{sio}{\ch{SiO}}%
        \case{h2}{\ch{H_2}}%
        \case{h2co}{\ch{H_2CO}}%
        \case{c2h}{\ch{C_2H}}%
        \case{13c18o}{\ch{^{13}C^{18}O}}%
    \end{switch}%
}%

\newcommand{\unit}[1]{\,\ensuremath{\mathrm{#1}}}

\newcommand{\pc}{\unit{pc}}

\newcommand{\microm}{\unit{\text\textmu m}}

\newcommand{\Kpccm}{\unit{K\,cm^{-3}}}

\usepackage{letltxmacro}
\LetLtxMacro{\originaleqref}{\eqref}

\renewcommand{\eqref}{Eq.~\ref}
\newcommand{\eqrefp}[1]{(Eq.~\ref{#1})}

\DeclareMathOperator*{\argmin}{arg\,min}

\newcommand{\R}{\mathbb{R}}

\newcommand{\noiseGeneral}{\mathcal{D}}

\newcommand{\multnoisefull}{\boldsymbol{\varepsilon}^{(m)}}
\newcommand{\multnoise}{\varepsilon_{n\ell}^{(m)}}
\newcommand{\multnoiseCovmat}{\mathbf{\Sigma}^{(m)}}

\newcommand{\addnoisefull}{\boldsymbol{\varepsilon}^{(a)}}
\newcommand{\addnoise}{\varepsilon_{n\ell}^{(a)}}
\newcommand{\addnoiseCovmat}{\mathbf{\Sigma}^{(a)}}

\newcommand{\obsfull}{\mathbf{Y}}
\newcommand{\obsvect}[1]{\mathbf{y}_{#1}}
\newcommand{\obselt}{y_{n\ell}}

\newcommand{\paramfull}{\boldsymbol{\Theta}}

\newcommand{\paramvect}[1]{\boldsymbol{\theta}_{#1}}
\newcommand{\paramelt}[1]{\theta_{#1}}

\newcommand{\paramfullEst}{\widehat{\boldsymbol{\Theta}}}

\newcommand{\parameltEst}[1]{\widehat{\theta}_{#1}}

\newcommand{\truef}{\mathbf{M}}

\newcommand{\truefell}{M_\ell}

\newcommand{\stepsize}{\eta}

\newcommand{\lognormal}{\ensuremath{\mathrm{log}\,\mathcal{N}}}
\newcommand{\logd}{\ensuremath{\log_{10}}}

\newcommand{\TMC}{T_\text{MC}}
\newcommand{\TBI}{T_\text{BI}}

\newcommand{\pmtm}{p_\text{MTM}}

\usepackage{ntheorem}
\theoremseparator{:}
\newtheorem{hyp}{Hypothesis}

\begin{document}

\title{
  \textsc{Beetroots}:
  spatially-regularized Bayesian inference of physical parameter maps - Application to Orion 
}

\author{%
  Pierre Palud\inst{\ref{LERMA/MEUDON},\ref{CRISTAL},\ref{APC}} %
  \and Emeric Bron\inst{\ref{LERMA/MEUDON}} %
  \and Pierre Chainais\inst{\ref{CRISTAL}} %
  \and Franck Le Petit\inst{\ref{LERMA/MEUDON}} %
  \and Pierre-Antoine Thouvenin\inst{\ref{CRISTAL}} %
  \and Miriam G. Santa-Maria\inst{\ref{CSIC},\ref{UF}} %
  \and Javier R. Goicoechea\inst{\ref{CSIC}} %
  \and David Languignon\inst{\ref{LERMA/MEUDON}} %
  \and Maryvonne Gerin\inst{\ref{LERMA/PARIS}} %
  \and J\'er\^ome Pety\inst{\ref{LERMA/PARIS},\ref{IRAM}} %
  \and Ivana Be\v{s}li\'c\inst{\ref{LERMA/PARIS}} %
  \and Simon Coud\'e\inst{\ref{WORCESTER},\ref{CfA}}%
  \and Lucas Einig\inst{\ref{IRAM},\ref{GIPSA-Lab}} %
  \and Helena Mazurek\inst{\ref{LERMA/PARIS}} %
  \and Jan H. Orkisz\inst{\ref{IRAM}} %
  \and L\'eontine S\'egal\inst{\ref{IRAM},\ref{Toulon}} %
  \and Antoine Zakardjian\inst{\ref{IRAP}}
  \and S\'ebastien Bardeau\inst{\ref{IRAM}} %
  \and Karine Demyk\inst{\ref{IRAP}} %
  \and Victor de Souza Magalh\~aes\inst{\ref{NRAO}} %
  \and Pierre Gratier \inst{\ref{LAB}} %
  \and Viviana V. Guzm\'an\inst{\ref{Catholica}} %
  \and Annie Hughes\inst{\ref{IRAP}} %
  \and François Levrier\inst{\ref{LPENS}} %
  \and Jacques Le Bourlot\inst{\ref{LERMA/MEUDON}} %
  \and Dariusz C. Lis\inst{\ref{JPL}} %
  \and Harvey S. Liszt\inst{\ref{NRAO}} %
  \and Nicolas Peretto\inst{\ref{UC}} %
  \and Antoine Roueff\inst{\ref{Toulon}} %
  \and Evelyne Roueff\inst{\ref{LERMA/MEUDON}} %
  \and Albrecht Sievers\inst{\ref{IRAM}} %
}

\institute{%
  LUX, UMR 8262, Observatoire de Paris, PSL Research University, CNRS, Sorbonne Universit\'es, 92190 Meudon, France.\label{LERMA/MEUDON} 
  \and Univ. Lille, CNRS, Centrale Lille, UMR 9189 - CRIStAL, 59651 Villeneuve d'Ascq, France.\label{CRISTAL} %
  \and Université Paris Cité, CNRS, Astroparticule et Cosmologie, F-75013 Paris, France. \email{palud@apc.in2p3.fr}\label{APC} %
  \and Instituto de Física Fundamental (CSIC), Calle Serrano 121, 28006, Madrid, Spain.\label{CSIC} %
  \and Department of Astronomy, University of Florida, P.O. Box 112055, Gainesville, FL 32611, USA.\label{UF} %
  \and LUX, UMR 8262, Observatoire de Paris, PSL Research University, CNRS, Sorbonne Universit\'es, 75014 Paris, France.\label{LERMA/PARIS} %
  \and IRAM, 300 rue de la Piscine, 38406 Saint-Martin-d'H\`eres, France.\label{IRAM} %
  \and Univ. Grenoble Alpes, Inria, CNRS, Grenoble INP, GIPSA-Lab, Grenoble, 38000, France.\label{GIPSA-Lab} %
  \and Univ. Toulon, Aix Marseille Univ., CNRS, IM2NP, Toulon, France.\label{Toulon} %
  \and Department of Earth, Environment, and Physics, Worcester State University, Worcester, MA 01602, USA.\label{WORCESTER} %
  \and Center for Astrophysics | Harvard \& Smithsonian, 60 Garden Street, Cambridge, MA 02138, USA.\label{CfA} %
  \and Institut de Recherche en Astrophysique et Planétologie (IRAP), Université Paul Sabatier, Toulouse cedex 4, France.\label{IRAP} %
  \and National Radio Astronomy Observatory, 520 Edgemont Road, Charlottesville, VA, 22903, USA.\label{NRAO} %
  \and Laboratoire d'Astrophysique de Bordeaux, Univ. Bordeaux, CNRS,  B18N, Allée Geoffroy Saint-Hilaire, 33615 Pessac, France.\label{LAB} %
  \and Instituto de Astrofísica, Pontificia Universidad Católica de Chile, Av. Vicuña Mackenna 4860, 7820436 Macul, Santiago, Chile.\label{Catholica} %
  \and Laboratoire de Physique de l'\'Ecole normale supérieure, ENS, Université PSL, CNRS, Sorbonne Université, Université de Paris, Sorbonne Paris Cité, Paris, France.\label{LPENS} %
  \and Jet Propulsion Laboratory, California Institute of Technology,  4800 Oak Grove Drive, Pasadena, CA 91109, USA.\label{JPL}
  \and School of Physics and Astronomy, Cardiff University, Queen's buildings, Cardiff CF24 3AA, UK.\label{UC} %
} %


\abstract
{
  The current generation of millimeter receivers is able to produce cubes of 800\,000 pixels by 200\,000 frequency channels to cover several square degrees over the 3\,mm atmospheric window.
  Estimating the physical conditions of the interstellar medium (ISM) with an astrophysical model on a such a dataset is challenging.
  Common approaches tend to converge to local minima and typically poorly reconstruct regions with low signal-to-noise ratio (S/N).
  This instrumental revolution thus calls for new scalable data analysis techniques with more advanced statistical modelling and methods.
}
{
  %
  We aim at designing a general method to reconstruct large maps of physical conditions from the rich datasets produced by new and future instruments.
  The method must scale to very large maps, be robust to varying S/N, and escape from local minima.
  In addition, we want to quantify the uncertainties associated with our reconstructions to produce reliable analyses.
}
{
  We present \textsc{Beetroots},
  a \textsc{Python} software that performs Bayesian reconstruction of maps of physical conditions from observation maps and an astrophysical model.
  It relies on an accurate statistical model, exploits spatial regularization to guide estimations, and uses state-of-the-art algorithms.
  It also assesses the ability of the astrophysical model to explain the observations, providing feedback to improve ISM models.
  We demonstrate the power of \textsc{Beetroots} with the Meudon PDR code on synthetic data,
  and then apply it to estimate physical condition maps in the full Orion molecular cloud 1 (OMC-1) star forming region based on \textit{Herschel} molecular line emission maps. 
  %
}
{
  The application to the synthetic case demonstrates that \textsc{Beetroots} can currently analyse maps with up to ten thousand pixels, addressing large variations of S/N within observations, escaping from local minima, and providing consistent uncertainty quantifications.
  %
  On a personal laptop, the inference runtime ranges from a few minutes for maps of 100 pixels to 28 hours for maps of 8100 pixels.
  Regarding OMC-1 maps, our reconstructions of the incident UV radiation field intensity $\Guv$ are consistent with those obtained from FIR luminosities, which demonstrates that the considered molecular tracers are able to constrain $\Guv$ over a wide range of environments.
  Besides, the reconstructed thermal pressures are high in all dense regions of OMC-1 and positively correlated with $\Guv$. 
  Finally, the Meudon PDR code successfully explains the observations and the obtained $\Guv$ values are reasonable, which shows that UV photons control the gas physics and chemistry across the rim of OMC-1.
}
{
  This work paves the way towards systematic and rigorous analyses of observations produced by current and future instruments.
  Work still needs to be invested in massive parallelization of the algorithm to gain two orders of magnitude in map sizes.
}


\maketitle{} %


\section{Introduction}%
\label{sec:introduction}

Several aspects of star and planet formation are only partially understood.
Major remaining questions include the impact of turbulence and feedback mechanisms on star formation efficiency, and the development of chemical complexity alongside the evolution of interstellar matter from diffuse clouds to planet-forming disks.
Recent large hyper-spectral surveys of giant molecular clouds and star forming regions are key to tackle these questions.
For instance, the IRAM-30m Large Program ``Orion B''~\citep{petyAnatomyOrionGiant2017} covered the full scale of a giant molecular cloud ($\sim 10$ pc size) at a dense core resolution ($< 0.1$ pc).
It produced a hyper-spectral image with a million pixels and $200\,000$ spectral channels, allowing to map the emission of dozens of molecules over the whole cloud.
More generally, recent instruments such as the IRAM-30m, ALMA, NOEMA and the James Webb spatial telescope, provide observation maps with thousands of pixels in multiple emission lines -- see, e.g.,~\citet{habartPDRs4AllIIJWSTs2024}. 

Astrophysical codes for ISM environments can model observed regions and link observables (e.g., lines intensities) to a few local physical parameters (e.g., the gas density).
For instance, radiative transfer codes can be used to relate gas density, temperature and column densities of detected species to observable line intensities.
Such codes include RADEX~\citep{vandertakComputerProgramFast2007},
RADMC-3D~\citep{dullemondRADMC3DMultipurposeRadiative2012},
LIME~\citep{brinchLIMEFlexibleNonLTE2010},
MCFOST~\citep{pinteMCFOSTRadiativeTransfer2022},
MOLPOP-CEP~\citep{asensioramosMOLPOPCEPExactFast2018},
or LOC~\citep{2020A&A...644A.151J}.
It is also possible to adopt a more holistic approach to model the ISM by taking multiple physical phenomena into account (e.g., large chemical networks, thermal balance, radiative transfer) as well as their coupling.
These more comprehensive codes are usually dedicated to a specific type of environment, described with specific parameters.
Shock models, such as the Paris-Durham code~\citep{godardModelsIrradiatedMolecular2019} and the MAPPINGS code~\citep{sutherlandMAPPINGSAstrophysicalPlasma2018},
compute observables in shock-dominated environments from shock speeds, pre-shock densities or magnetic field.
Photodissociation regions (PDR) models such as the Meudon PDR code~\citep{lepetitModelAtomicMolecular2006} and H\textsc{ii} region models such as \textsc{Cloudy}~\citep{ferland2017RELEASECloudy2017} describe ultraviolet (UV) irradiated regions of the ISM and predict observables from the incident UV flux properties and gas properties (e.g., density or metallicity).
%
%
%
For each model, small changes in the physical parameters can cause large changes in the predicted observables~\citep{RolligPDRComparisonStudy2007}.

Inference consists in finding the physical parameters leading to predicted observations consistent with available observations.
%
The most widespread approach for this adjustment is to solve a chi-square minimization problem, that is, a least square problem (see e.g.,~\citealt{wuConstrainingPhysicalConditions2018}).
%
%
However, this approach relies on the assumption of a Gaussian additive noise which is applicable to many ISM observations.
Besides, it produces physical parameters maps that may not be physically consistent due to the existence of multiple solutions reconstructing the observations equally well.
These issues are particularly present in low signal-to-noise ratio (S/N) regions, where observations mostly provide upper bounds on the true observable.
Neighboring pixels in reconstructed maps thus might contain large physically-inconsistent differences.
However, a chi-squared minimization approach does not ensure a satisfactory reconstruction even at high S/N regimes, due to local minima in the cost function.
These specificities become more problematic as the size of the observation map increases.
Most of the estimation methods currently in use in ISM studies do not scale to observation maps of more than a few hundreds pixels.

In this work, we introduce \textsc{Beetroots} (BayEsian infErence with spaTial Regularization of nOisy multi-line ObservaTion mapS), a \textsc{Python} software that performs Bayesian reconstruction of maps of physical parameters from a set of observation maps and an astrophysical model.
Our approach is based on a finer statistical modelling of the noise affecting the observations.
Identifiability issues are mitigated using a spatial regularization that enables pixels to exploit the information contained in their neighboring pixels.
This regularization favors smooth maps that are physically consistent, and is particularly helpful for large diffuse regions in which parameters are assumed to vary smoothly.
\textsc{Beetroots} runs the sampling algorithm introduced in~\citet{paludEfficientSamplingNon2023} and can thus provide much more information than optimization-based approaches, such as credibility intervals, covariance estimations and detection of multiple solutions or degeneracies.
Finally, it automatically quantifies and assesses the ability of the astrophysical model to explain the observations, using the Bayesian hypothesis testing approach introduced in~\citet{paludProblemesInversesTest2023a}.
This model checking permits users to determine whether reconstructions are trustworthy.
\textsc{Beetroots} is publicly available as a \textsc{Python} package on PyPi%
\footnote{
  \url{https://pypi.org/project/beetroots/}. All results in this work are obtained with version 1.1.0.
}.


Section~\ref{ssec:statistical-modelling-to-study-the-ISM} reviews the general framework that we adopt for statistical modelling. 
Section~\ref{ssec:methods-estimation} describes the inference and model checking methods implemented in \textsc{Beetroots}. 
The contributions in \textsc{Beetroots} are highlighted and compared with common choices in ISM studies.
%
%
Section~\ref{sec:application-set-up} describes the specificities considered to illustrate the proposed package, such as the astrophysical model and noise properties.
Section~\ref{sec:application-to-synthetic-observation-maps} applies the method to a synthetic yet realistic case at different spatial resolutions.
Section~\ref{sec:application-to-real-data} applies the method to Orion molecular cloud~1~(OMC-1) observation maps. 
Section~\ref{sec:conclusion} provides concluding remarks.
Appendix~\ref{sec:how-to-use-beetroots-in-practice} explains how to use \textsc{Beetroots}. 






\section{Statistical modelling to study the ISM}%
\label{ssec:statistical-modelling-to-study-the-ISM}

This work considers observations $\obsfull$ that correspond to a set of integrated intensities of emission lines, their ratios, or full line profiles.
An astrophysical model $\truef$ simulates the physics of the observed region.
It is assumed to predict the observables $\obsfull$ from a set of physical parameters $\paramfull$, such as the volume density, the thermal pressure, the visual extinction, or the radiative field intensity.
The goal of \textsc{Beetroots} is to infer the physical parameters $\paramfull$ and the associated uncertainties from the observation~$\obsfull$ and the astrophysical model~$\truef$, and to assess whether the reconstruction actually explains the observation.
To do so, we resort to a Bayesian framework.

This section presents the key components of statistical modelling and standard choices in ISM studies.
First, we introduce fundamental notions of Bayesian statistics and draw their relative links -- see~\cite[chapter 1]{robertMonteCarloStatistical2004} for a more complete introduction.
Then, we discuss the main advantages and drawbacks of the standard choices in ISM studies for the astrophysical model, the uncertainties on observations, and the prior distribution.
For more details on the statistical model adopted in \textsc{Beetroots}, see~\citet{paludSamplingMethodsStatistical2023}.

\subsection{Bayesian modelling}%
\label{sssec:bayesian-modelling}

The Bayesian framework is a statistical approach that encodes uncertainty using random variables and probability distributions.
It models the physical parameters to infer $\paramfull$ and the observation $\obsfull$ as random variables.
In a nutshell, Bayesian statistical modelling aims at building a posterior distribution $\pi (\paramfull \vert \obsfull)$.
It involves the definition of a likelihood function $\pi (\obsfull \vert \paramfull)$ -- from a deterministic astrophysical model $\truef$ and a deterioration $\noiseGeneral$ that includes noise, for instance -- and the choice of a prior distribution $\pi (\paramfull)$.

The likelihood function $\pi( \obsfull \vert \paramfull)$ evaluates how well the parameter $\paramfull$ reconstructs the observations $\obsfull$.
It is the Probability Density Function (PDF) of the observations $\obsfull$ given a fixed value for the parameters $\paramfull$, according to an observation model
%
\begin{align}\label{eq:general_obs_model}
  \obsfull = \noiseGeneral \left[
    \truef(\paramfull)
  \right]
  ,
\end{align}
where $\truef$ is the deterministic astrophysical model and $\noiseGeneral$ is a deterioration that can include measurement noise, calibration errors, modelling error from the choice of astrophysical model, or upper limits on observables.
%
%
%

The prior distribution $\pi \left(\paramfull \right)$ encodes prior information on the physical parameters of interest $\paramfull$.
It may be non-informative (e.g., a uniform distribution) or informative (e.g., to include the results of other trusted studies or to constrain acceptable solutions to be physically meaningful).
For informative priors, the negative log-prior, that is, the function $\paramfull \mapsto - \log \pi (\paramfull)$, is called regularization function.

In contrast, the distribution of interest for inference is the posterior distribution, that is, the probability of $\paramfull$ knowing the observation $\obsfull$.
It combines the likelihood function and the prior distribution following the Bayes theorem:
\begin{align}\label{eq:posterior_full}
  \pi \left(\paramfull \vert \obsfull \right)
  =
  \frac{
    \pi \left(\obsfull \vert \paramfull \right)
    \pi \left(\paramfull \right)
  }{
    \pi \left( \obsfull \right)
  }
  .
\end{align}
The Bayesian evidence $\pi \left(\obsfull \right)$ is a normalization constant that does not depend on $\paramfull$.
In inference,~\eqref{eq:posterior_full} is thus simplified to
\begin{align}\label{eq:posterior_def}
    \pi \left( \paramfull \vert \obsfull \right)
    \propto
    \pi \left(\obsfull \vert \paramfull \right) \; \pi \left(\paramfull \right)
    .
\end{align}
Estimations of the physical parameters $\paramfull$ are derived from the likelihood function, $\pi(\obsfull \vert \paramfull)$, or from the posterior, $\pi \left( \paramfull \vert \obsfull \right)$.
Uncertainty quantifications are extracted from the posterior.
Before describing how in Sect.~\ref{ssec:methods-estimation}, we discuss the choices of astrophysical model $\truef$, random deterioration $\noiseGeneral$ and prior $\pi(\paramfull)$.

\subsection{The astrophysical model}%
\label{sssec:the-astrophysical-model}

The astrophysical model $\truef$ predicts observables from physical parameters.
In ISM studies, some relatively simple models such as \textsc{RADEX}%
~\citep{vandertakComputerProgramFast2007}
and \textsc{UCLCHEM}%
~\citep{holdshipUCLCHEMGasgrainChemical2017} are sometimes used directly in inference~\citep{behrensTracingInterstellarHeating2022,keilUCLCHEMCMCMCMCInference2022}.
More comprehensive codes provide more reliable predictions at the cost of longer evaluation times.
Such codes are usually dedicated to a specific physical regime, such as photodissociation regions (PDRs) for
Kosma-$\tau$%
~\citep{rolligKOSMAPDRModel2022}
or the Meudon PDR code%
~\citep{lepetitModelAtomicMolecular2006},
\textsc{Hii} regions for \textsc{Cloudy}%
~\citep{ferland2017RELEASECloudy2017},
or shock-dominated regions for the Paris-Durham code%
~\citep{godardModelsIrradiatedMolecular2019}.
These codes usually take as input parameters at the cloud scale, and return emissions integrated over the full cloud.
It is common in ISM studies to associate these inputs and outputs with pixels in an observation map -- see, e.g.,~\citet{wuConstrainingPhysicalConditions2018}.
That is, for a map of $N$ pixels, the full set of physical parameters $\paramfull$ can be broken down into $N$ vectors $\paramvect{n}$ (each corresponding to a pixel) of $D$ physical parameters (e.g., gas density).
Similarly, the full set of observations $\obsfull$ can be broken down into $N$ vectors $\obsvect{n}$ of $L$ observables.
For each pixel, the astrophysical model outputs $\truef(\paramvect{n}) = (\truefell(\paramvect{n}))_{\ell = 1}^L$, which is to be compared with the observation of the corresponding pixel $\obsvect{n} = (\obselt)_{\ell=1}^L$.
Therefore, for a map with 100 pixels, comparing physical parameters and observations requires 100 evaluations of the astrophysical model, that is, one evaluation per pixel.
In addition, as will be shown in Sect.~\ref{ssec:methods-estimation}, extracting estimators from the posterior distribution requires many evaluations of the likelihood function, and thus of the astrophysical model for each pixel.
The astrophysical model thus needs to be fast while providing accurate predictions.

\textsc{Beetroots} was implemented assuming that an evaluation of the astrophysical model is fast (less than 0.1 second) and can be processed in batch.
These specifications are nearly impossible to satisfy with a comprehensive numerical code, and require the use of a fast and accurate emulator.
Using emulators built from precomputed model grids is common practice in ISM studies, using
interpolation methods~\citep{gallianoDustSpectralEnergy2018,wuConstrainingPhysicalConditions2018,ramambasonInferringHIIRegion2022},
artificial neural networks (ANN)~\citep{GrassiMaNN2011,demijollaIncorporatingAstrochemistryMolecular2019,holdshipChemulatorFastAccurate2021,grassiReducingComplexityChemical2022,paludNeuralNetworkbasedEmulation2023}
or other machine learning approaches~\citep{smirnov-pinchukovMachineLearningacceleratedChemistry2022,bronTracersIonizationFraction2021}.

\subsection{The random deterioration and noise model}%
\label{sssec:the-noise-model}

To compare the observations $\obsfull$ with the predictions of the astrophysical model $\truef(\paramfull)$, it is necessary to describe the random deteriorations (i.e., noise) that deteriorate the observations.
This section reviews common noise models $\noiseGeneral$ in the ISM community and those implemented in \textsc{Beetroots}.

The most widespread random deterioration in ISM studies is Gaussian and additive~\citep{gallianoISMPropertiesLowmetallicity2003, chevallardInsightsContentSpatial2013,perez-monteroDerivingModelbasedTeconsistent2014, chevanceMilestoneUnderstandingPDR2016a,
wuConstrainingPhysicalConditions2018,leeRadiativeMechanicalFeedback2019,roueffC18O13CO12CO2021,keilUCLCHEMCMCMCMCInference2022,behrensTracingInterstellarHeating2022}.
For a pixel $1 \leq n \leq N$ and an observable $1 \leq \ell \leq L$, it reads
\begin{align}\label{eq:obsmodel_pure_gauss_model}
  \obselt = \truefell( \paramvect{n} ) + \addnoise,
  \quad \addnoisefull \sim \mathcal{N}(0, \addnoiseCovmat)
  ,
\end{align}
where $\addnoisefull$ is a measurement noise (typically thermal) with zero mean and covariance matrix $\addnoiseCovmat$.
The Gaussian noise is sometimes considered on the log of the observations $\log \obselt$~\citep{valeasariBONDBayesianOxygen2016}, which is equivalent to a multiplicative error $\multnoisefull$ following a lognormal distribution.
In all these works, the covariance matrix $\addnoiseCovmat$ is assumed to be diagonal, that is, all the noise components are independent.
An additive Gaussian model is often a good approximation thanks to the central limit theorem.
Besides, it is simple to manipulate.
However, it fails to accurately capture all uncertainties associated with ISM observations.

\begin{figure}
  \centering
  \includegraphics[width=0.9\linewidth]{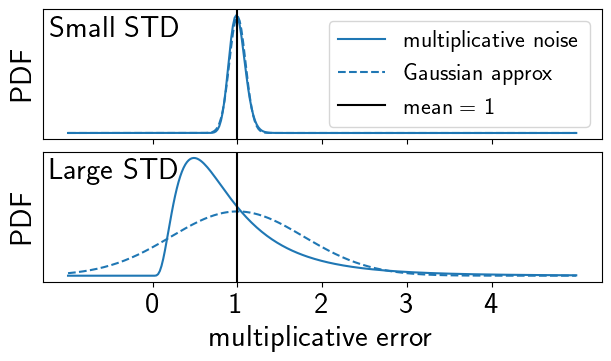}
  \caption{
    Quality of a Gaussian approximation to a lognormal distribution for low and high STD.
    Top: multiplicative noise associated with an average error of 10\%.
    The lognormal distribution is well approximated by a normal distribution.
    Bottom: multiplicative noise associated with an average error of a factor of two.
    The PDF of the lognormal distribution is significantly asymmetric and thus poorly approximated by a normal distribution.
  }%
  \label{fig:gaussian_approx_mult_noise}
\end{figure}

To consider a more realistic random deterioration, some ISM studies combine a Gaussian additive noise $\addnoisefull$ and a Gaussian multiplicative noise $\multnoisefull$.
This multiplicative noise represents calibration errors~\citep{gordonDustGasMagellanic2014,ciurloHotMolecularHydrogen2016,gallianoDustSpectralEnergy2018,gallianoNearbyGalaxyPerspective2021} or modelling errors in the astrophysical model~$\truef$~\citep{blancIZIInferringGas2015,valeasariBONDBayesianOxygen2016,johannessonBayesianAnalysisCosmic2016}.
The observation model then reads
\begin{align}\label{eq:obsmodel_add_mult}
  \obselt = \multnoise \truefell( \paramvect{n} ) + \addnoise, \,
   \multnoisefull \sim \mathcal{N}(1, \multnoiseCovmat),
   \addnoisefull \sim \mathcal{N}(0, \addnoiseCovmat)
  .
\end{align}
%
%
Using a Gaussian model for multiplicative noise simplifies computations, as the overall uncertainty model is Gaussian.
%
%
%
%
Although practical, a Gaussian model may not accurately account for multiplicative noise depending on the standard deviation (STD) of the multiplicative error.
In particular, a Gaussian distribution allows unphysical negative multiplicative factors.
Figure~\ref{fig:gaussian_approx_mult_noise} illustrates that the Gaussian approximation for multiplicative errors is relevant for small errors, but inappropriate for large ones.
In statistics, a typical model for multiplicative noise is the lognormal distribution, that is, a Gaussian distribution for the multiplying factor on the log scale.
%
%
%
A random deterioration combining an additive Gaussian noise and a multiplicative lognormal noise has already been used in ISM~\citep{kellyDUSTSPECTRALENERGY2012}.
See~\citet[chapter 3]{paludSamplingMethodsStatistical2023} for more details on how this random deterioration arises in the context of integrated intensities.

In \textsc{Beetroots}, the user can choose a random deterioration involving an additive Gaussian uncertainty only~\eqrefp{eq:obsmodel_pure_gauss_model} or along with multiplicative noise, Gaussian~\eqrefp{eq:obsmodel_add_mult} or lognormal~\eqrefp{eq:obs_model_practice}.
It can also handle upper limits on observations -- see Appendix~\ref{sec:upper-limits}~\eqrefp{eq:censorship_correct} for more details on upper limits.
The current implementation assumes that additive and multiplicative errors are independent.

\subsection{The prior distribution}%
\label{sssec:the-prior-distribution}

The prior distribution $\pi \left(\paramfull \right)$ encodes a priori information on the maps of physical parameters that we wish to reconstruct.
In \textsc{Beetroots}, we combine a uniform prior on validity intervals with a spatial regularization prior.
The first guarantees that the estimated values lie in a meaningful range.
The second guides reconstructions towards spatially smooth maps and removes spurious local minima.

Using validity intervals on the components of $\paramfull$ is common in ISM studies -- see e.g.,~\citet{behrensTracingInterstellarHeating2022,blancIZIInferringGas2015,thomasInterrogatingSeyfertsNebulaBayes2018,holdshipBayesianInferenceRates2018,wuConstrainingPhysicalConditions2018}.
Validity intervals are also implicitly involved when working on a lattice dataset. 
%
%
In a Bayesian framework, wide validity intervals are usually used to define a weakly informative uniform prior that does not bias estimations.
In \textsc{Beetroots}, a uniform prior can be set on the linear or log scale, depending on the physical parameter and its dynamic range.


Spatial regularizations have been used in ISM studies to fit Gaussian line profiles on hyperspectral observations~\citep{paumardRegularizedOSIRIS3D2014,ciurloHotMolecularHydrogen2016,marchalROHSARegularizedOptimization2019,paumardRegularized3DSpectroscopy2022}.
Assuming that spatial variations of physical parameters in low S/N regions are smooth, we use an $L_2$ penalty on the discrete gradient of each parameter map, as in \textsc{Rohsa}~\citep{marchalROHSARegularizedOptimization2019}.
For each physical parameter map, this spatial regularization is weighted by a hyperparameter $\tau_d > 0$.
We consider equal values $\tau_d$ for all $d$, since they all describe the same cloud and thus correspond to the same typical size.
Such a regularization improves reconstructions particularly in the low S/N regions, where the observations mostly contain noise.
Concretely, spatial regularization enables pixels to exploit the information contained in their neighbors, which improves estimations.
In low S/N regions, that typically contain more diffuse gas with larger spatial scales, the regularization orients the likelihood and can remove unphysical solutions.
In high S/N pixels, the likelihood overcomes the regularization and dominates the posterior.

Informative priors such as spatial regularization require careful hyperparameters tuning, as different values yield different trade-offs between prior and likelihood.
For instance,~\citet{ciurloHotMolecularHydrogen2016} presents a manual setting of the six hyperparameters of its spatial regularization.
Hierarchical Bayesian models infer the prior parameters from the data along with the physical parameters of interest $\paramfull$.
In ISM studies, hierarchical models are mostly used in dust studies~\citep{kellyDUSTSPECTRALENERGY2012,juvelaDegeneracyDustColour2013,venezianiBAYESIANMETHODANALYSIS2013,gallianoDustSpectralEnergy2018,gallianoNearbyGalaxyPerspective2021}.
In this work, we estimate the regularization weights $\tau_d$ along with physical parameters $\paramfull$ with sampling methods (see Sect.~\ref{sssec:sampling-based-estimation}) by maximizing the Bayesian evidence $\pi(\obsfull)$~\eqrefp{eq:posterior_def} as proposed in~\citet{vidalMaximumLikelihoodEstimation2020a}.
For optimization methods (Sect.~\ref{sssec:optimization-based-estimation}), we manually set them.
In practice, additive error bars in integrated intensity observations sometimes contain multiplicative noise, resulting in overestimated additive standard deviations $\sigma_{a,n\ell}$.
For instance, the STDs $\sigma_{a,n\ell}$ considered in~\citet{joblinStructurePhotodissociationFronts2018} or~\citet{wuConstrainingPhysicalConditions2018} are correlated with the intensities, which indicates that these STDs include multiplicative errors.
In such context, adjusting automatically $\tau_d$ can lead to over-smoothed maps.


\section{Beetroots: performing estimation and model checking}%
\label{ssec:methods-estimation}

Building on the described statistical model, this section details some methods to infer physical parameters $\paramfull$ from observations $\obsfull$, with a focus on tendencies in ISM studies.
%
%
We first depict optimization methods that are usually fast.
Then, we present sampling methods that are slower but more informative, as they natively provide uncertainty quantifications along with estimates.
Both approaches can be used with \textsc{Beetroots}.
For more details on the estimation method adopted in \textsc{Beetroots}, see~\citet{paludEfficientSamplingNon2023}.
Finally, we describe the model checking approach implemented in \textsc{Beetroots}, first presented in~\citep{paludProblemesInversesTest2023a}.

\subsection{Optimization-based estimation}%
\label{sssec:optimization-based-estimation}

Physical parameter estimations (i.e., reconstructions) can be extracted either from the likelihood or the posterior distribution.
As their name suggests, the maximum likelihood estimator (MLE) $\widehat{\paramfull}_{\text{MLE}}$ and the maximum a posteriori (MAP) $\widehat{\paramfull}_{\text{MAP}}$ are the values of $\paramfull$ that achieve the maximum value for the likelihood and for the posterior probability density function, respectively.
%
These estimators are often evaluated by minimizing the negative logarithm of these functions:
\begin{align}\label{eq:mle_neglog_posterior}
  \widehat{\paramfull}_{\text{MLE}}
  & =
  \argmin_{\paramfull} \left[ -\log \pi \left(\obsfull \vert \paramfull \right) \right]
  , \\
  \label{eq:map_neglog_posterior}
  \widehat{\paramfull}_{\text{MAP}}
  & =
  \argmin_{\paramfull} \left[ -\log \pi \left(\paramfull \vert \obsfull \right) \right]
  ,
\end{align}
with $-\log \pi \left(\paramfull \vert \obsfull \right)
= -\log \pi \left(\obsfull \vert \paramfull \right) - \log \pi \left(\paramfull \right)$ up to an additive constant~\eqrefp{eq:posterior_def}.
In this context, the negative log-likelihood and the negative log-posterior are often called ``loss function'' or ``cost function''.
The MLE is by definition the value of $\paramfull$ that leads to the predictions $\truef(\paramfull)$ that are most compatible with observations $\obsfull$.
It is sensitive to noise and hence unstable in low S/N regions or when the observables are not good tracers of the physical parameters of interest.
The MAP defines a trade-of between fitting the observation and satisfying prior knowledge.
It leads to more robust results.
Both the MLE and MAP are widespread estimators in ISM studies.

Solving the optimization problems in~\eqref{eq:mle_neglog_posterior} and~\eqref{eq:map_neglog_posterior} is challenging due to the potential existence of multiple local minima in the cost function.
Three main strategies prevail in ISM studies:
a discrete search within a grid of models~\citep{shefferPDRMODELMAPPING2011,shefferPDRMODELMAPPING2013,joblinStructurePhotodissociationFronts2018,leeRadiativeMechanicalFeedback2019},
meta-heuristics~\citep{mollerModelingAnalysisGeneric2013},
and gradient descent algorithms~\citep{schilkeHerschelObservationsOrtho2010,gallianoISMPropertiesLowmetallicity2003,paumardRegularized3DSpectroscopy2022,wuConstrainingPhysicalConditions2018} such as Levenberg-Marquardt.
Gradient descent algorithms fail to escape local minima and may return unphysical results.
Discrete searches and meta-heuristics approaches may escape from local minima, but their computational cost becomes prohibitive when the dimension of $\paramfull$ exceeds 10.
When used for optimization, \textsc{Beetroots} combines a gradient descent for fast convergence with a global exploration step to escape from local minima -- see Sect.~\ref{sssec:estimation-methods-in-beetroots}.

The main limitation of optimization methods is that they do not natively provide uncertainty quantifications associated with estimations.
In this case, producing uncertainty quantifications requires the use of additional methods.
The Cramér-Rao bound is sometimes used~\citep{roueffC18O13CO12CO2021}, but it is only relevant when the posterior is well approximated by a Gaussian at its mode, which is often not the case in astrophysics as noted by~\citet{panterStarFormationMetallicity2003}.
In contrast, sampling-based approaches natively provide uncertainty quantifications on the physical parameter $\paramfull$ along with point estimates for general posterior distributions.

\subsection{Sampling-based estimation}%
\label{sssec:sampling-based-estimation}

A Bayesian framework usually exploits the full posterior distribution to define estimators.
It is the case of the posterior mean, also called minimum mean squared error (MMSE), $\widehat{\paramfull}_{\text{MMSE}}
= \mathbb{E} \left[ \paramfull \vert \obsfull \right]$. 
%
%
%
Uncertainties on physical parameters -- that can help to identify the existence of multiple solutions or degeneracies among physical parameters -- can be quantified in multiple ways.
The covariance matrix provides a typical error for each parameter, and the correlations between pairs of parameters.
However, it is mostly relevant for nearly Gaussian distributions.
Credibility intervals yield lower and upper limits on each physical parameter, without any assumption on their probability distribution.
Both quantities require evaluating integrals over multiple variables.
In ISM studies, these integrals are sometimes evaluated with an integration on a discrete grid~\citep{dacunhaSimpleModelInterpret2008,pacificiRelativeMeritsDifferent2012,perez-monteroDerivingModelbasedTeconsistent2014,blancIZIInferringGas2015,valeasariBONDBayesianOxygen2016,thomasInterrogatingSeyfertsNebulaBayes2018,villa-velezFittingSpectralEnergy2021}, which does not scale to high dimensions.
An alternative approach resorts to Monte Carlo (MC) estimators, computed from samples of the posterior distribution.
Noting $\TMC$ the number of samples $\paramfull^{(t)} \sim \pi \left(\paramfull \vert \obsfull \right)$, the MMSE is estimated as
\begin{align}\label{eq:mmse_estim}
  \widehat{\paramfull}_{\text{MMSE}}
  \simeq \frac{1}{\TMC} \sum_{t=1}^{\TMC} \paramfull^{(t)}
  .
\end{align}
The generation of these samples is often performed iteratively with a Markov chain Monte Carlo (MCMC) algorithm~\citep{robertMonteCarloStatistical2004}.
At each iteration, a transition kernel first generates a candidate from a proposal distribution, and then accepts of rejects this candidate with a certain probability that involves the ratio of the posterior PDF of the candidate and of the previous iterate.
Candidates with low posterior PDF are thus likely to be rejected.
Therefore, the proposal distribution needs to generate candidates with high posterior PDF while sufficiently exploring the parameter space to be able escape from a local mode.

MCMC algorithms were first popularized in astronomy through cosmology in~\citet{christensenBayesianMethodsCosmological2001}.
The first public MCMC code, \textsc{CosmoMC}, was published in~\citet{lewisCosmologicalParametersCMB2002}.
Both articles used random walk Metropolis-Hastings (RWMH) to generate posterior samples, arguably the most widespread transition kernel.
%
%
%
The RWMH kernel was also applied in ISM studies~\citep{makrymallisUnderstandingFormationEvolution2014,paradisVariationsSpectralIndex2010}.
Although it can be applied to maps of physical parameters, this transition kernel becomes quickly inefficient in dimensions higher than about 20.
In \textsc{Beetroots}, we implement a new and efficient MCMC algorithm that scales to large maps.
Other sampling methods are also already popular in the ISM community.
Appendix~\ref{sec:other-sampling-methods-used-in-ism-studies} lists such algorithms and their applications in ISM studies, and discusses their limitations in comparison with the sampler implemented in \textsc{Beetroots} (see Sect.~\ref{sssec:estimation-methods-in-beetroots}).

Describing exhaustively a quantification of uncertainties can be quite tedious.
For instance, for $N \times D$ physical parameters, a complete covariance matrix would contain $ND (ND - 1) / 2$ different terms.
%
%
We favor credibility intervals (CIs), that permit to reduce the number of uncertainty quantification terms to $2ND$, that is one map for the CI lower bound, and one map for its upper bound.
To further reduce to a single map, we quantify the CI size.
For a 95\% CI, $[q^\text{(l,95)}, q^\text{(u,95)}]$, a natural description of the CI size would involve the difference $0.5 ( q^\text{(u,95)} - q^\text{(l,95)} )$, where the $0.5$ factor returns the distance from the interval center.
However, in ISM studies, the range of physical parameters such as the thermal pressure $\Pth$ covers numerous orders of magnitude, such that the difference $0.5 ( q^\text{(u,95)} - q^\text{(l,95)} ) \simeq 0.5 q^\text{(u,95)}$ is not informative.
Therefore, for strictly positive physical parameters, using this difference in logarithmic scale is more relevant.
As we describe all our estimators in linear scale in the following applications, we summarize the CI size using a quantity we call ``uncertainty factor'' (UF), defined as the exponential of this difference in logarithmic scale
%
%
\begin{align}
  \text{UF} \text{ 95\%}
  =
  \exp \left\{
    \frac{1}{2} \left( \log q^\text{(u,95)} - \log q^\text{(l,95)} \right)
  \right\}
  =
  \sqrt{
    \frac{
      q^\text{(u,95)}
    }{
      q^\text{(l,95)}
      }
  }.
\end{align}
For instance, an uncertainty factor of two indicates that an estimation is associated with an uncertainty of a factor of two.

\subsection{Estimation methods in \textsc{Beetroots}}%
\label{sssec:estimation-methods-in-beetroots}

\begin{figure}
  \centering
    \includegraphics[width=0.7\linewidth]{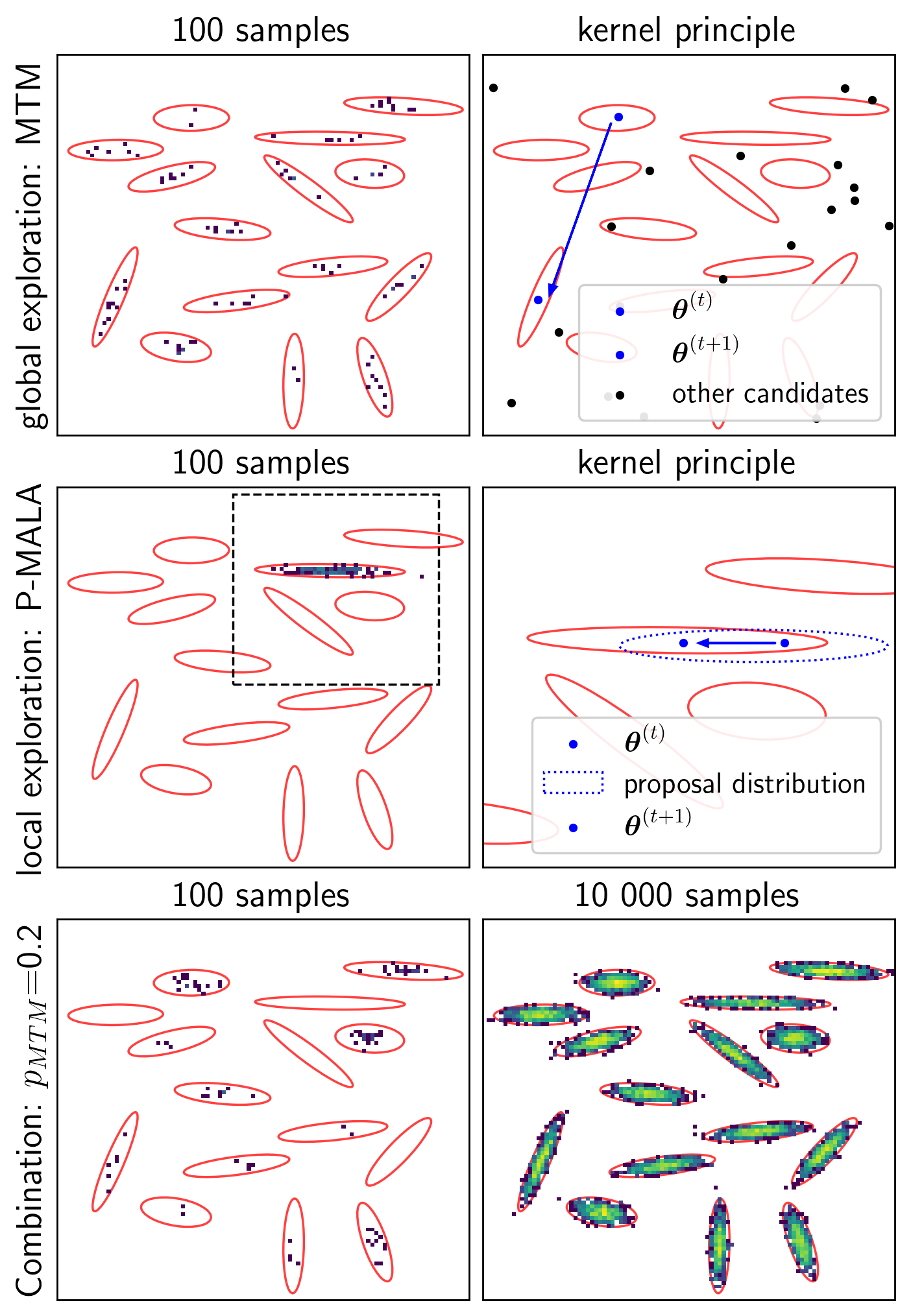}
  \caption{
    Illustration of the two transition kernels PMALA and MTM on a two-dimensional probability distribution: a Gaussian mixture model (shown with the red ellipses) restricted to validity intervals (shown with the large black square).
    (Top) MTM update rule (on the right), and a histogram after 100 sampling steps (on the left).
    With this update rule only, the sampling visits all the distribution modes at the cost of computationally heavier individual steps.
    (Middle) PMALA update rule (on the right), and a histogram after 100 sampling steps (on the left).
    With this cheaper update rule only, the sampling fails to visit all the distribution modes.
    (Bottom) Two histograms obtained using both kernels after 100 steps (left) and 10\,000 steps (right).
    The histograms converge to the correct distribution.
  }%
  \label{fig:sampling_kernels}
\end{figure}

In ISM studies, the numerical model is almost always non-linear, which can causes the negative log-likelihood~\eqrefp{eq:mle_neglog_posterior} and negative log-posterior~\eqrefp{eq:map_neglog_posterior} to have many local minima.
Common samplers such as RWMH usually fail to escape from a local minimum and should be avoided.
Besides, with many pixels, the dimension of the parameter space becomes very large, and usual quadrature-based numerical integration become intractable. 
%

\textsc{Beetroots} implements an MCMC algorithm introduced in~\citet{paludEfficientSamplingNon2023} that specifically addresses these two difficulties.
It relies on two transition kernels: multiple-try Metropolis kernel (MTM), that performs a ``global exploration'', and preconditioned Metropolis-adjusted Langevin algorithm (PMALA), that performs a ``local exploration''.
At each step $t$, the transition kernel is randomly selected with probability $\pmtm$ for MTM and $1-\pmtm$ for PMALA.
Figure~\ref{fig:sampling_kernels} illustrates the principle of these two kernels and applies them to a two-dimensional Gaussian mixture model restricted to a large square.
%


The MTM kernel allows escaping from local minima.
For each pixel, it randomly proposes $K \geq 1$ candidates, and selects one with a high probability density.
The proposal distribution (used to generate candidates) is defined from the prior.
When the prior includes spatial regularization, the proposal distribution for each pixel is a Gaussian mixture defined from the neighboring pixels.
This choice guides the reconstruction towards smooth maps.
When the prior is a uniform distribution on validity intervals, this proposal distribution is set to the same uniform distribution.
This is the case in single-pointing observations (i.e., one-pixel maps), or when no spatial regularization is considered.
This is also the case on Fig.~\ref{fig:sampling_kernels} where candidates are generated from a uniform distribution on the large black square.
Even with 100 iterations, this kernel manages to visit all the modes.
However, generating many candidates at each iteration is expensive.


The PMALA kernel efficiently explores the posterior distribution locally by exploiting its geometry.
Similarly to a gradient descent algorithm, it relies on a step-size $\stepsize > 0$.
It is faster than the MTM kernel, as it only generates one candidate.
However, with local exploration only, the sampling fails to visit other modes, and stays trapped in the mode closest to the initialization.


Combining PMALA with MTM exploits the best of the two methods.
As shown on Fig.~\ref{fig:sampling_kernels}, the combination leads to sample histograms converging to the correct distribution, at a cheaper computational cost than with MTM alone.

The implementation of \textsc{Beetroots} is general enough so that it can be used for optimization, to return either the MAP or the MLE.
In this case, PMALA boils down to an efficient preconditioned gradient descent update step, and MTM to keeping the candidate with the lowest cost only if this cost is lower than that of the current iterate.
When $0 < \pmtm < 1$, the optimization combines a fast convergence to a local minimum with the ability to escape from local minima.
However, the parameter $\pmtm$ can also be set to $0$ or $1$ to exploit only one of the two transition kernels.
In practice, setting $\pmtm$ to 0.1, 0.2 or 0.5 exploits well both kernels.
For the MTM kernel, we recommend setting the number of candidates to a few tens if considering spatial regularization and to hundreds or thousands if not, as the spatial regularization guides the generation of candidates.
For the PMALA kernel, the step size $\stepsize$ should be set so that the sampling has an accept rate of about 60\%  -- $\stepsize = 10^{-2} - 10^{-3}$ has been observed to yield such an accept rate in the experiments reported in Sect~\ref{sec:application-to-synthetic-observation-maps}.


\subsection{Methods: Model checking}%
\label{ssec:methods-model-checking}

Once the inference is performed, the quality of the reconstruction needs to be assessed.
Model selection is a quite common approach in astrophysics~\citep{ramambasonInferringHIIRegion2022,chevallardModellingInterpretingSpectral2016,zuckerThreedimensionalStructureLocal2021,kamenetzkySurveyMolecularISM2014}.
However, it only compares different models, assessing their relative success, as the absolute values of the associated criteria are not interpretable.
%
Conversely, model checking approaches assess a model individually, and yield interpretable diagnoses on whether a model can explain the observations.

In ISM studies, the obtained loss function value in a non-linear least squares problem, often noted $\chi^2$, is used in a three-case interpretation.
When $\chi^2 > 1$, the estimated parameters are judged not able to reproduce observations.
A $\chi^2 \ll 1$ suggests an overfit or an overestimation of uncertainties on observations.
The ideal case, $\chi^2 \simeq 1$, indicates that the physical parameters reasonably reproduce the observations.
This interpretation appears for instance in \citet{chevanceMilestoneUnderstandingPDR2016a,joblinStructurePhotodissociationFronts2018,villa-velezFittingSpectralEnergy2021}.
However, it comes with limitations and was criticized in the astrophysics community~\citep{andraeDosDonTs2010}.
In particular, the degree of freedom is challenging to estimate for non-linear astrophysical models and when a prior distribution is considered.
Besides, this $\chi^2$ rule only applies to Gaussian noise.
Finally, when $\paramfull$ is described by a posterior distribution and not by a single estimation $\paramfullEst$, the obtained $\chi^2$ value is approximated with an MC estimator.
The error associated with this approximation is seldom accounted for in astrophysics.

Bayesian model checking~\citep{gelmanPosteriorPredictiveAssessment1996}, also known as posterior predictive checking, is a statistically principled model checking technique that was also considered in ISM studies~\citep{chevallardModellingInterpretingSpectral2016,gallianoNearbyGalaxyPerspective2021,lebouteillerTopologicalModelsInfer2022}.
It is a hypothesis testing method that tests the following hypothesis:
\begin{hyp}[H\ref{hyp:first}]\label{hyp:first}
  The astrophysical model $\truef$, random deterioration $\noiseGeneral$ and prior $\pi(\paramfull)$ can reproduce the observations $\obsfull$.
\end{hyp}
To do so, reproduced observations are generated from posterior samples $\paramvect{}^{(t)}$ using the observation model~\eqrefp{eq:general_obs_model}.
The distribution of these noisy predicted observations is called the ``posterior predictive distribution''.
The goal is then to compare the posterior predictive distribution and the true observations $\obsfull$ with respect to some discrepancy measure $T$ on $(\obsfull, \paramfull)$.
The results of this comparison are summarized in a $p$-value defined as the probability that the parameters $\paramfull$ yield observations that are worse (in the sense of the chosen discrepancy measure) than the one we have observed.
H\ref{hyp:first} is then rejected with a confidence level $1 - \alpha$  when $p \leq \alpha$.
The posterior predictive distribution can also be visualized to get a more detailed diagnosis, for instance to identify the observables that led to a model rejection.
%

In \textsc{Beetroots}, to ensure that the $p$-value is consistent with the likelihood function in general cases, the discrepancy measure is set to the negative log-likelihood $T(\obsfull, \paramfull) = -\log \pi(\obsfull \vert \paramfull)$. 
As all noise realizations are assumed independent, we can test H\ref{hyp:first} for each pixel and provide maps of $p$-values.
This approach helps to identify regions that are poorly modelled by the astrophysical model~$\truef$.
The implemented approach also accounts for the uncertainties on the $p$-value that come from the MC evaluation.
This avoids rejections that are due to chance and caused by insufficient number of samples.
More details on the implemented model checking approach can be found in~\citet{paludProblemesInversesTest2023a}.


\section{Experimental setting}%
\label{sec:application-set-up}

This Section describes two applications: a synthetic case, where the ground truth is known, and observations of the Orion molecular cloud 1 (OMC-1).
In both applications, we consider observation maps of integrated intensities $\obsfull \in \R^{N \times L}$ that contain $N$ pixels and $L$ channels, and analyse them with the Meudon PDR code~$\truef$.
For a pixel $n$, we assume the Meudon PDR code to be able to reproduce the line integrated intensities $\obsvect{n}$ from a set of $D$ physical parameters $\paramvect{n} \in \R^D$ such as the thermal pressure.
The goal of both applications is to reconstruct physical parameter maps $\paramfull = \left( \paramvect{n} \right)_{n=1}^N$ and the associated uncertainties. 

This Section presents the likelihood elements that we use in the two applications, namely, the Meudon PDR code, its ANN emulator, and the random deterioration.
It also introduces the estimators to be evaluated and compared.

\subsection{The Meudon PDR code}%
\label{subsec:meudon_pdr}

The Meudon PDR code%
\footnote{\url{https://ism.obspm.fr/pdr.html}}%
~\citep{lepetitModelAtomicMolecular2006}
is a one-dimensional stationary code that simulates a PDR, that is, neutral interstellar gas illuminated with a UV radiation field.
It permits the investigation of the radiative feedback of a newborn star on its parent molecular cloud, for instance.
Although primarily designed for PDRs, it can also simulate a wide variety of environments such as diffuse clouds, nearby galaxies, damped Lyman alpha systems and protoplanetary disks.

%
The user specifies physical conditions such as the intensity of the incident UV radiation field~\Gnaught,
the elemental abundances,
and the depth of the slab of gas expressed as its total visual extinctions~\AV.
The user can also choose the thermal pressure in the cloud~\Pth (for isobaric models) or the gas volume density (for constant density models).
The code then solves multiphysics coupled balance equations of radiative transfer, thermal balance, and chemistry on an adaptive spatial grid over a one-dimensional slab of gas.
For the radiative transfer equation, the code considers absorption in the continuum by dust and gas and in the lines of key atoms and molecules such as \ch{H} and \HH~\citep{2007A&A...467....1G}.
For thermal balance, it computes the gas and grain temperatures from the specific intensity of the radiation field.
The code accounts for many heating and cooling processes, in particular photoelectric and cosmic ray heating, and line cooling.
For chemistry, the code provides the densities of about 200 species at each position.
About $3\,000$ reactions are considered, both in the gas phase and on the grains.
The chemical reaction network was built combining sources such as the KIDA database%
~\citep{wakelamKIneticDatabaseAstrochemistry2012},
the UMIST database%
~\citep{mcelroyUMISTDatabaseAstrochemistry2013},
and articles.
For key photoreactions, cross sections are taken from~\citet{heaysPhotodissociationPhotoionisationAtoms2017} and from Ewine van Dishoeck's photodissociation and photoionization database\footnote{\url{https://home.strw.leidenuniv.nl/~ewine/photo/index.html}}.
The successive resolution of these three coupled aspects is iterated until reaching a global stationary state.

The code yields 1D-spatial profiles of the volume density of all chemical species and of temperature of both grains and gas as a function of depth in the PDR.
From these spatial profiles, it also computes the line integrated intensities emerging from the cloud that can be compared to observations.
As of version 7 (released in 2024), thousands line intensities are predicted from species such as H$_2$, HD, H$_2$O, C$^+$, C, CO, \latexmol{13co}, \latexmol{c18o}, \latexmol{13c18o}, SO, \latexmol{hcop}, OH,  \latexmol{hcn}, \latexmol{hnc}, CH$^+$, CN or CS.

\subsection{The neural network emulator}%
\label{ssec:the-nnbma-emulator}

A single full run of the Meudon PDR code typically lasts a few hours for one input vector $\paramvect{n}$.
Running a full reconstruction with the original code would thus be very slow, as it requires thousands of astrophysical model evaluations~$\truefell(\paramvect{n})$.
In this work, we use the fast, light (memory-wise) and accurate ANN approximation of the Meudon PDR code proposed in~\citet{paludNeuralNetworkbasedEmulation2023}.
This ANN emulator was built with the \textsc{nnbma} software%
\footnote{\url{https://github.com/einigl/ism-model-nn-approximation}}
(neural-network-based model approximation).

This approximation was generated for isobaric models.
It is valid for \mbox{$\logd \Pth \in [5, 9]$}, \mbox{$\logd \Gnaught \in [0, 5]$}, \mbox{$\logd \AV \in [0, \logd(40)]$}.
As ANNs can process multiple inputs at once in batches, the evaluation of $10^3$ input parameter vectors with this approximation lasts about 10\,ms on a personal laptop.
For the lines studied in this paper and chosen validity intervals, the emulator has an average error of 3.5\%, which is three time lower than the typical calibration errors considered for our applications. 
In the remainder of this paper, to simplify notation, we denote $\truef$ this ANN approximation.

Some secondary parameters are set to standard values in the Meudon PDR grid used to train our ANN emulator.
For instance, we consider standard galactic grains, the average galactic dust extinction curve and set $R_V = 3.1$~\citep{fitzpatrickAnalysisShapesInterstellar2007}, $N_H / E(B - V) = 5.8 \times 10^{21}$ cm$^{-2}$~\citep{bohlinSurveyInterstellarLalpha1978} and the cosmic ray ionization rate to $10^{-16}$ s$^{-1}$ per H$_2$~\citep{lepetitH3OtherSpecies2004,indrioloDiffuseInterstellarClouds2007}.
%
Finally, the interstellar standard radiation field is set to that of~\citet{mathisInterstellarRadiationField1983}.
%
Although these standard values might not be the most adequate for a specific environment, they represent a relevant starting point to infer the considered main parameters over wide maps that cover dense and diffuse gas regions%
\footnote{%
A better approach would include these secondary parameters as free parameters in the grid used for training, and then infer them with \textsc{Beetroots}.
However, including many free parameters is expensive, as the number of Meudon PDR evaluations necessary to train a neural network emulator scales exponentially with the number of free parameters.
Besides, the considered observations of OMC-1 (Sect.~\ref{sec:application-to-real-data}) only contain 5 lines per pixel, and we chose to keep the number of parameters to estimate per pixel lower than the number of lines.
Therefore, we focused on three physical parameters from the Meudon PDR code, namely $\Pth$, $\Gnaught$ and $\Av$, and set the others to standard values from the literature.%
}.
This is the case for instance for the synthetic case (Sect.~\ref{sec:application-to-synthetic-observation-maps}) and for the OMC-1 map (Sect.~\ref{sec:application-to-real-data}).

\subsection{Noise model and characteristics}%
\label{ssec:noise-model-and-characteristics}

As indicated in Sect.~\ref{sssec:the-noise-model}, we consider a combination of additive Gaussian and multiplicative lognormal uncertainties.
The additive noise contains thermal noise.
The multiplicative noise contains calibration error and model mis-specification.
We assume no model mis-specification for synthetic observations, since the data are generated with the Meudon PDR code emulator.
For real observations, the model mis-specification STD $\sigma_\text{model}$ is set so that a $3\sigma$ error corresponds to an error of a factor of $3$, that is, $\sigma_\text{model} = \frac{1}{3} \log 3$. 
All noise realizations are assumed independent.

To account for beam dilution and geometrical effects (viewing angle, irradiation angle, multiple overlapping PDR fronts), we consider a last multiplicative parameter $\boldsymbol{\scalingParam} = (\scalingParam_n)_{n=1}^N$ that affects all lines identically in a pixel.
This scaling parameter $\scalingParam_n$ is introduced in~\citet{shefferPDRMODELMAPPING2013} as a product of four terms representing the four effects listed above, which can be smaller than 1 (e.g. if beam dilution dominates) or larger than 1 (e.g. if limb brightening due to a very oblique viewing angle dominates).
%
\citet{joblinStructurePhotodissociationFronts2018} considered a similar multiplicative parameter.
Like~\citet{joblinStructurePhotodissociationFronts2018}, we infer this parameter.
We consider that only values of $\scalingParam$ of the order of unity are physically reasonable, and thus assume that $\logd \scalingParam_n \in [-1, 1]$.
Using a single multiplicative parameter is only an approximation: for limb brightening, it assumes that all lines are optically thin, while for beam dilution, it assumes that they are all emitted in a similar fraction of the beam.
In the following, the observation model from~\eqref{eq:general_obs_model} can be written for any line~$\ell$ and pixel~$n$
\begin{align}\label{eq:obs_model_practice}
  \obselt
  \; = \;
  \multnoise \scalingParam_n \truefell( \paramvect{n} ) + \addnoise
  ,
\end{align}
with $\multnoise \sim \lognormal(-\sigma_m^2 / 2, \sigma_m^2)$ and $\addnoise \sim \mathcal{N}(0, \sigma_{a,n\ell}^2)$.
To simplify notation, this parameter $\scalingParam_n$ is added to the parameter vector $\paramvect{n}$.
In other words, in the following, each pixel is described with the physical parameter vector $\paramvect{n} = (\scalingParam, \Pth, \Gnaught, \AV) \in \R^4$.


Finally, in the synthetic observation test, we also include the case where only upper limits are available, that is, when the line is undetected.
More precisely, for a pixel $n$ and a line $\ell$, when the simulated intensity $\multnoise \scalingParam_n \truefell( \paramvect{n} ) + \addnoise$ is smaller than $3\sigma_{a,n\ell}$, we only keep the upper limit $\omega_{nl} = 3 \sigma_{a,n\ell}$ in our synthetic observations for this line in this pixel.

\subsection{Considered reconstruction methods}%
\label{sssec:considered methods}

In the following applications, we demonstrate the potential of \textsc{Beetroots} by computing multiple estimators.
First, we consider the MMSE computed with the MCMC algorithm presented in Sect.~\ref{sssec:estimation-methods-in-beetroots}.
The algorithm is also used in optimization mode, to return the MAP.
In the remainder, we call these estimators ``MMSE (ge)'' and ``MAP (ge)'' to highlight the use of the MTM kernel (i.e., of global exploration), together with the PMALA kernel.
For each application, the ``MAP (ge)'' is evaluated for multiple values of the regularization weight $\tau_d$.
For the ``MMSE (ge)'', the regularization weight $\tau_d$ is estimated along with the physical parameters $\paramfull$.
This estimation relies on the approach from~\citet{vidalMaximumLikelihoodEstimation2020a} that maximizes the Bayesian evidence~$\pi(\obsfull)$~\eqrefp{eq:posterior_def}.

In addition to the MMSE (ge) and MAP (ge), we consider two approaches to evaluate the maximum likelihood estimator (MLE) to highlight the critical roles of the spatial regularization and global exploration.
The first exploits global exploration, and is thus called ``MLE (ge)''.
The second approach, that we call ``MLE (le)'', uses local exploration only by setting $\pmtm = 0$.
This approach is the closest to standard gradient descent estimations used in ISM studies.
We use it to demonstrate the importance of the MTM kernel to escape from local minima.
To achieve better results with local exploration only, a common strategy consists in running the optimization procedure multiple times with different initial values (e.g., 100 times for each pixel in~\citealt{wuConstrainingPhysicalConditions2018}).
This strategy assumes that at least one initialization is close to the global minimum, which is not guaranteed. 
For fair comparisons, for each application in the following, all estimators are evaluated only once with the same random initialization, using CPU only.
All the experiments presented in this work ran on a personal laptop with an Apple M3 Pro chip with 11 cores for CPU and a 18\,GB memory.


\section{Application to synthetic observation maps}%
\label{sec:application-to-synthetic-observation-maps}

This section introduces a synthetic yet realistic reconstruction problem, and applies the estimation algorithm presented in Sect.~\ref{sssec:estimation-methods-in-beetroots}.
From synthetic true physical parameter maps $\paramfull^*$, synthetic observations~$\obsfull$ are generated.
The goal in this experiment is to provide an estimator of the physical parameter maps $\paramfullEst$ from $\obsfull$ as close as possible to the true values~$\paramfull^*$.
To demonstrate the power of the presented algorithm, we consider three sets of observations of the same cloud with different spatial resolutions.

First, the synthetic molecular cloud structure and the observation generation method are presented.
Then, the metrics used to assess the performance of the reconstruction are described.
Finally, the reconstructions for the three cases are described and a
comparison between the proposed methods is detailed.

\subsection{Generation of synthetic observation maps}%
\label{ssec:generation-of-synthetic-observation-maps}

\begin{figure}
  \centering
  \begin{subfigure}{1\linewidth}
    \centering
    \includegraphics[width=0.6\linewidth]{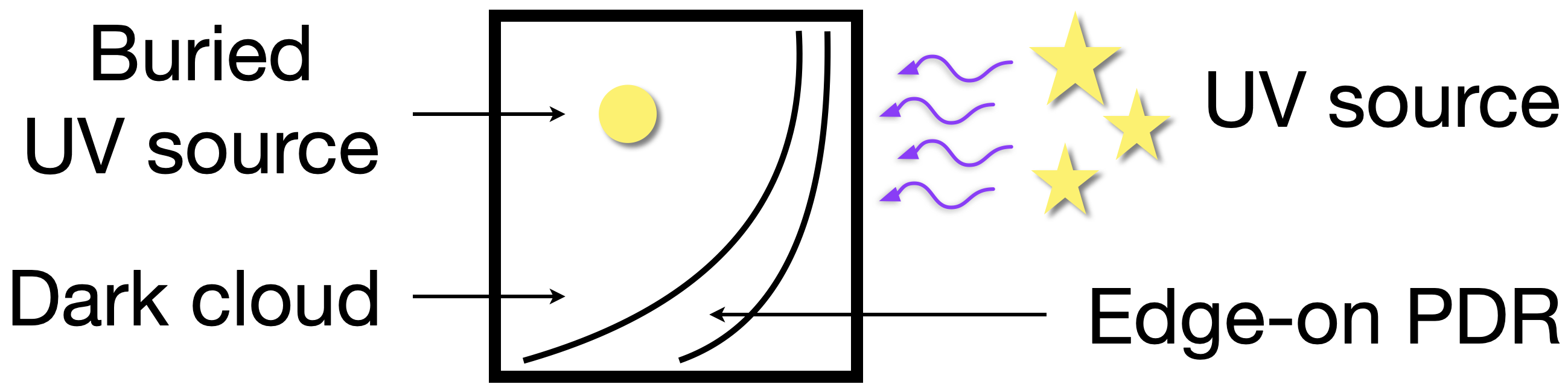}
    \caption{}%
    \label{fig:synthetic_appli_setup:cloud_struct}
  \end{subfigure}
  \begin{subfigure}{1\linewidth}
    \centering
    \includegraphics[width=0.75\linewidth]{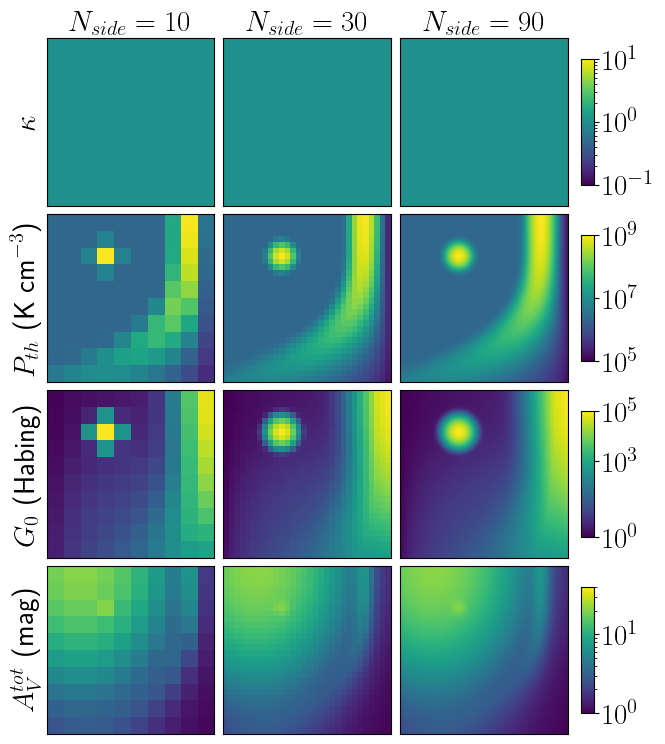}
    \caption{}%
    \label{fig:synthetic_appli_setup:true_params}
  \end{subfigure}
  \begin{subfigure}{1\linewidth}
    \centering
    \includegraphics[width=0.75\linewidth]{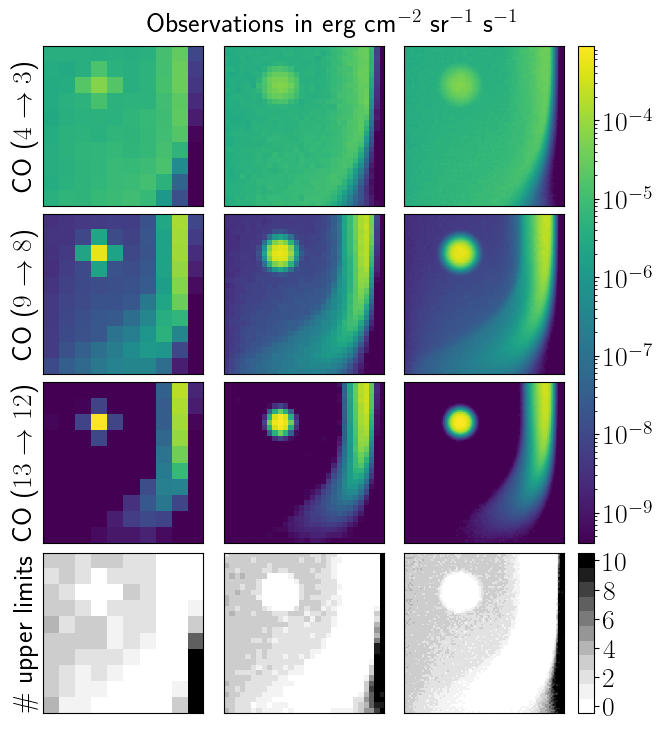}
    \caption{}%
    \label{fig:synthetic_appli_setup:obs}
  \end{subfigure}
  \caption{
    Synthetic case set up.
      (a)~Structure of the synthetic cloud.
      (b)~True physical parameter maps at resolution $N_\text{side} = 10$, $30$, and $90$ (from left to right).
      The rows show the maps of the scaling factor $\scalingParam$,
      the thermal pressure $\Pth$,
      the intensity of the radiation field $\Guv$,
      and the visual extinction $\Av$.
      (c)~Simulated observation maps of three out of the ten considered lines, and map of number of lines with upper limits per pixel.
  }%
  \label{fig:synthetic_appli_setup}
\end{figure}

This section describes the maps of true physical parameter $\paramfull^* \in \R^{N \times D}$ and of simulated observations $\obsfull \in \R^{N \times L}$. 
Figure~\ref{fig:synthetic_appli_setup:cloud_struct} shows the structure of the considered fictitious cloud, representative of a plausible astrophysical scenario. 
%
%
It represents a roughly spherical dark cloud, illuminated from the right by nearby massive stars creating a high-UV, high thermal pressure surface layer, as well as a buried source inside the cloud (on the left).


Figure~\ref{fig:synthetic_appli_setup:true_params} presents true physical parameters maps $\paramfull^*$.
%
%
The same scenario is considered at multiple spatial resolutions, with $N_\text{side} = 10$, $30$ and $90$.
The corresponding total number of pixels for each case is $N = N_\text{side}^2$, that is, $100$, $900$ and $8\,100$, respectively.
In this synthetic application, the line intensities in each pixel are taken as the face-on intensities from the model and we do not include any beam dilution, so that the true map of the scaling parameter $\scalingParam$ is set to $1$ everywhere.


From the true maps $\paramfull^*$, the emulator $\truef$ of the Meudon PDR code generates observation maps of $L = 10$ $^{12}$CO emission lines of mid-$J$ rotational transitions, from $J = 4-3$ to $J= 13 - 12$.
These lines are selected because their emission in PDRs observed with \textit{Herschel} SPIRE FTS has been well studied with PDR models before (see, e.g.,~\citealt{wuConstrainingPhysicalConditions2018}).
The physics governing their emission is thus relatively well understood.
For each line $\ell$, the noiseless integrated intensities $\truefell(\paramvect{n})$ range from $10^{-18}$ to $10^{-2}$ erg\,cm$^{-2}$\,s$^{-1}$\,sr$^{-1}$.
As detailed in Sect.~\ref{ssec:noise-model-and-characteristics}, these maps are deteriorated with a Gaussian additive noise and a lognormal multiplicative noise, and can contain upper limits of the observations.
%
The standard deviation of the multiplicative noise is set to $\sigma_m = \log(1.1)$, which roughly represents a $10$\% alteration in average due to calibration errors (model mis-specification noise is not considered in this example).
The additive noise is set constant on the map for all lines to $\sigma_{a,n\ell} = \sigma_{a} = 1.39 \times 10^{-10}$ erg\,cm$^{-2}$\,s$^{-1}$\,sr$^{-1}$, so that the S/N varies between $10^{-8}$ and $10^8$. 
This noise level is typical for CO observations with \textit{Herschel}~\citep{joblinStructurePhotodissociationFronts2018,wuConstrainingPhysicalConditions2018}.

Figure~\ref{fig:synthetic_appli_setup:obs} shows the observation maps of three of the ten considered lines and the maps of number of upper limits considered in each pixel.
For instance, in the harsh conditions (high $\Guv$, low $\Av$) of the rightmost region of the map, no molecule can survive: all ten lines are thus undetected and only provided as upper limits.
Besides, to be bright, the $J = 13 \to 12$ transition requires both a high pressure and a high UV irradiation for $^{12}$CO to form in a layer warm enough to populate such an excited rotation level.
This condition is only met in the edge-on PDR and around the buried source.
In the rest of the map, this line only has an upper limit.
Similarly, in the deep cloud region and far from the buried source, $\Pth$ and $\Guv$ are low ($\Pth \sim 10^7$ K cm$^{-3}$, $\Guv \lesssim 10^1$), causing the two or three line intensities with the highest energies to be upper limits in this region.

\begin{figure*}
  \centering
  \includegraphics[width=0.7\linewidth]{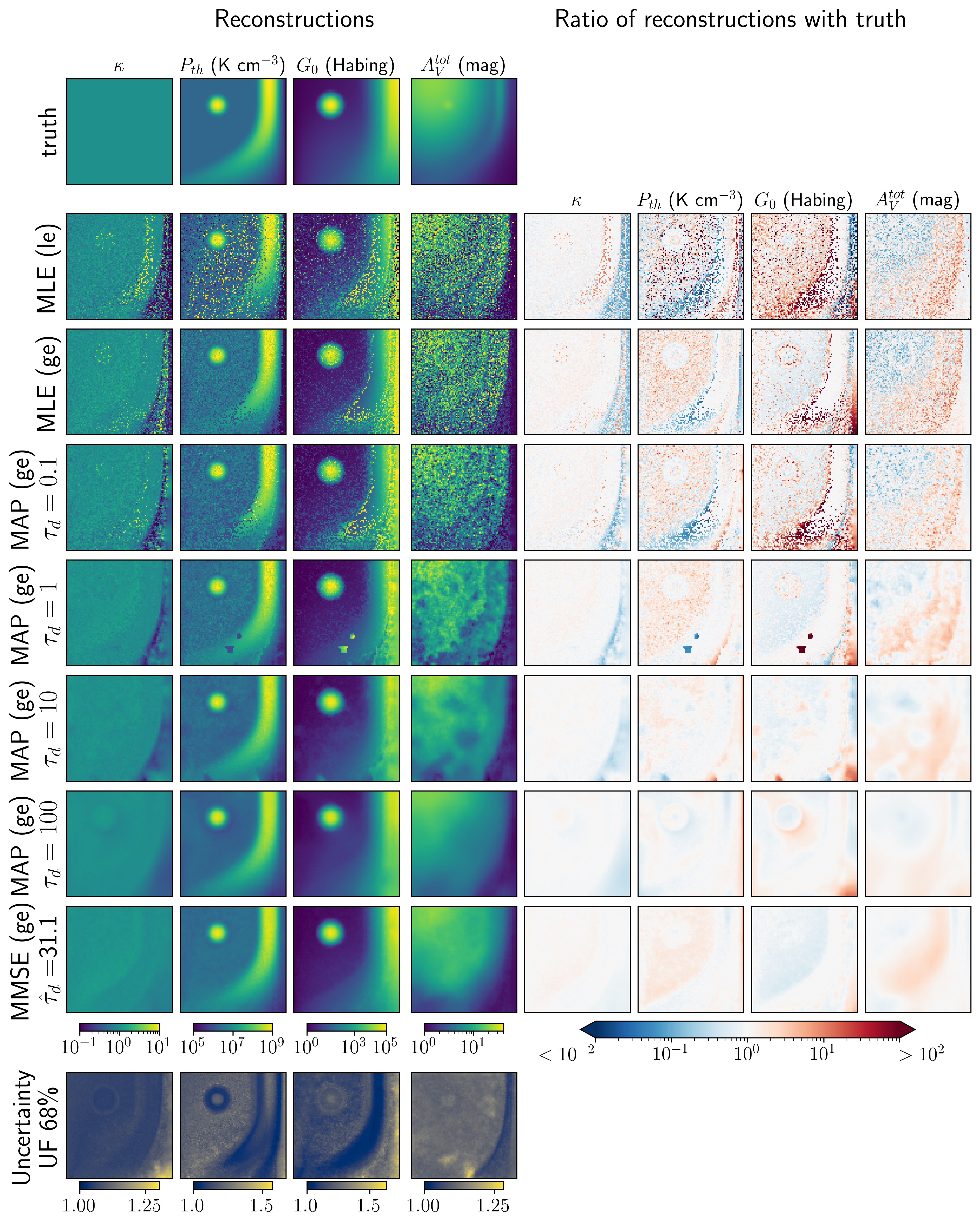}
  \caption{
    Reconstruction results for the $N_\text{side} = 90$ spatial resolution.
    The first row shows the $D=4$ true maps that each estimator tries to reconstruct.
    The obtained reconstructions are shown below the true maps.
    The first estimator is ``MLE (le)'': the MLE evaluated using the PMALA kernel only, that is, only local exploration.
    The second estimator is ``MLE (ge)'': the MLE evaluated using both the PMALA and MTM kernels, that is, a combination of local and global exploration.
    The third to fifth estimators are MAP estimators (i.e., including spatial regularization) with different regularization weight $\tau_d$ values.
    The sixth estimator is the MMSE, obtained with sampling.
    The associated regularization weight $\tau_d$ is inferred along with the physical parameters $\paramfull$.
    The last row, ``UF 68\%'', quantifies the uncertainties associated with the MMSE by indicating the size of the 68\% CI.
    The maps on the right display the ratios between the estimated and true maps, to better assess the quality of each estimation.
  }%
  \label{fig:synthetic_case_estimation_90}
\end{figure*}

\subsection{Comparison metrics}%
\label{sssec:comparison-metrics}

We assess the performance of the reconstruction methods in terms of speed and quality of the resulting estimate.
The reconstruction quality is evaluated with the error factor (EF), introduced in~\citet{paludNeuralNetworkbasedEmulation2023}, computed for each estimated physical parameter on each pixel $\parameltEst{nd}$ as
\begin{align}
  \label{eq:error-factor}
  \text{EF}\left(
    \paramelt{nd}^*, \parameltEst{nd}
  \right)
  =
  10^{\left\vert \logd \paramelt{nd}^* - \logd \parameltEst{nd} \right\vert}
  =
  \max \left\{
    \frac{\paramelt{nd}^*}{\parameltEst{nd}}, \;
    \frac{\parameltEst{nd}}{\paramelt{nd}^*}
  \right\}
  .
\end{align}
Similarly to the UF for uncertainty quantification, the EF provides a multiplicative description of the error in linear scale.
This is more relevant than using a distance in linear scale, as the range of a physical parameter typically covers many orders of magnitude.
We also find it more intuitive than an error in logarithmic scale, although equivalent.
For instance, $\text{EF} = 2$ indicates that the reconstruction $\parameltEst{nd}$ has an error of a factor of two with the truth $\paramelt{nd}^*$.
A perfect reconstruction corresponds to $\text{EF} = 1$.
%
We quantitatively compare estimators with the mean EF, averaged over each physical parameter map.

The inference speed is quantified with the complete runtime of each inference procedure and with the mean runtime per iteration.
The complete runtime allows to assess the time required to run an estimation with \textsc{Beetroots}.
The mean runtime per iteration provides a fair comparison of speed.
%

\subsection{Discussion of experimental results}%
\label{sssec:results-comparison}

\setlength{\tabcolsep}{4pt}
\renewcommand{\arraystretch}{1.03}
\begin{table}[t]
  \centering
  \caption{
    Performance of inference methods on a synthetic case.
  }%
  \label{tab:results_synthetic}
  \resizebox{\linewidth}{!}{%
  \begin{tabular}{ccc|cccc|cc}
    & \multirow{2}{*}{estimator} & \multirow{2}{*}{$\tau_d$} & \multicolumn{4}{c|}{mean error factor} & \multicolumn{2}{c}{runtime} \\
    &           &          & $\scalingParam$ & $\Pth$ & $\Guv$ & $\Av$ & total & per iter. \\
    \hline
    \multirow{7}{0.5em}{\rotatebox[origin=c]{90}{$N_\text{side} = 10$}}
    & MLE (le) & -
    & 1.9 & 31.1 & 56.9 & 3.1
    & 4m 09s& 0.06s \\
    & MLE (ge) & -
    & 1.6 & 2.3 & 15.6 & 2.8
    & 5m 11s& 0.06s \\
    & MAP (ge) & $10^{-1}$
    & 1.4 & 1.8 & 11.8 & 2.1
    & 5m 24s& 0.07s \\
    & MAP (ge) & $10^0$
    & \textbf{1.1} & 1.4 & \textbf{1.4} & \textbf{1.3}
    & 5m 27s& 0.07s \\
    & MAP (ge) & $10^1$
    & \textbf{1.1} & \textbf{1.3} & \textbf{1.4} & 1.5
    & 5m 21s& 0.06s \\
    & MAP (ge) & $10^2$
    & 1.5 & 4.7 & 2.2 & 4.8
    & 4m 50s& 0.06s \\
    & MMSE (ge) & 1.1
    & \textbf{1.1} & 1.5 & 1.9 & \textbf{1.3}
    & 27m 49s & 0.17s \\[0.2em]
    \hline
    \multirow{7}{0.5em}{\rotatebox[origin=c]{90}{$N_\text{side} = 30$}}
    & MLE (le) & -
    & 2.3 & 20.2 & 1\,231 & 2.9
    & 24m 17s & 0.29s \\
    & MLE (ge) & -
    & 1.7 & 2.4 & 31.4 & 2.9
    & 29m 48s & 0.36s \\
    & MAP (ge) & $10^{-1}$
    & 1.4 & 2.0 & 9.5 & 2.1
    & 31m 45s & 0.38s \\
    & MAP (ge) & $10^0$
    & \textbf{1.2} & 1.7 & 3.2 & 1.6
    & 31m 21s & 0.38s \\
    & MAP (ge) & $10^1$
    & \textbf{1.2} & \textbf{1.3} & \textbf{1.3} & \textbf{1.3}
    & 34m 14s & 0.41s \\
    & MAP (ge) & $10^2$
    & \textbf{1.2} & 1.5 & 1.5 & 1.5
    & 32m 14s & 0.39s \\
    & MMSE (ge) & 4.7
    & \textbf{1.2} & \textbf{1.3} & 1.4 & \textbf{1.3}
    & 2h 09m 44s & 0.78s \\[0.2em]
    \hline
    \multirow{7}{0.5em}{\rotatebox[origin=c]{90}{$N_\text{side} = 90$}}
    & MLE (le) & -
    & 2.3 & 21.4 & 1\,223 & 3.0
    & 3h 01m 23s & 2.18s \\
    & MLE (ge) & -
    & 1.7 & 2.4 & 22.8 & 2.8
    & 4h 53m 46s & 3.53s \\
    & MAP (ge) & $10^{-1}$
    & 1.5 & 2.6 & 32.8 & 2.1
    & 6h 10m 57s & 4.45s \\
    & MAP (ge) & $10^{0}$
    & 1.3 & 1.6 & 5.5 & 1.6
    & 7h 13m 01s & 5.20s \\
    & MAP (ge) & $10^{1}$
    & 1.2 & \textbf{1.3} & 1.5 & 1.4
    & 7h 30m 36s & 5.41s \\
    & MAP (ge) & $10^{2}$
    & 1.2 & \textbf{1.3} & \textbf{1.3} & \textbf{1.2}
    & 6h 15m 13s & 4.50s \\
    & MMSE (ge) & 31.1
    & \textbf{1.1} & \textbf{1.3} & \textbf{1.3} & 1.3
    & 28h 43m 02s & 10.34s
  \end{tabular}
  }
  %
  \tablefoot{
    le: local exploration, that is, with MTM kernel disabled ($\pmtm = 0$).
    ge: global exploration, that is, with MTM kernel enabled ($\pmtm > 0$).
    In each case, the number of pixels in each map is $N = N_\text{side}^2$.
    The three considered cases thus contain $N=100$, $900$ and $8\,100$ pixels, respectively.
    For the MMSE (ge), the regularization weight $\tau_d$ is estimated along with the physical parameters $\paramfull$.
    For each spatial resolution and physical parameter, the lowest mean error factor is highlighted in bold.
  }
\end{table}

Optimization algorithms are run for $\TMC = 5\,000$ steps, and the sampling method for $\TMC = 10\,000$ iterations.
Since the goal of sampling -- for the evaluation of the MMSE (ge) -- is to approximate the posterior distribution instead of only reaching a mode, significantly more iterations are required.
In all cases involving the MTM kernel, its selection probability is set to $\pmtm = 20$\% and the number of candidates $K$ to $25$.

Table~\ref{tab:results_synthetic} compares the performance of the considered estimators for all spatial resolutions.
Figure~\ref{fig:synthetic_case_estimation_90} shows the estimation results on the $N_\text{side} = 90$ case.
The reconstructed maps for the $N_\text{side} = 10$ and $30$ cases are displayed in Appendix~\ref{ssec:estimation-results-for-more-spatial-resolutions}.

\subsubsection{Maximum likelihood: Effect of global exploration}%
\label{sssec:maximum-likelihood-effect-of-global-search}

This section analyses the two presented evaluations of the MLE, shown in the second and third rows in Fig.~\ref{fig:synthetic_case_estimation_90}.
The first evaluation, ``MLE (le)'', only exploits the PMALA kernel, that is, local exploration.
The second evaluation, ``MLE (ge)'', uses both PMALA and MTM kernels, that is, local and global exploration.
The ``MLE (le)'' leads to a very poor reconstruction for all spatial resolutions.
It yields mean EFs of up to about $10^2$ for the thermal pressure and about $10^4$ for the intensity of the radiation field, that is, the estimation misses the true values in average by a factor of $10^2$ for $\Pth$ and of $10^4$ for $\Guv$%
\footnote{
  The validity interval of $\Guv$, $[10^0, 10^5]$, covers five orders of magnitude.
  In contrast, the validity interval of $\Pth$, $[10^5, 10^9]$ K cm$^{-3}$, only covers four,
  that of $\scalingParam$, $[10^{-1}, 10^1]$, only two and that of $\Av$, $[1, 40]$ mag, less than two.
  In addition, $\Guv$ has the largest spatial variations in the true maps -- see Fig.~\ref{fig:synthetic_appli_setup:true_params}.
  The mean EF of $\Guv$ is therefore expected to be the largest for most estimators.
}%
.
Although they display the main regions of the synthetic cloud, the reconstructed maps contain many uphysical large variations and are thus not exploitable for physical interpretation.
%
%
With these reconstructions and only one initialization, it is difficult to assess whether poor reconstructions missed the global minimum or whether the model itself defines an unphysical global minimum.

Using global exploration answers this question: with only one initialization, the maps obtained with ``MLE (ge)'' 
%
miss the true values in average by a factor of up to $2.4$ for $\Pth$ and of up to $31$ for $\Guv$, which represents a significant improvement.
It is safe to assume that this reconstruction procedure attains the global mode of the likelihood function.
However, some pixels still have unphysical values for $\Pth$ and $\Guv$ because of an identifiability issue.
Besides, as we only consider rotationally excited $^{12}$CO transitions that originate from a surface layer of the cloud, both MLE evaluations poorly reconstruct $\Av$.
Additional information is thus required to further improve the reconstructions.

\subsubsection{Effect of spatial regularization}%
\label{sssec:effect-of-spatial-regularization}

Spatial regularization can provide this additional information.
For large enough regularization weight $\tau_d$, the ``MAP (ge)'' estimator systematically produces lower mean EF than the ``MLE (ge)''.
For $\Pth$ and $\Guv$, for each pixel, it selects the mode that is consistent with the pixel observations and with those in its neighbors.
In addition, it compensates the likelihood inability to reconstruct $\Av$ by returning smooth maps that are physically exploitable.
However, the regularization weight needs careful tuning.
On the one hand, using too low regularization weight can yield artifacts in the reconstructions, due to a local minimum.
This happened for instance in the $N_\text{side} = 90$ case with $\tau_d = 1$, with two sets of neighboring pixels staying in a local minimum (visible in the bottom middle part of the map as a small contiguous group of high $\Guv$, low $\Pth$ pixels contrasting with the surrounding pixels).
%
%
%
%
%
%
%
On the other hand, in the $N_\text{side} = 10$ case where true maps are not smooth, using a too large regularization weight $\tau_d$ biases the estimations.
It erases the buried source, and under-estimates the spatial variations of $\Pth$ and $\Guv$ in the transition from the edge-on PDR to the deep cloud.
Appendix~\ref{ssec:effect-of-the-regularization-weight} provides more quantitative results on the  regularization weight influence on the mean EF achieved with MAP (ge).

\subsubsection{Sampling-based approach}%
\label{sssec:sampling-based-approach-getting-more-information}

A sampling-based approach provides more information and can further improve estimation.
We find that the MMSE (ge) can be a better estimator than MAP (ge), that is, with lower mean EF.
This is particularly true for loosely constrained observations, such as the rightmost region of the map that contains mostly upper limits of the considered $^{12}$CO transitions.
It is also true for loosely constrained physical parameters such as the visual extinction $\Av$ in the deep molecular cloud, or the scaling parameter $\scalingParam$ in the rightmost part of the maps.
As the MMSE (ge) is the mean of the posterior samples (i.e., that explain the observations while respecting the prior), it can be closer to the true value.
On the $N_\text{side} = 90$ use case, the MMSE (ge) yields a mean EF of only 1.1 for $\scalingParam$, 1.3 for $\Pth$, 1.3 for $\Guv$ and 1.3 for $\Av$, that is, mean errors of at most 30\%.
Achieving such low errors is remarkable, especially considering the variety of environments and of S/N in the considered scenario.

In addition to the MMSE, the sampling approach quantifies the uncertainty associated with the estimation.
The obtained CIs and UFs are consistent with the physical intuition.
In all three cases, the UFs of $\scalingParam$, $\Pth$ and $\Guv$ are highest in the rightmost part of the maps.
%
%
In particular, it only yields upper bounds on $\scalingParam$ and $\Av$, and a lower bound on~$\Guv$.
Beyond these obtained bounds, most molecules are photodissociated, which explains the observation upper limits.
For the deep molecular cloud (on the left), the reconstruction only yields lower bounds on $\Av$ as the considered lines only trace the surface of the cloud.
For the PDR (on the right) and the buried source (on the left), the observations are in high S/N and do not include upper limits on observations.
The reconstructions of these two regions are thus associated with the smallest UFs for $\Pth$ and $\Guv$.
Finally, $\scalingParam$ is well reconstructed in all three cases when at least one transition is in a high S/N regime.
The estimated values are indeed close to the true value (i.e., no bias), and the UFs are close to one (i.e., very small uncertainty) everywhere but in the rightmost part of the maps.
%
%

The sampling-based approach allows for a model checking accounting for as many uncertainty sources as possible.
Appendix~\ref{ssec:sampling-approach-model-checking-results} provides the results of the model checking implemented in \textsc{Beetroots} applied to these three synthetic cases.
Here, this assessment confirms that the model is fully compatible with the observations, as expected for these synthetic observations.

This application shows that \textsc{Beetroots} produces significantly more accurate estimators compared to the usual MLE approach, both in optimization (with MAP (ge) estimator) and in sampling (with MMSE estimator).
The automatically tuning of the regularization weight prooves to yield relevant values.
%
%
The uncertainty quantification results are interpretable and consistent with the physics of the considered $^{12}$CO rotational emission lines.

\subsubsection{Comparison of inference runtimes}%
\label{ssec:comparison-of-duration}

Table~\ref{tab:results_synthetic} shows the total inference runtime for each estimation.
It shows that for small maps (with up to $10^3$ pixels), running an optimization (MLE or MAP) or a sampling procedure takes less than two hours.
It shows that \textsc{Beetroots} can analyse most past observations similar to this synthetic case.
For larger maps with about $10^4$ pixels, reconstructions are longer: from 3 to 8 hours for optimization, and about 28 hours for sampling.
As inference runtime does not scale linearly, we believe that \textsc{Beetroots} currently can manage maps of up to $N \simeq 30\,000$ pixels.

The runtime per iteration provides a more fair comparison of the actual speed of each estimation, as sampling was run for twice as many iterations as optimization.
As PMALA only considers one candidate at each iteration, a local exploration only (le, with only PMALA) is faster than a global exploration (ge, combining MTM and PMALA).
Because of the way the spatial regularization is used to generate candidates in MTM, larger values of $\tau_d$ lead to faster iterations (this effect should be removed in a future version).
Finally, a sampling iteration is slower than an optimization iteration, as it requires more intermediate steps.




\section{Application to Orion molecular cloud 1 (OMC-1)}%
\label{sec:application-to-real-data}

In the previous section, we demonstrated the potential of \textsc{Beetroots} on three synthetic cases.
In this section, we apply our inference procedure on $\sim$7.5' $\times$ $\sim$11.5' far-infrared and (sub)millimeter line emission maps of  Orion molecular cloud 1 core (OMC-1 core) described in~\citet{goicoecheaMolecularTracersRadiative2019}.
To the best of our knowledge, this inference is the first performed on a large map (that is, with $1\,000$ pixels or more) for multiple physical parameters from molecular tracers.
It is also the first time that one quantifies the uncertainties in an inference of maps of physical parameters from molecular tracers.
Unlike the synthetic applications, it is here not clear a priori whether the observed lines are sufficient to constrain the physical parameters.

First, we describe the region and the observation map.
Then, we set up the inversion procedure.
The results are discussed and compared with the literature, including the analysis of the Orion Bar from~\citet{joblinStructurePhotodissociationFronts2018}.
Third, we compare our estimates of
$\Guv$ with those derived by
~\citet{goicoecheaVelocityresolvedCIIEmission2015}
from total far infrared (FIR) luminosities.
%
%
Finally, we explore the relationship between $\Guv$ and $\Pth$ that appears in our results.

\subsection{Cloud, observation and noise description}%
\label{ssec:cloud-and-observation-description}

\begin{figure}[t]
  \centering
  \begin{subfigure}{0.47\linewidth}
    \centering
    \includegraphics[width=0.95\linewidth]{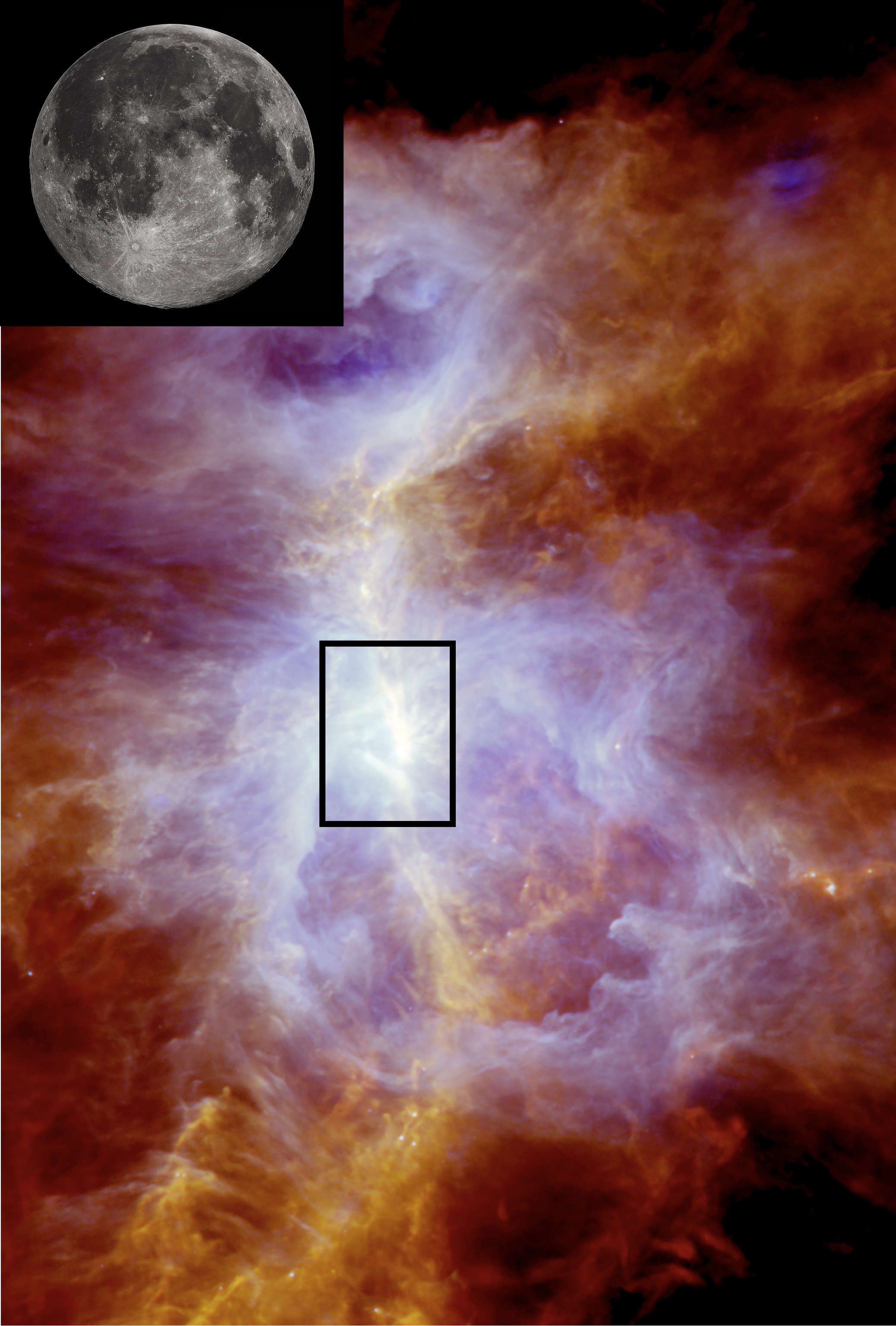}
    \caption{}%
    \label{fig:omc1_struct:obs}
  \end{subfigure}
  \begin{subfigure}{0.47\linewidth}
    \centering
    \includegraphics[width=0.95\linewidth]{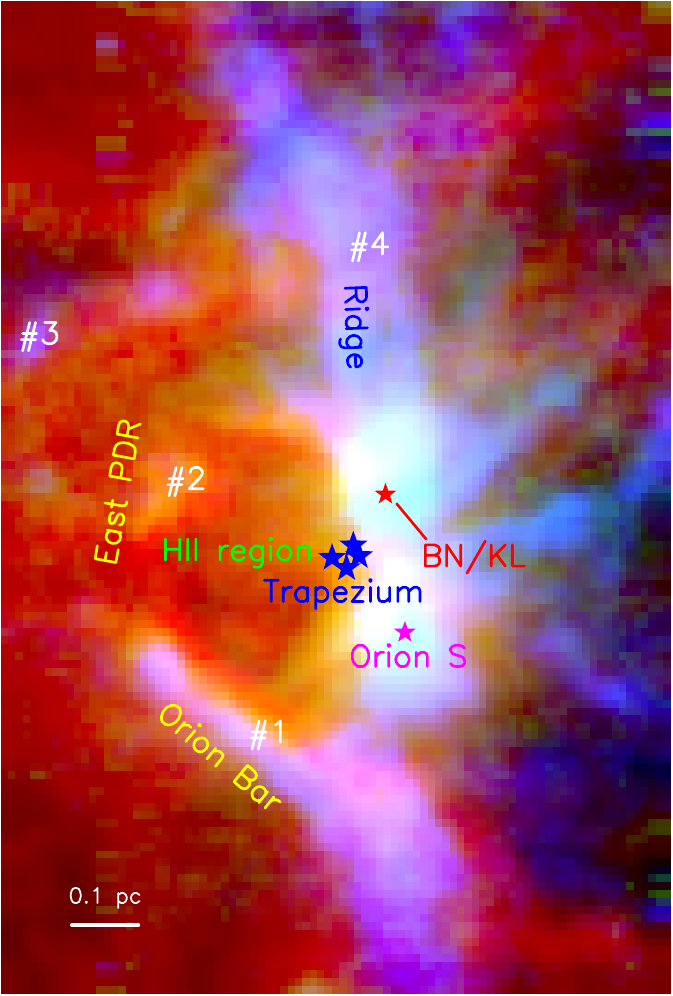}
    \caption{}%
    \label{fig:omc1_struct:annotated}
  \end{subfigure}
  \caption{
    Structure of
    OMC-1.
    %
    (a) The Orion A star-forming cloud seen by \textit{Herschel} with dust thermal emission.
    The black rectangle outlines the observed region of OMC-1, centered on the Trapezium star cluster.
    The image is a composite of $70 \microm$ (blue), $160 \microm$ (green) and $250 \microm$ (red) emission.
    It spans about $1.3 \times 1.6$ deg$^2$.
    The moon is shown for scale. 
    %
    %
    %
    %
    (b) OMC-1 composite image covering $\sim 85$ arcmin$^2$ at $\sim 12$'' with three emission lines:
    the [CII] $158 \microm$ line which traces the FUV-illuminated surface of the molecular cloud (red),
    the C$^{18}$O, $J = 2 \to 1$ line which traces cold dense gas (blue) and
    the HCO$^+$ $J=3 \to 2$ line (green).
    Reconstruction results will be detailed for positions 1 to 4.
    Adapted from~\citet{goicoecheaMolecularTracersRadiative2019}.
  }%
  \label{fig:omc1_struct}
\end{figure}

OMC-1
is a bright region of the Orion A cloud.
It is located at about 414\pc\ from the Sun~\citep{mentenDistanceOrionNebula2007}, making it the closest region of intermediate- and high-mass star formation.
%
Figure~\ref{fig:omc1_struct} shows its position in Orion A and its general structure.
The surface of OMC-1 is illuminated by the Trapezium star cluster, which contains a few massive stars.
%
%
The FUV radiation from these massive stars heats and photodissociates the molecular surface of OMC-1, at the border of the central H\textsc{ii} region.
The Orion Bar and the East PDR, first reported by~\citet{goicoecheaVelocityresolvedCIIEmission2015}, are prominent examples of dense PDRs located at this interface.
%
The two major star-forming sites are the Becklin-Neugebauer/Kleinmann-Low region (BN/KL) and Orion South (Orion S).
Sect.~\ref{sssec:analysis-of-highlighted-pixels} will detail reconstruction results for the positions highlighted in Fig.~\ref{fig:omc1_struct:annotated}.

The observation $\obsfull$ contains $N = 2\,475$ pixels and $L = 5$ molecular emission lines:
CH$^+$ $(J=1 \to 0)$,
CO $(J=2 \to 1)$,
CO $(J=10 \to 9)$,
$^{13}$CO $(J = 2 \to 1)$, and
HCO$^+$ $(J=3 \to 2)$.
%
A [CII] $158 \microm$ emission map was also presented in~\citet{goicoecheaMolecularTracersRadiative2019} but was left aside in this study.
Indeed, our attempts at simultaneously fitting [CII] with the molecular lines resulted in an incompatibility between the models and observations -- identified with our model checking procedure.
%
%
This incompatibility when trying to simultaneously fit [CII] emission and warm molecular emission with the Meudon PDR code was also observed in~\citet{joblinStructurePhotodissociationFronts2018} in the Orion Bar, with a systematic underestimation of the [CII] intensity, which was proposed to be caused by the larger beam filling factor of [CII] emission in Herschel observations compared to warm molecular tracers that originate in denser filamentary structures.
%
%
%

\begin{figure}[t]
  \centering
    \includegraphics[height=9.91em]{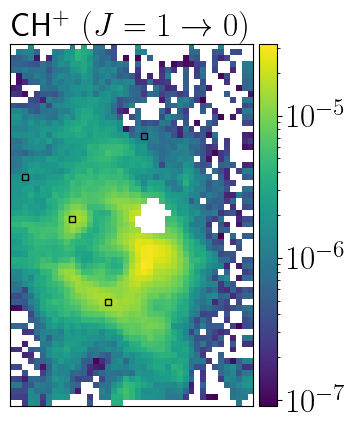}
    \includegraphics[height=9.91em]{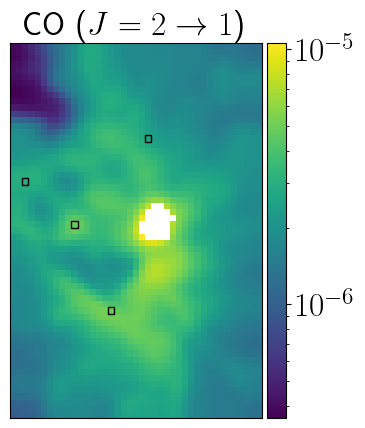}
    \includegraphics[height=9.91em]{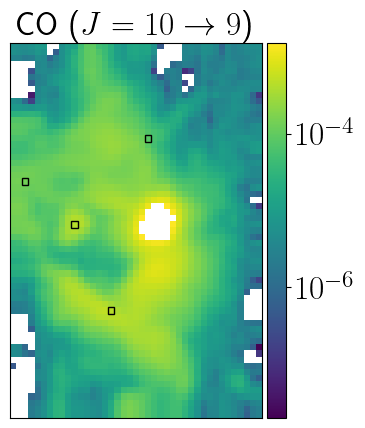} \\
    \includegraphics[height=9.91em]{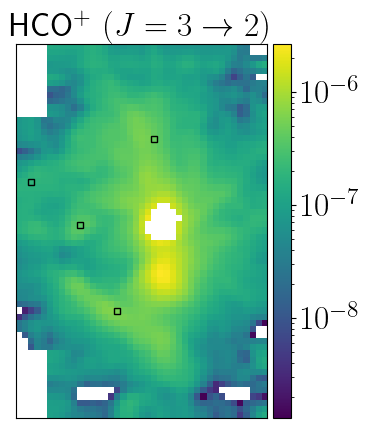}
    \includegraphics[height=9.91em]{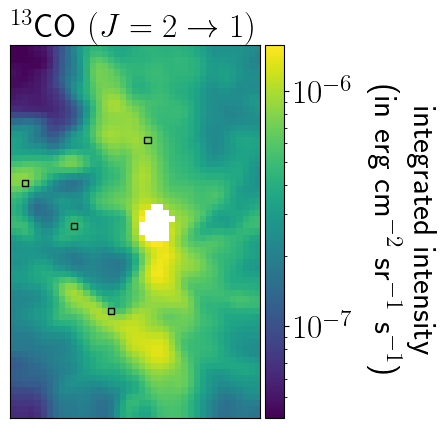}
  \caption{
    OMC-1 core observed integrated intensity maps of the lines used for inversion. 
    The white region is the middle of the maps is the BN/KL region, dominated by shocks, and not considered in the inversion.
    The remaining white pixels correspond to negative integrated intensities due to the additive Gaussian noise.
    The inference results for the four pixels highlighted with a black square are detailed in Sect.~\ref{sssec:analysis-of-highlighted-pixels}.
  }
  \label{fig:omc1_observations}
\end{figure}

Figure~\ref{fig:omc1_observations} shows the observation maps associated with the five lines used in the inversion.
The associated STD maps are displayed in Appendix~\ref{ssec:standard-deviation-maps-of-the-additive-noise}.
Four pixels are highlighted in this figure, corresponding to the Orion Bar PDR (lowest square, labelled \#1 in Fig~\ref{fig:omc1_struct:annotated}), the East PDR (mid-height square, labelled \#2), the North-East border of the map (top left square, labelled \#3) and the North-West ridge (top right square, labelled \#4).
Results for these pixels are detailed in Sect.~\ref{sssec:analysis-of-highlighted-pixels}.
The BN/KL region is masked (in white, in the center of the maps).
Since its emission contains a major contribution from shocks, and since such environments are not described by PDR models, reconstruction results on this region would not be relevant.
Dedicated models such as the Paris-Durham code~\citep{godardModelsIrradiatedMolecular2019} would better simulate this region.
Removing this region avoids biasing the neighboring pixels with the spatial regularization.
%

\subsection{Inversion setup}%
\label{ssec:inversion-setup}

The integrated intensity maps $\obsfull$ and the associated additive noise STDs $(\sigma_{a,n\ell})_{n\ell}$ are computed directly from the hyperspectral cubes.
The multiplicative noise is modelled as a lognormal distribution that combines two terms.
As indicated in Sect.~\ref{ssec:noise-model-and-characteristics}, the model mis-specification STD $\sigma_\text{model}$ is set so that a $3\sigma$ error corresponds to an error of a factor of three, that is, $\sigma_\text{model} = \frac{1}{3} \log 3$. 
An estimated 8\% calibration error is considered for the whole map%
\footnote{
  See~\citet{roelfsemaInorbitPerformanceHerschel2012} for the calibration error of \textit{Herschel}-HIFI and \url{https://www.iram.fr/GENERAL/calls/s23/30mCapabilities.pdf} for that of the IRAM-30m.
}, that is, $\sigma_c = \log 1.08$.
The total multiplicative STD is $\sigma_m = ( \, \sigma_\text{model}^2 + \sigma_c^2 \, )^{0.5} \simeq \log 1.452$. 
%
%
The angle $\paramAngle$ between the cloud surface and the line of sight varies in the observed map.
For instance, the Orion S region is considered to be mostly face-on ($\paramAngle \sim 0$ deg) while the Orion Bar is edge-on ($\paramAngle \simeq 90$ deg).
To avoid inferring more parameters, we set $\paramAngle = 0$ deg and assume inclination effects to be captured by the scaling factor $\scalingParam$.

We compare two estimators, namely ``MLE (ge)'' and ``MMSE (ge)'', both evaluated with \textsc{Beetroots}.
The MLE (ge) (Sect.~\ref{sssec:optimization-based-estimation}) is obtained with the optimization algorithm described in Sect.~\ref{sssec:estimation-methods-in-beetroots}, run for $5\,000$ iterations.
The preconditioned gradient descent update (i.e., local exploration) step size is set to $\stepsize = 0.01$.
The MMSE (ge) is evaluated from a Markov chain of $\TMC = 10^4$ iterates, including $\TBI = 2\,000$ of burn-in.
%
%
The PMALA step size is set to $\stepsize = 0.1$.
%
For both estimators and at each iteration, MTM (i.e., global exploration) has a probability $\pmtm = 0.2$ of being used, and generates $K = 25$ candidates per pixel.

\subsection{Inference results}%
\label{ssec:inference-results}

\begin{figure}
  \centering
    \includegraphics[height=12em]{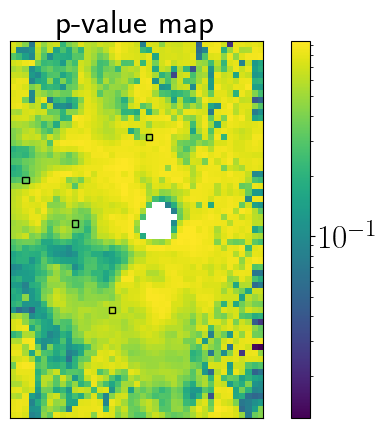}
    %
  %
    \includegraphics[height=12em]{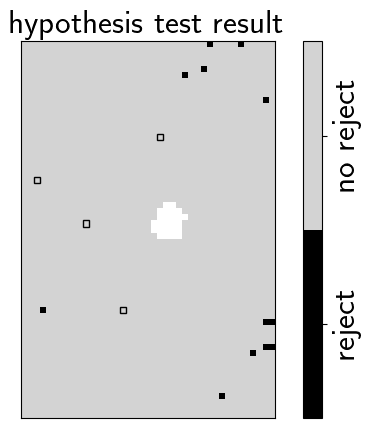}
    %
  %
  \caption{
    Model checking for OMC-1.
    (left) Estimated $p$-value map.
    (right) $p$-value-based decision for a 95\% confidence level.
    Only twelve pixels (i.e., 0.5\%) are rejected, and are not in the regions of interest.
    %
    %
    Inference results are thus trustworthy.
  }
  \label{fig:omc1_pval}
\end{figure}

\begin{figure*}
  \centering
      \includegraphics[width=0.79\linewidth]{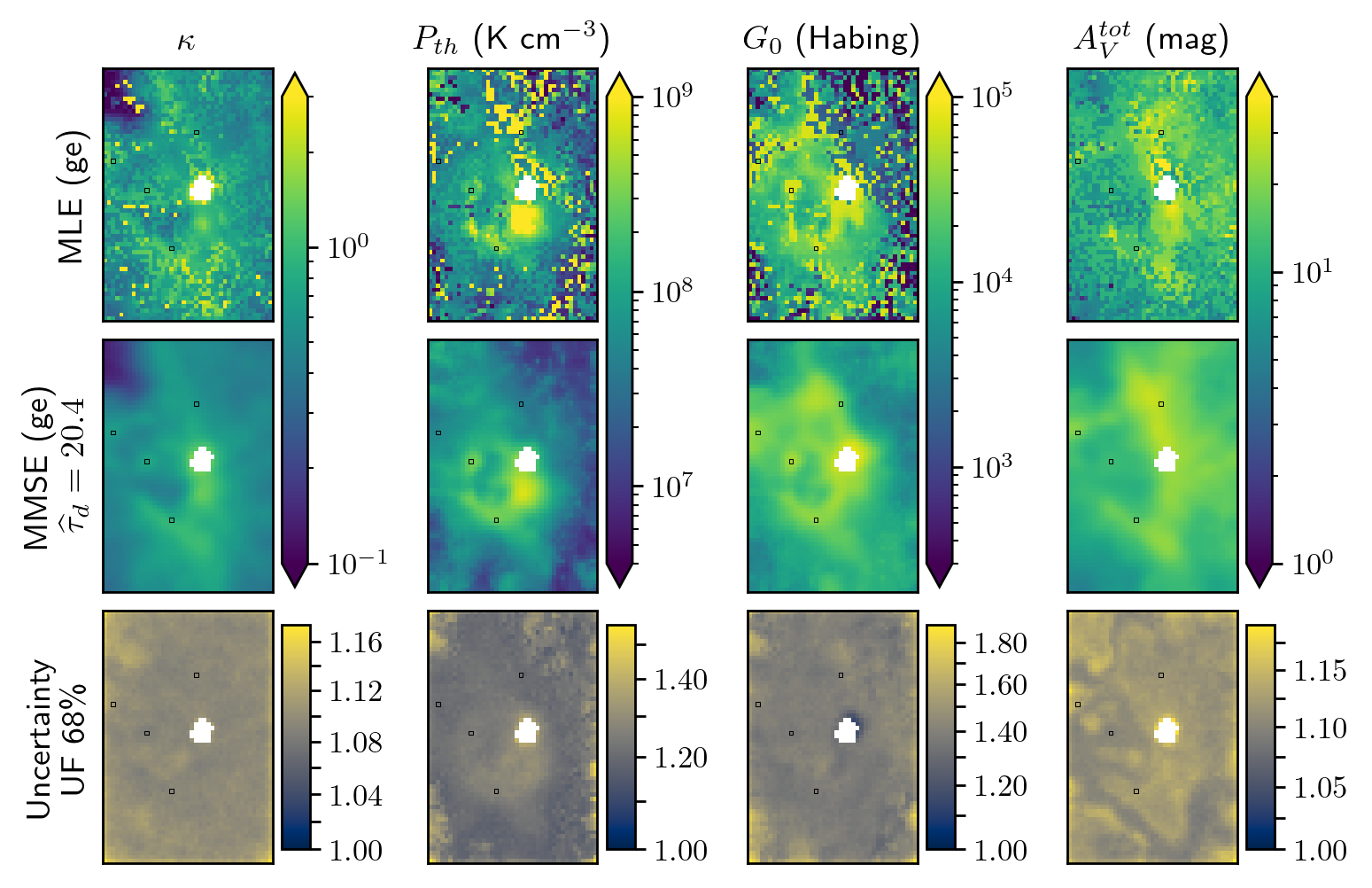}
  \caption{
    Inference results for OMC-1.
    From top to bottom, the rows show the maximum likelihood estimator (MLE), the MMSE estimator, and then the maps of uncertainty factor (UF) for the 68\% credibility interval (CI).
    From left to right, the rows show results on the scaling factor $\scalingParam$, thermal pressure $\Pth$, intensity of the radiation field $\Guv$, and the visual extinction $\AV$.
  }
  \label{fig:omc1_inference_results}
\end{figure*}

This section presents the inference results.
After confirming that the Meudon PDR code successfully reproduces the observations -- and thus that the reconstructed physical parameter maps are trustworthy --, we describe the reconstructed maps and explore the results more in detail for four selected pixels.

\subsubsection{Model checking and Bayesian $p$-value}%
\label{sssec:model-checking-and-Bayesian-p-value}

Before presenting the reconstructions themselves, we assess their relevance with the model checking approach presented in Sect.~\ref{ssec:methods-model-checking}.
As a reminder, this approach computes a $p$-value for each pixel to test the hypothesis H\ref{hyp:first}, that is ``the astrophysical model, the random deterioration and the prior can reproduce the observations $\obsfull$''.
Figure~\ref{fig:omc1_pval} shows the map of estimated $p$-values.
%
%
We find that the estimated $p$-values are larger than 0.5 in all the regions of interest of the maps.
As a result, only 12 pixels out of $2\,475$ -- about 0.5\% -- lead to a model rejection.
In particular, H\ref{hyp:first} is not rejected for any of the four highlighted positions.
This indicates that the reconstructions to be presented next are trustworthy.
One might expect the proportion of rejected pixels to be closer to 5\% when H\ref{hyp:first} is true.
This low rejection rate might be due to the fact that only $L = 5$ lines are used to reconstruct $D = 4$ physical parameters, or to the large considered model mis-specification STD $\sigma_\text{model}$.
%

As mentioned in Sect.~\ref{ssec:methods-model-checking}, this model validation is based on the posterior predictive distribution, that is, the distribution of predicted integrated intensities $\truef(\paramvect{})$ affected by noise~\eqrefp{eq:obs_model_practice}.
%
%
Visualizing this distribution is therefore a powerful diagnosis tool to identify lines that are not well predicted by the astrophysical model.
Appendix~\ref{ssec:model-checking-posterior-predictive-distribution} shows this distribution for the four highlighted positions.

\subsubsection{Inferred parameter maps}
\label{sssec:MLE-MMSE-and-UF-maps}

Figure~\ref{fig:omc1_inference_results} shows inference results for both likelihood minimization (top row) and posterior sampling (middle row).
The spatial structures visible in the observations are recovered with the MMSE when including the spatial smoothness prior, but hardly visible with the MLE.
As there are few lines and thus few constraints for the astrophysical model, the likelihood is very sensitive to noise and the MLE presents unphysical variations.
The spatial regularization thus plays a key role in recovering the correct spatial structure.

We now focus on the MMSE and UF maps.
%
%
The Orion Bar, the \textsc{Hii} region, the Orion S region, the East PDR and the North-West ridge are visible in each of the four reconstructed maps.
The scaling factor $\scalingParam$ is close to $1$ in the bright regions.
In the Orion S region, in the East PDR and in the Orion Bar, the thermal pressure $\Pth$ is estimated at around $3 \times 10^8 \Kpccm$, and the incident UV radiation field intensity $\Gnaught$ at around $4 \times 10^4$.
In these regions, the recovered visual extinction $\AV$ is roughly $10$ mag.
These values are consistent with the literature. 
In particular, our results robustly confirm that the high thermal pressures found in the Orion Bar are widespread in the PDRs of OMC-1~\citep{goicoecheaMolecularTracersRadiative2019}.
Moreover, we will show in Sect.~\ref{ssec:comparison-with-estimation-of-G0-from-far-infrared-luminosities} that the full $\Gnaught$ MMSE map shows very good agreement, both in terms of structure and values, with the map derived by~\citet{goicoecheaVelocityresolvedCIIEmission2015} from FIR luminosities. 

Over the whole map, the UF is lower than 1.2 for the scaling factor $\scalingParam$ and the visual extinction $\AV$.
This means that the typical $1\sigma$ error is at most 20\% for $\scalingParam$ and $\AV$.
Similarly, over most of the map, the UF is lower than 1.5 for the thermal pressure $\Pth$ and lower than 2 for the incident UV radiation field intensity $\Gnaught$.
These four parameters are thus relatively well constrained considering their large validity intervals and the fact that we use only $L = 5$ lines to infer $D = 4$ parameters for each pixel.

\subsubsection{Analysis of the highlighted positions}%
\label{sssec:analysis-of-highlighted-pixels}

\begin{figure*}
  \centering
  \begin{subfigure}{0.49\linewidth}
    \centering
    \includegraphics[width=0.8\textwidth]{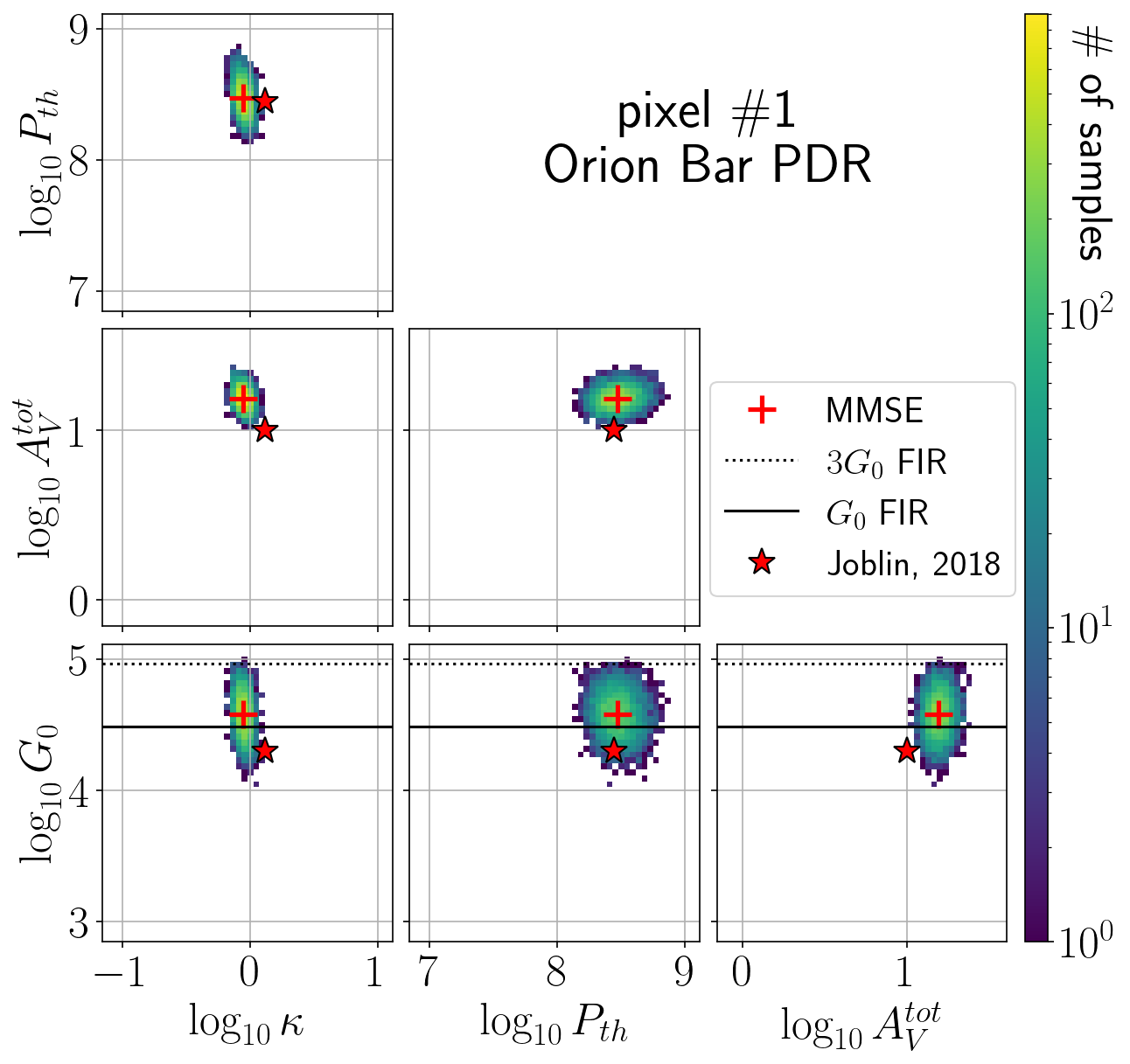}
    \caption{}
    \label{fig:omc1_hists_orionbar}
  \end{subfigure}
  \begin{subfigure}{0.49\linewidth}
    \centering
    \includegraphics[width=0.8\textwidth]{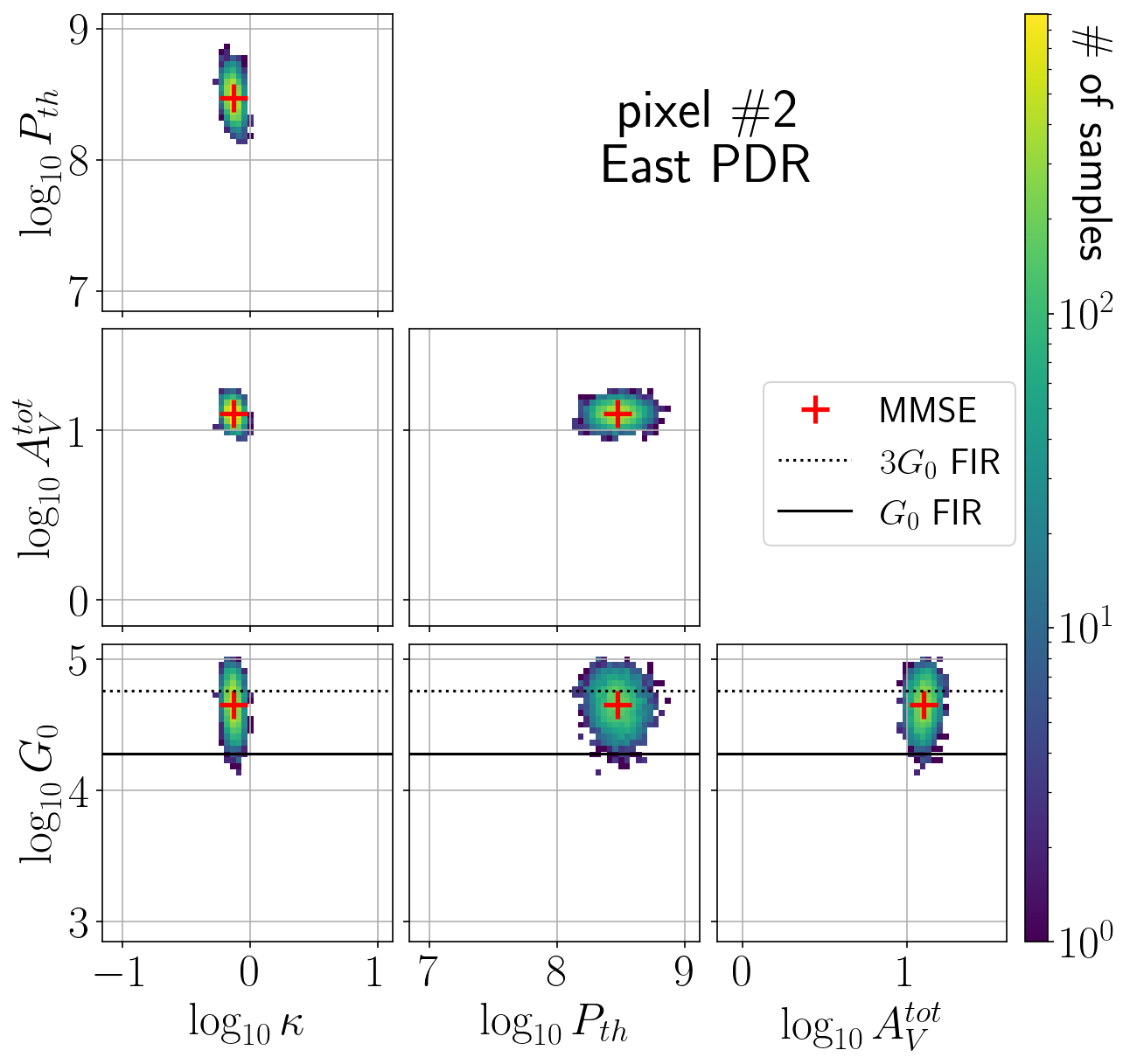}

    \caption{}%
    \label{fig:omc1_hists_east_pdr}
  \end{subfigure}
    \begin{subfigure}{0.49\linewidth}
      \centering
      \includegraphics[width=0.8\textwidth]{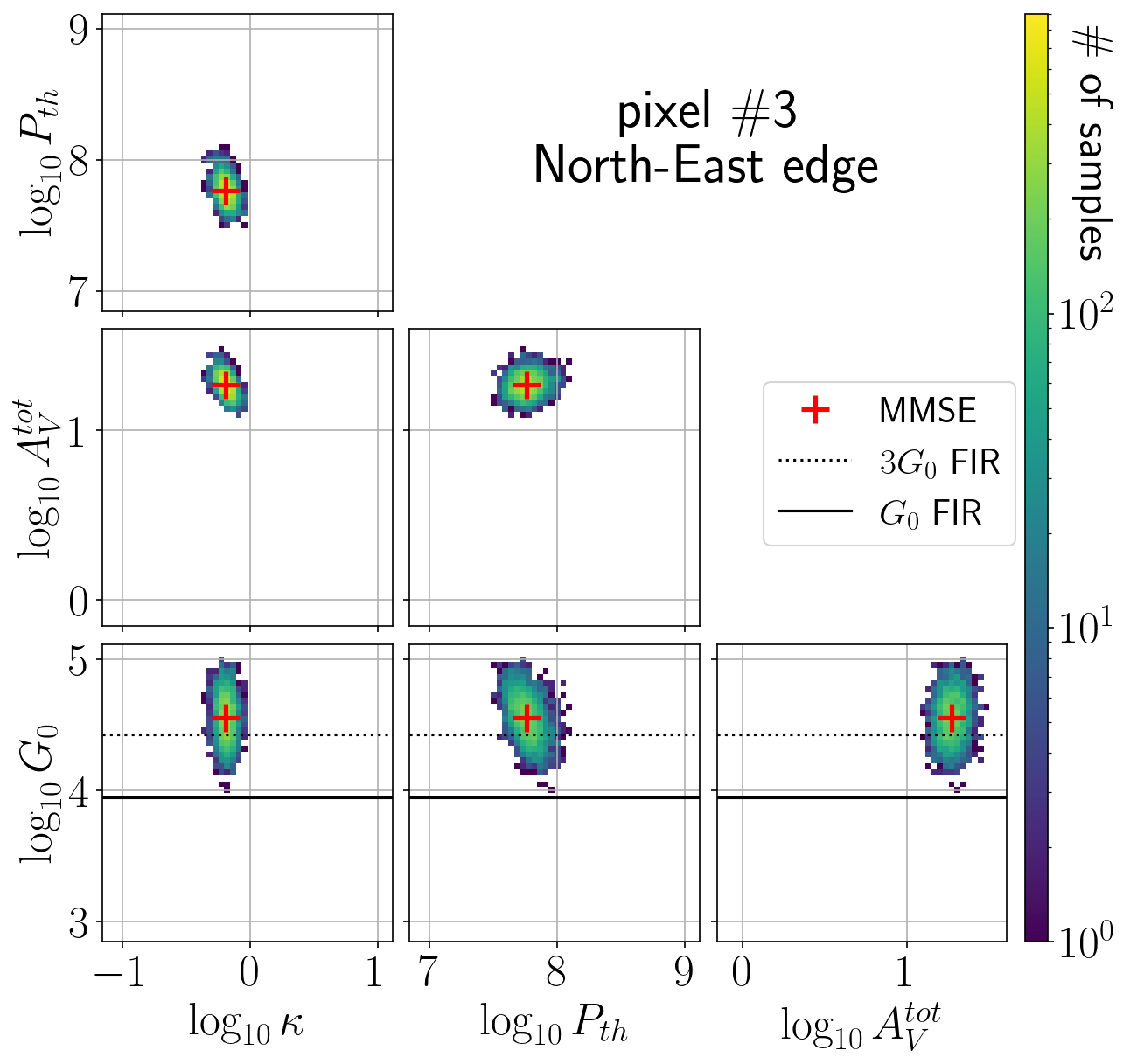}
      \caption{}%
      %
      \label{fig:omc1_hists_northeast_edge}
    \end{subfigure}
    \begin{subfigure}{0.49\linewidth}
      \centering
      \includegraphics[width=0.8\textwidth]{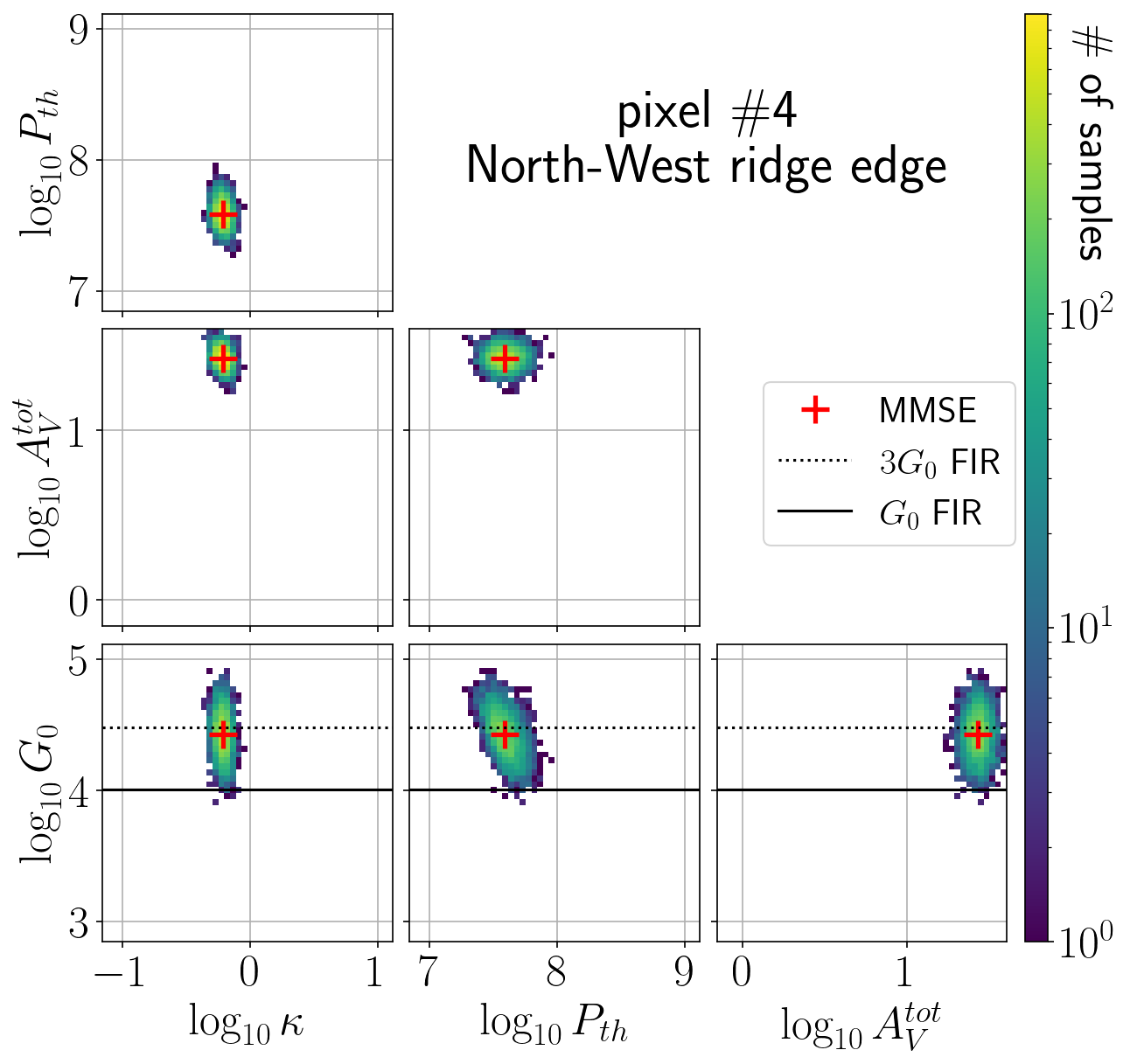}
      \caption{}%
      \label{fig:omc1_hists_northwest_ridge_edge}
    \end{subfigure}
    \caption{
      Inference results for OMC-1 on four specific pixels, presented as two-dimensional marginal histograms in the physical parameters $\paramvect{n}$ space.
      (a) Orion Bar PDR pixel (along with the estimation from~\citealt{joblinStructurePhotodissociationFronts2018}),
      (b) East PDR pixel,
      (c) North-East edge pixel.
      (d) North-West ridge edge pixel,
      %
      %
      %
      The horizontal black line indicates the $\Guv$ values estimated from FIR luminosities from~\citet{goicoecheaVelocityresolvedCIIEmission2015}.
    }
    \label{fig:omc1_hists}
\end{figure*}

This subsection discusses the detailed reconstruction results on the four highlighted positions.
For each of these pixels, Fig.~\ref{fig:omc1_hists} shows the posterior PDF for pairs of parameters, that is, pairwise histograms of samples.

Figure~\ref{fig:omc1_hists_orionbar} show the results for the Orion Bar pixel (lowest of the four squares)
It shows that the marginal posterior distributions for this pixel are simple, as they could be well approximated by a Gaussian distribution with a diagonal covariance matrix.
Consistently with the UF maps in Fig.~\ref{fig:omc1_inference_results}, the uncertainties on $\scalingParam$ and $\Av$ are small, and a bit larger on $\Pth$ and $\Guv$.
These histograms also show that there is no degeneracy between parameters.
The figure compares our estimation results with those from~\cite{joblinStructurePhotodissociationFronts2018} (red star in the figure), that analysed the Orion Bar with an observation of $L = 24$ lines in $N = 1$ pixel.
The observed lines they used for inversion include
$^{12}$CO lines (from $J= 11 \to 10$ to $J = 23 \to 22$),
rotational H$_2$ lines (from S(0) to S(5))
and low level CH$^+$ rotational lines (from $J=1 \to 0$ to $J=6 \to 5$).
The noise on the observation $\obsvect{}$ was assumed additive, Gaussian and uncorrelated with known STDs $(\sigma_{a,\ell})_{\ell=1}^L$.
The authors inferred $\paramvect{} = (\scalingParam, \Pth)$, while fixing the observation angle to $\paramAngle = 60$ deg and the total extinction to $\AV = 10$ mag -- as the observed lines did not provide sufficient constraint on $\AV$.
Based on~\citet{tielensPhotodissociationRegionsBasic1985} and~\citet{marconiInfraredSpectraOrion1998}, they set $\Gnaught = 2 \times 10^4$.
The fit was performed with a grid search on $\Pth$ and a continuous optimization on $\scalingParam$.
They obtained $\Pth = 2.8 \times 10^8$ K cm$^{-3}$ and $\scalingParam = 1.3$.
Overall, Fig.~\ref{fig:omc1_hists_orionbar} shows compatible results between the two estimations.
It also displays the $\Guv$ value estimated in~\cite{goicoecheaVelocityresolvedCIIEmission2015} from FIR luminosities (black line in the figure).
Again, the two $\Guv$ reconstructions are compatible.
These agreements between our estimations and these two independent reconstructions from the literature are remarkable given that they all resort to very different tracers.

Figure~\ref{fig:omc1_hists_east_pdr} shows the results for the East PDR pixel (mid-height square).
These estimations are very similar to those of the Orion Bar, which is remarkable as the two pixels are distant on the map.
It shows that these two pixels, located on the cavity edge, share very similar physical conditions.
This figure also shows that the estimated $\Guv$ value is still compatible with the FIR estimation, although less than for the Orion Bar.

Figures~\ref{fig:omc1_hists_northeast_edge} and~\ref{fig:omc1_hists_northwest_ridge_edge} show the North-East pixel (top left square) and the North-West ridge edge pixel (top right square), respectively.
These two pixels and their surrounding regions yield a more moderate $\Pth$ ($\simeq 3-6 \times 10^7 \Kpccm$) while $\Gnaught$ remains similar to what is found in the Orion Bar ($\simeq 3 \times 10^4$), which seems in contradiction with the positive correlation between $\Pth$ and $\Gnaught$ described in~\citet{joblinStructurePhotodissociationFronts2018} and~\citet{wuConstrainingPhysicalConditions2018} (see discussion
in Sect.~\ref{ssec:a-G0-Pth-relation}).
%
We note that the histograms show a slight negative correlation between $\Pth$ and $\Guv$, which indicates that the actual $\Guv$ for these pixels might be lower and the actual $\Pth$ might be higher.
%
In addition, our $\Guv$ estimation in these two pixels appears to be only marginally compatible (or even clearly incompatible for the North East edge pixel) with the FIR estimation.
This might indicate that additional processes, not included in the models, are increasing the molecular emission in these pixels, as could for instance be the case if additional energy injection processes, besides FUV irradiation, are contributing to the warm molecular emission.
Further investigation is required to better understand these regions.

\subsection{Comparison with estimation of $\Guv$ from far infrared luminosities}%
\label{ssec:comparison-with-estimation-of-G0-from-far-infrared-luminosities}

\begin{figure}
  \centering
  \begin{subfigure}{0.8\linewidth}
    \centering
    \includegraphics[width=\linewidth]{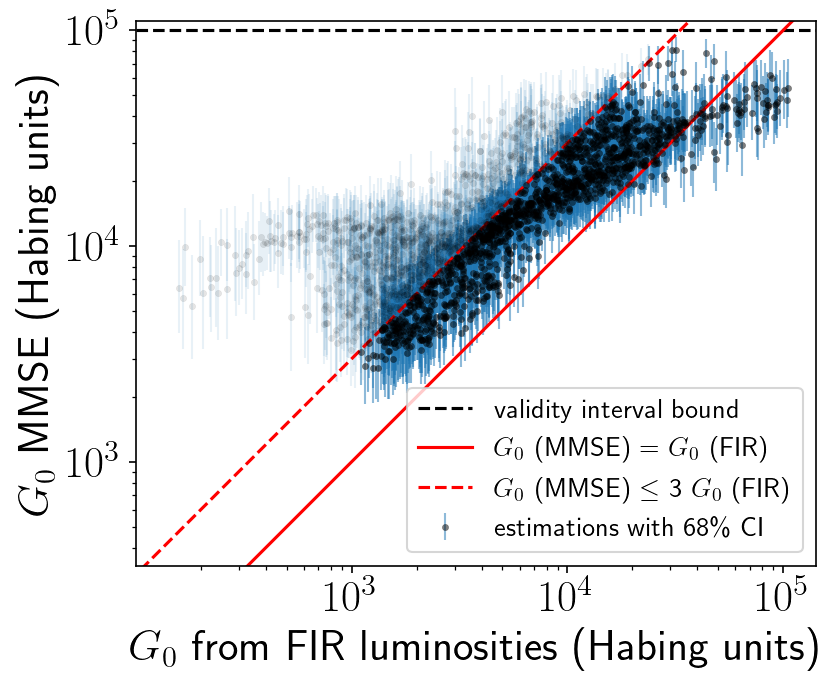}
    \caption{}%
    \label{fig:compare-javier:scatter}
  \end{subfigure}
  \begin{subfigure}{0.49\linewidth}
    \centering
    \includegraphics[height=13.5em]{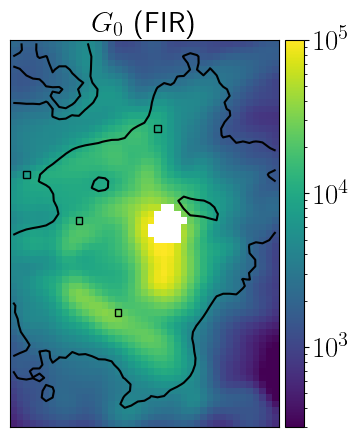}
    \caption{}%
    \label{fig:compare-javier:preds_fir}
  \end{subfigure}
  \begin{subfigure}{0.49\linewidth}
    \centering
    \includegraphics[height=13.5em]{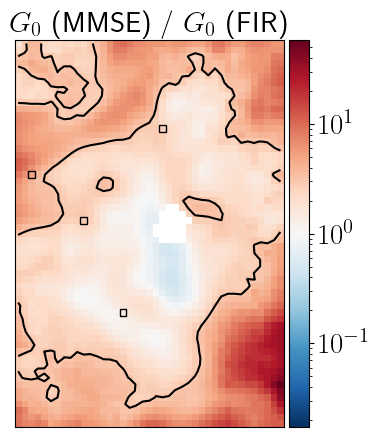}
    \caption{}%
    \label{fig:compare-javier:ratio}
  \end{subfigure}
  %
  %
  \caption{
    Comparison of our estimations of $\Guv$ with those of~\citep{goicoecheaVelocityresolvedCIIEmission2015} obtained from FIR luminosities.
    (a)~Scatter plot of estimated $\Guv$, from our reconstruction and from FIR luminosities.
    (b)~Map of estimated $\Guv$ with FIR luminosities, adapted from~\citet{goicoecheaMolecularTracersRadiative2019}.
    (c)~Map of the ratio between our $\Guv$ reconstructions and those based on FIR luminosities.
    In both maps, the black contour identifies the regions where $\Gnaught \text{ (MMSE) } \leq 3 \, \Gnaught \text{ (FIR)}$.
  }%
  \label{fig:compare-javier}
\end{figure}

This section compares the $\Guv$ reconstructions from~\citet{goicoecheaVelocityresolvedCIIEmission2015} with our $\Guv$ estimations (MMSE and CIs) from Sect.~\ref{sssec:MLE-MMSE-and-UF-maps}.
Figure~\ref{fig:compare-javier:scatter} shows an overall good agreement between these independent reconstructions, although the MMSE estimate tends to be larger than the estimations from FIR luminosities.
For most pixels with high FIR $\Guv$, this difference is lower than a factor of three.
However, for pixels with lower $\Guv$ from FIR luminosities, our reconstructions have a systematic bias much larger than the 68\% CIs.

To better understand this discrepancy, we show in Figure~\ref{fig:compare-javier:preds_fir} the $\Guv$ map from~\citet{goicoecheaMolecularTracersRadiative2019} estimated using FIR luminosities, and in Figure~\ref{fig:compare-javier:ratio} the map of the ratio between our MMSE estimate of $\Guv$ and this FIR luminosity-based estimation.
In both maps, the black contour identifies the regions where $\Gnaught \text{ (MMSE) } \leq 3 \, \Gnaught \text{ (FIR)}$.
It shows that over all the central PDR regions bordering the walls of the cavity around the Trapezium stars, the two estimation are compatible within a factor of 3.
In contrast, the region where the discrepancy is larger than a factor of 3 are all located in the outer regions, where the physics and chemistry might not be dominated by the FUV illumination from the Trapezium cluster.
In addition, these outer regions of the map also present lower S/N in the observations.
Interestingly, the only region where our $\Guv$ are lower than those based on FIR luminosities are in Orion S, where the $\Guv$ is highest.
This might be due to the upper bound of the validity interval on $\Guv$, at $10^5$ Habing units, constraining our MCMC posterior samples to remain below this threshold, and thus possibly biasing the MMSE towards lower values.

This comparison to an independent measurement of $\Gnaught$ indicates that despite inferring four parameters from five molecular lines only, our results are reliable over a large area of the OMC-1 map.
Adding to that the fact that our PDR model succeeds in reproducing the observed line intensities over the full map, this provides evidence that FUV irradiation controls the physics and the chemistry of the warm molecular tracers over most of the central regions of OMC-1.

\subsection{Relationship between $\Guv$ and $\Pth$}%
\label{ssec:a-G0-Pth-relation}

\begin{figure}
  \centering
  \begin{subfigure}{0.8\linewidth}
    \centering
    \includegraphics[width=1\linewidth]{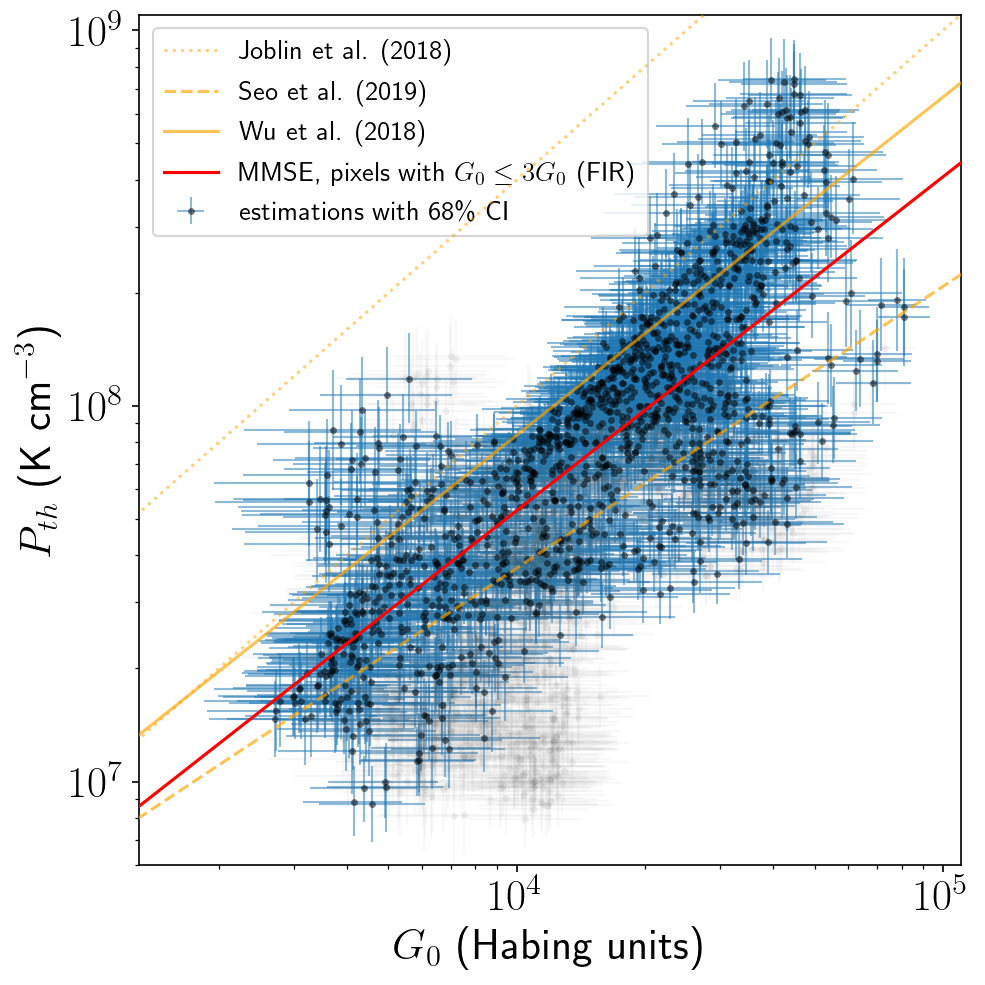}
    \caption{}%
    \label{fig:relation_G0_Pth:scatter}
  \end{subfigure}
  %
  %
  \begin{subfigure}{0.45\linewidth}
    \centering
    \includegraphics[height=13.5em]{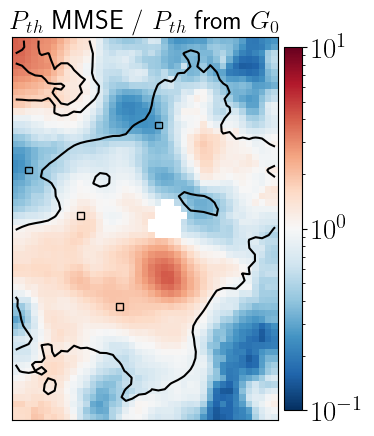}
    \caption{}%
    \label{fig:relation_G0_Pth:residual}
  \end{subfigure}
  \caption{
    Analysis of the relationship between $\Guv$ and $\Pth$.
    (a)~Scatter plot of the MMSE pixel values for $\Guv$ and $\Pth$, that shows a strong positive correlation.
    The pixels such that $\Gnaught \text{ (MMSE) } \leq 3 \, \Gnaught \text{ (FIR)}$ are highlighted in blue, and the other pixels are in light grey.
    A power law fit on those highlighted pixels is shown in red, compared with previous power law relations proposed in the literature (in yellow).
    %
    %
    (b)~Map of ratio of the MMSE for $\Pth$ and the $\Pth$ predicted from $\Guv$ with our power law fit.
    In both maps, the black contour identifies the regions where $\Gnaught \text{ (MMSE) } \leq 3 \, \Gnaught \text{ (FIR)}$, as in Fig.~\ref{fig:compare-javier}.
  }%
  \label{fig:relation_G0_Pth}
\end{figure}

The existence of a positive correlation in PDRs between the incident UV radiation field intensity $\Guv$ and the thermal pressure $\Pth$ has been discussed in multiple ISM studies~\citep{joblinStructurePhotodissociationFronts2018,wuConstrainingPhysicalConditions2018,2019ApJ...878..120S,2022A&A...658A..98P,2022ARA&A..60..247W}.
%
This relationship has been found to follow a linear relationship (with $\Pth/ \Guv \in [1, 4] \times 10^4$,~\citealt{joblinStructurePhotodissociationFronts2018})
or near linear relationship (exponent of $0.9$ in~\citet{wuConstrainingPhysicalConditions2018} and of $0.75$ in~\citet{2019ApJ...878..120S}).
A similar relationship has been suggested in~\citet{wuSpatiallyResolvedPhysical2015} from extragalactic observations in M83.
This relationship suggests that the radiative feedback of massive stars can induce compression of the gas at the edge of PDRs~\citep{bronPhotoEvaporation2018}.
Whether this could trigger additional star formation remains an open question.


Figure~\ref{fig:relation_G0_Pth:scatter} shows the MMSE pixel values along with the 68\% CI (i.e., a $1\sigma$ error) for our OMC-1 estimation results.
A similar positive correlation between $\Guv$ and $\Pth$ can be seen.
%
%
As the joint $(\Pth, \Gnaught)$ histograms in Fig.~\ref{fig:omc1_hists} show null or negative correlations, we confirm that this positive correlation among multiple sources is not due to a degeneracy between the two parameter uncertainties and has a physical origin, as hypothesized in~\citet{joblinStructurePhotodissociationFronts2018}.
%
%
%
In the following, we focus on the pixels that satisfy $\Guv \text{ (MMSE) } \leq 3 \Guv \text{ (FIR)}$ (in blue in Fig.~\ref{fig:relation_G0_Pth:scatter}), as these are the most likely to be adequately described by our PDR models (see previous discussion in Sect.~\ref{ssec:comparison-with-estimation-of-G0-from-far-infrared-luminosities}).
A weighted least square fit on $\logd \Pth = a \logd (\Guv) + b$ leads to $a = 0.89 \pm 0.02$ and $b = 4.2 \pm 0.1$ (results are shown with mean $\pm 1\sigma$), for a determination coefficient $R^2 = 0.62$ (in red in the figure).
%
%
This fit is quite close to the relationship presented in~\citet{wuConstrainingPhysicalConditions2018}.
The small set of points in blue lying above the rest at moderate $\Guv$ correspond to the map North-West corner.
In this region, the S/N is much lower than on the rest of the map (see Fig.~\ref{fig:omc1_observations}).
Although the $\Guv$ MMSE estimate is close to the FIR luminosity-based $\Guv$ estimate in this region, the uncertainties on both $\Pth$ and $\Guv$ are large, as the UF maps from Fig.~\ref{fig:omc1_inference_results} clearly show.
These pixels thus have a small weight in our weighted least square fit.
%

We observe that the $\Pth-\Guv$ relation found from warm molecular tracers (this work, \citealt{joblinStructurePhotodissociationFronts2018}, \citealt{wuConstrainingPhysicalConditions2018}) appears shifted towards higher thermal pressures compared to the theoretical relation of~\citet{2019ApJ...878..120S}.
Moreover, \citet{2022ARA&A..60..247W} suggests that magnetic and turbulent pressure contribution in the PDR should be accounted for, which, assuming equipartition, would result in thermal pressures a factor of 3 lower than the~\citet{2019ApJ...878..120S} curve.
This would further increase the discrepancy with our estimations.
The theoretical law of ~\citet{2019ApJ...878..120S} is based on an argument of pressure equilibrium between the PDR and ionized gas at the Strömgren radius.
In addition, it attempts to account for the photoevaporation flows present in blister-type H\textsc{ii} regions by equating the PDR pressure to \emph{twice} the ionized gas pressure (which corresponds to a D-critical ionization front, e.g., \citealt{drainePhysicsInterstellarIntergalactic2011}, Chap. 37).
However, the ionized gas pressure in this D-critical relation should be that at the base of the so-called ionized boundary layer (\citealt{Bertoldi1989}), which is higher than the global average thermal pressure in the H\textsc{ii} region (especially near the densest parts of the surrounding neutral wall).
Using a global Strömgren model is thus not justified for a blister-type H\textsc{ii} region, and will tend to underestimate thermal pressure in the PDRs.
The fact that the relationship observed from molecular tracers is higher than the Strömgren-based law of~\citet{2019ApJ...878..120S} is thus compatible with a pressure unbalance, with the global thermal pressure in the H\textsc{ii} region unable to contain the thermal pressure of the molecular PDR, and is thus consistent with the expected photoevaporation flows from dense PDRs bordering blister-type H\textsc{ii} regions, as initially proposed in~\citet{joblinStructurePhotodissociationFronts2018}.

%
%
%
%
%
Figure~\ref{fig:relation_G0_Pth:residual} shows the map of the ratio between our pressure MMSE estimates and the pressures expected from our $\Guv$ estimates according to the power law fit.
This allows to visualize which regions of the map are adequately described by this power law relation.
%
In most of the PDRs bordering the central cavity, including the Orion Bar and the East PDR, $\Pth$ from the MMSE is consistent within a factor of two with the power law fit.
%
It reaches a factor of about three above the value expected from $\Guv$ in the Orion S region.
Larger discrepancies are found in the outer regions of the map, with the power law fit generally predicting too high $\Pth$ values compared to the MMSE estimate.
Values of $\Pth$ from the MMSE reach a factor of up to five lower than expected from $\Guv$ in the South West and North East corners of the map.
In particular, a ratio below one is found in the ridge North to the BN/KL region and the bar that goes East, each already mentioned in the histogram analysis in Fig.~\ref{fig:omc1_hists}.
As previously discussed, these discrepancies are consistent with the explanation that the physics and chemistry of molecular emission in these outer regions might not be solely driven by FUV irradiation from the Trapezium cluster.



\section{Conclusion}%
\label{sec:conclusion}

This paper introduced \textsc{Beetroots}, an open-source software that reconstructs maps of physical parameters from observation maps and an astrophysical model.
This software leverages a realistic noise model and exploits spatial regularization to guide the estimations towards physically consistent reconstructions.
It uses state-of-the-art reconstruction algorithms, both for optimization (for faster results) and for MCMC sampling (to produce uncertainty quantifications), that can escape from local modes thanks to global exploration steps.
These algorithms are able to scale to high dimensions to analyse large observation maps in a reasonable time.
Finally, the software quantifies and assesses the ability of the astrophysical model to explain the observations, providing feedback to improve ISM models.

After demonstrating the power of \textsc{Beetroots} on synthetic data, we used it to deduce physical parameter maps (thermal pressure $\Pth$,  $\Guv$, total visual extinction $\AV$) from the large observation map of the Orion molecular cloud 1 (OMC-1) of~\citet{goicoecheaMolecularTracersRadiative2019}.
To the best of our knowledge, this inference is the first performed on such a large map (that is, with $1\,000$ pixels or more) for multiple physical parameters from molecular tracers.
It is also the first time that the uncertainties are quantified in an inference of maps of physical parameters from molecular tracers.
%
%
%
We show that our PDR model succeeds to reproduce the observed line intensity maps, with parameters maps that are spatially coherent and in which all of the main features of the region can be recognized.
We demonstrate in particular that our spatially regularized Bayesian approach is key in permitting this success, as simpler approaches yield unusable noise-dominated parameter maps. Our uncertainty quantification also shows that the five considered molecular lines are able to constrain $\Guv$ and the thermal pressure over a wide range of conditions.
Moreover, we demonstrate that our $\Guv$ reconstructions are consistent with those obtained from FIR luminosities (i.e., dust thermal emission) over most of the central part of the map.
This indicates that warm molecular emission over a large area of the map is controlled by UV irradiation from the Trapezium cluster.
We find high $\Pth$ values in a significant portion of the map, including the full Orion Bar region as well as the East PDR, confirming that the high pressure values found by \citealt{joblinStructurePhotodissociationFronts2018} at one position in the Orion Bar are widespread in the dense PDRs of OMC-1.
In addition, we find the thermal pressures to be correlated to the incident $\Guv$ with a near-linear relationship, confirming and extending a finding of several previous studies~\citep{joblinStructurePhotodissociationFronts2018,wuConstrainingPhysicalConditions2018,2019ApJ...878..120S,2022A&A...658A..98P}, and which we ascribe to photoevaporation-driven compression of the PDRs surrounding a blister-type H\textsc{ii} region.
%
%
%

This work demonstrates how much can be gained by using specially-built state-of-the-art Bayesian inversion methods to derive physical parameter maps from spatially resolved line intensity maps of interstellar clouds,
and paves the way to a systematic and rigorous analysis of observations produced by past, current and future instruments.
Several further improvements of the method proposed here can also be foreseen.
As shown in Sect.~\ref{sec:application-to-synthetic-observation-maps},  the current implementation of \textsc{Beetroots} can perform inference on maps of up to about $10^4$ pixels on a personal laptop.
However, it does not scale to larger Galactic observations containing hundreds of thousands or millions of pixels, such as the Orion B dataset~\citep{petyAnatomyOrionGiant2017} obtained at IRAM.
%
Analysing such datasets at once would require the implementation of a distributed version of \textsc{Beetroots}.
Besides, such observation maps often cover a variety of environments that cannot be accurately simulated with a single astrophysical model -- for instance, some areas might contain astrophysical shocks while other dense and dark regions might be governed by slow, time-dependent chemical evolution with little influence from UV irradiation.
To analyse such clouds, \textsc{Beetroots} would need to combine models describing multiple environments (e.g., PDR models, shock models, H\textsc{ii} region models, time-dependent chemistry models, etc.).
It would determine which environment best describes each pixel and infer the associated physical parameters.
Both improvements are left for future work.




\begin{acknowledgements}
  This work received support from the French Agence Nationale de la Recherche through the DAOISM grant ANR-21-CE31-0010,
  and from the Programme National ``Physique et Chimie du Milieu Interstellaire'' (PCMI) of CNRS/INSU with INC/INP, co-funded by CEA and CNES.
  It also received support through the ANR grant ``MIAI $@$ Grenoble Alpes'' ANR-19-P3IA-0003.
  This work was partly supported by the CNRS through 80Prime project OrionStat, a MITI interdisciplinary program,
  by the ANR project ``Chaire IA Sherlock'' ANR-20-CHIA-0031-01 held by P. Chainais,
  and by the national support within the {\em programme d'investissements d'avenir} ANR-16-IDEX-0004 ULNE and Région HDF.
  JRG and MGSM thank the Spanish MCINN for funding support under grants PID2019-106110G-100 and PID2023-146667NB-I00.
  MSGM acknowledges support from the NSF under grant CAREER 2142300.
  Part of the research was carried out at the Jet Propulsion Laboratory, California Institute of Technology, under a contract with the National Aeronautics and Space Administration (80NM0018D0004).
  D.C.L. acknowledges financial support from the National Aeronautics and Space Administration (NASA) Astrophysics Data Analysis Program (ADAP).
\end{acknowledgements}


\bibliographystyle{aa} %
\bibliography{main.bib} %

\begin{appendix}

\section{How to use \textsc{Beetroots} in practice?}%
\label{sec:how-to-use-beetroots-in-practice}

  \begin{figure}[t]
    \centering
    \includegraphics[width=0.49\textwidth]{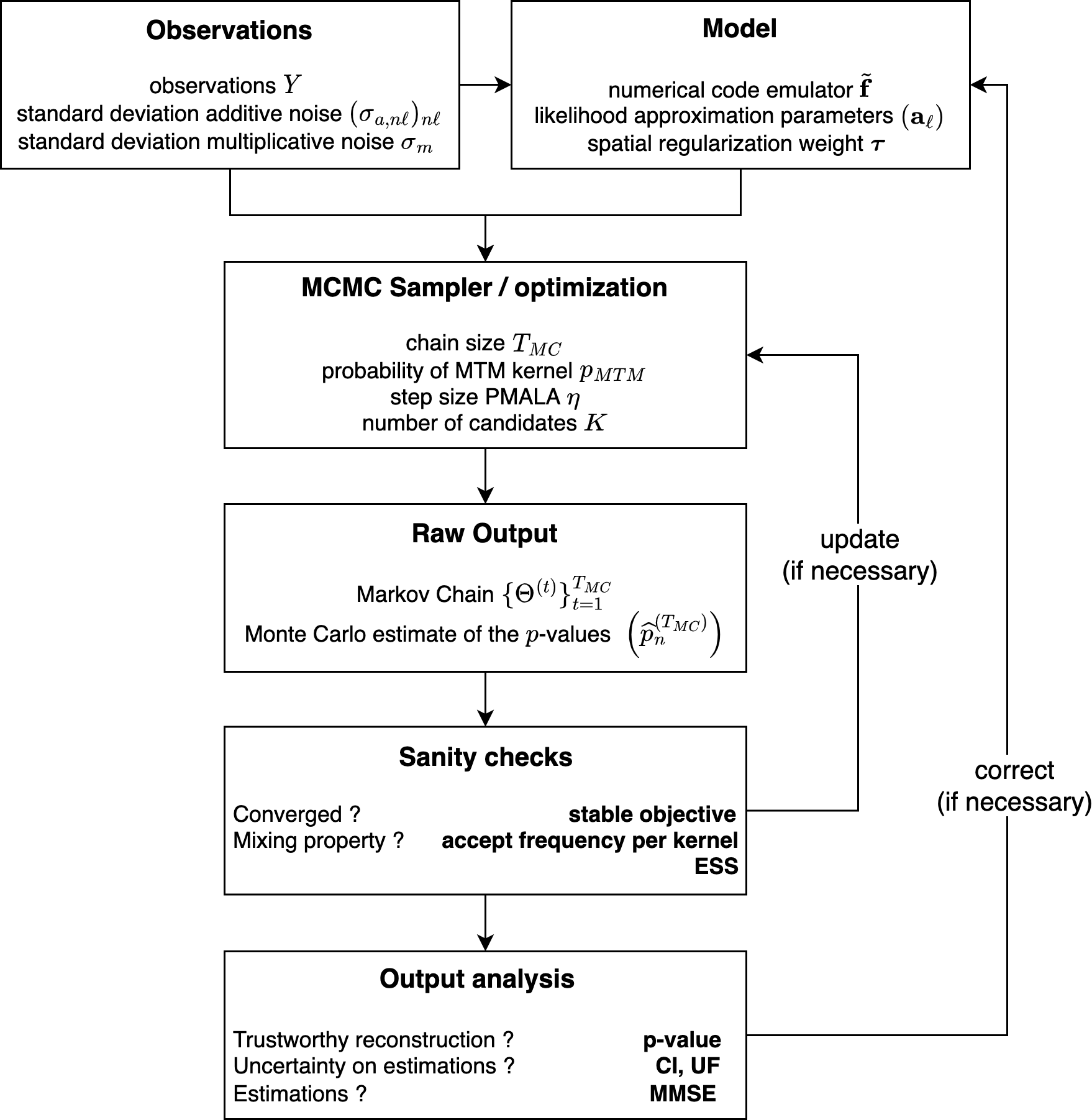}
    \caption{
      Full inference workflow with \textsc{Beetroots}.
    }%
    \label{fig:process}
  \end{figure}

This section describes the steps to take in practice to run \textsc{Beetroots} on an observation $\obsfull$.
Figure~\ref{fig:process} summarizes the full process in the case of an observation map affected by both a multiplicative noise and an additive noise, regularized with a spatial prior.
The STD $\sigma_m$ of the multiplicative noise and the STD map $(\sigma_{a,n\ell})$ of additive noise in the observation are assumed provided by observers, which is generally the case in ISM.

Before performing the inference, the user needs to choose an astrophysical model that can predict the observables contained in $\obsfull$ from the physical parameters of interest $\paramfull$.
To fully exploit the potential of \textsc{Beetroots}, this model should be fast to run and give access to its gradient~$\nabla \truef$.
With ANNs, this gradient is simple to obtain using autodifferentiation techniques.
To fully exploit the ANN characteristic (speed, accuracy and gradient), \textsc{Beetroots} has an interface with \textsc{nnbma}\footnote{\url{https://github.com/einigl/ism-model-nn-approximation}}~\citep{paludNeuralNetworkbasedEmulation2023}.
Other astrophysical models can also be interfaced with \textsc{Beetroots}, even those that do not allow gradient evaluation.
Then, the user should select the noise model to be used -- see discussion in Sect.~\ref{sssec:the-noise-model}.
%
%
In a first run, a purely Gaussian approximation can be used as a default choice.
Finally, the user can manually set the value of the spatial regularization prior $\tau_d$, or automatically adjust it during sampling.
If manually set, low values should initially be preferred to avoid biasing the estimation towards over-smoothed maps.

The sampling or optimization parameters are introduced in Sect.~\ref{sssec:estimation-methods-in-beetroots}.
\textsc{Beetroots} produces a list of posterior samples (in sampling mode) or iterates (in optimization mode) $\paramfull^{(t)}$, which are used to derive estimators.
To avoid mis-interpretations, we recommend performing some sanity checks before analysing the reconstructions.
First, one should check whether the algorithm converged using the figure of the loss function evolution with iterations (produced automatically).
For optimization, the loss function should be constant during the last iterations.
For a sampling procedure, it should look noisy around a constant value.
Then, for the sampling procedure, the exploration of the posterior can be checked with the average acceptance frequencies of the kernels.
Ideally, this acceptance frequency should be around 60\% for PMALA and non-zero for MTM~\citep{paludEfficientSamplingNon2023} (the higher the better).
The effective sample size (ESS) quantifies how efficiently the sampler explored the posterior distribution by indicating the effective number of independent samples within the returned Markov chain (ESS maps are produced automatically).
For a sampling procedure with $\TMC = 10^4$ iterations, the ESS should ideally be larger than one hundred on most pixels of the map.
See~\citet[chapter 12]{robertMonteCarloStatistical2004} for more information on MCMC convergence diagnosis
%
%
If the posterior samples pass the sanity checks, then the ability of the model to reproduce the observations should be tested, using the model checking approach presented in Sect.~\ref{sssec:model-checking-and-Bayesian-p-value} (maps of $p$-value and of results of the pixelwise hypothesis testing are produced automatically).
If the observations are well reproduced, then the reconstructions and associated uncertainty quantifications can be analysed from automatically produced maps of MMSE and UFs.

\textsc{Beetroots} was built to automatically perform all these steps for each new run.
For this reason, we recommend it as an all-in-one tool for future analyses of ISM observations.


\section{More on statistical modelling and estimation}%
\label{sec:more-on-statistical-modelling-estimation-and-model-checking}

\subsection{Likelihood for upper limits}%
\label{sec:upper-limits}

Upper bounds on observations are quite common in ISM studies.
The corresponding observables are sometimes discarded in the inversion process.
Omitting observables leads to a loss of information that could damage the inference results.
Including these upper bounds on observations in the likelihood function permits to account for all constraints provided by the observations.
In statistics, censorship permits to include such upper limits $\omega \geq 0$.
In case of Gaussian noise, the observation model is
\begin{align}\label{eq:obs_model_censor}
  \obselt
  = \max \left\{ \omega, \; \truefell (\paramvect{n}) + \addnoise \right\}
  = \begin{dcases}
    \omega \; \text{ if } \truefell (\paramvect{n}) + \addnoise \leq \omega \\
    \truefell (\paramvect{n}) + \addnoise \: \text{ otherwise}
  \end{dcases}
\end{align}
In~\citep{ramambasonInferringHIIRegion2022}, censorship is modelled with a half-normal distribution $\mathcal{N}_{-}(\omega, \sigma^2)$.
This approach prohibits all values above the censoring threshold $\omega$, while values slightly above $\omega$ should also be allowed.
Besides, it biases $\truefell(\paramvect{n})$ towards values close to the threshold $\omega$, while $\truefell(\paramvect{n}) \ll \omega$ also satisfies the upper bound constraint.
Statistically, a better encoding of censorship in the likelihood is~\citep[chapter 1]{robertMonteCarloStatistical2004}
\begin{align}\label{eq:censorship_correct}
  - \log \pi(\omega \vert \paramvect{n}) =
  - \log \int_{-\infty}^\omega \pi ( \obselt \vert \paramvect{n}) \; \diff \obselt
  .
\end{align}
The integral covers all the integrated intensities $\obselt \leq \omega$, and sums the associated uncensored likelihoods $\pi ( \obselt \vert \paramvect{n})$.

\begin{figure}
  \centering
  \includegraphics[width=0.95\linewidth]{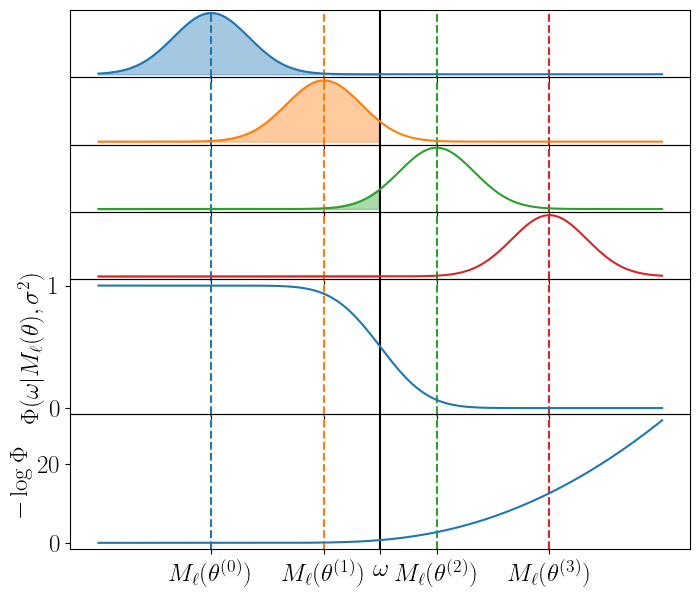}
  \caption{
    Illustration of censorship in likelihood for Gaussian additive random deteriorations. 
    The first four panels show four values of $\truefell(\paramvect{}^{(i)})$, $i= 0, \ldots, 3$ and the associated Gaussian additive PDF.
    The censored portion, below the censorship threshold $\omega$, is highlighted.
    The fifth panel shows the likelihood function taking censorship into account, that is, the Gaussian CDF $\Phi$.
    The sixth panel shows the corresponding negative log-likelihood.
  }%
  \label{fig:censorship}
\end{figure}

Figure~\ref{fig:censorship} illustrates this censorship modelling for a Gaussian additive random deterioration $\addnoise \sim \mathcal{N}(0, \sigma^2)$.
In this case, the integral in~\eqref{eq:censorship_correct} can be evaluated efficiently for Gaussian noise using the Gaussian cumulative density function (CDF) $\Phi(\omega \vert \truefell(\paramvect{}), \sigma^2)$.
Unlike the half Gaussian mentioned above, it does not forbid all physical parameters $\paramvect{}$ such that $ \truefell(\paramvect{}) > \omega$.
Instead, it smoothly and increasingly penalizes them as $\truefell(\paramvect{})$ gets farther to the threshold limit, that is, as the probability of $\obselt \leq \omega$ becomes smaller.
Some ISM inversion codes resort to this censorship modelling (see, e.g.,~\citealt{blancIZIInferringGas2015,holdshipBayesianInferenceRates2018,thomasInterrogatingSeyfertsNebulaBayes2018}).


\subsection{Other sampling methods used in ISM studies}%
\label{sec:other-sampling-methods-used-in-ism-studies}

Among ISM studies, inference is often performed with one of four classes of methods in addition to RWMH.

The affine-invariant MCMC sampler~\citep{goodmanEnsembleSamplersAffine2010} and the associated \textsc{emcee} package~\citep{foreman-mackeyEmceeMCMCHammer2013} is a very popular MCMC algorithm in astronomy~\citep{thraneIntroductionBayesianInference2019}.
In ISM studies, it was applied for instance in astrochemistry~\citep{gratierNewReferenceChemical2016,holdshipBayesianInferenceRates2018,keilUCLCHEMCMCMCMCInference2022} and extragalactic molecular gas~\citep{yangMolecularGasHerschelselected2017}.
However, this sampler exclusively addresses low dimensional problems.
Besides, it fails to explore multimodal distributions and requires to be initialized close to a mode. 

Nested sampling~\citep{skillingNestedSampling2004} is also a very popular Bayesian framework in astrophysics and ISM studies.
Popular algorithms include \textsc{ }~\citep{ferozMULTINESTEfficientRobust2009}, \textsc{Dynesty}~\citep{speagleDynestyDynamicNested2020} and \textsc{UltraNest}~\citep{buchnerUltraNestPythonicNested2016,buchnerUltraNestRobustGeneral2021}, that were developed by and for astrophysicists.
\textsc{MultiNest} was applied to study extragalactic molecular clouds~\citep{kamenetzkySurveyMolecularISM2014,chevallardModellingInterpretingSpectral2016} and cosmic ray propagation~\citep{johannessonBayesianAnalysisCosmic2016}.
\textsc{Dynesty} was applied to evaluate distances between the Earth and nearby molecular clouds~\citep{zuckerMappingDistancesPerseus2018,zuckerLargeCatalogAccurate2019} or to reconstruct their 3D structure~\citep{zuckerThreedimensionalStructureLocal2021}.
\textsc{UltraNest} was applied on extragalactic spectral energy densities (SEDs)~\citep{behrensTracingInterstellarHeating2022}.
These codes handle multimodal distribution but are limited to low dimensional distributions -- up to 30 in~\citet{ferozMULTINESTEfficientRobust2009} and~\citet{johannessonBayesianAnalysisCosmic2016}.

Sequential MC~\citep{delmoralSequentialMonteCarlo2006} is quite rare in ISM.
It was applied in~\citet{ramambasonInferringHIIRegion2022} in an inversion of integrated intensities of ionic, atomic and molecular emission lines.
In that work, it is chosen for its ability to sample from multimodal distributions in low dimension settings -- up to 28 in the paper -- and to compute the Bayesian evidence $\pi(\obsfull)$~\eqrefp{eq:posterior_full}.

One sampling algorithm based on a combination of Metropolis-Hastings, Gibbs sampling and the ancillarity-sufficiency interweaving strategy (ASIS)~\citep{yuCenterNotCenter2011} is widespread in dust studies that involve a hierarchical model.
ASIS is a sampling strategy that relies on complementary data augmentation schemes to better explore the posterior, including along strongly correlated directions.
This sampler was first used in a dust study in~\citet{kellyDUSTSPECTRALENERGY2012}. 
It was then coupled with a Gibbs sampling approach to exploit the structure in the physical parameter $\paramfull$
in very high dimensional applications.
For instance,~\citet{gallianoDustSpectralEnergy2018} involve a few thousand dimensions, and~\citep{gallianoNearbyGalaxyPerspective2021} about $10^4$.

\section{More results for synthetic case}%
\label{sec:more-results-for-synthetic-case}

This section provides more details on the synthetic case application from Sect.~\ref{sec:application-to-synthetic-observation-maps}.
First, we display the results on the $N_\text{side} = 10$ and $30$ use cases.
Then, the effect of the regularization weight is studied in more details.
Finally, the model checking results on the three cases are shown and discussed.

\subsection{Estimation results for more spatial resolutions}%
\label{ssec:estimation-results-for-more-spatial-resolutions}

\begin{figure*}[p]
  \centering
    \includegraphics[width=0.9\linewidth]{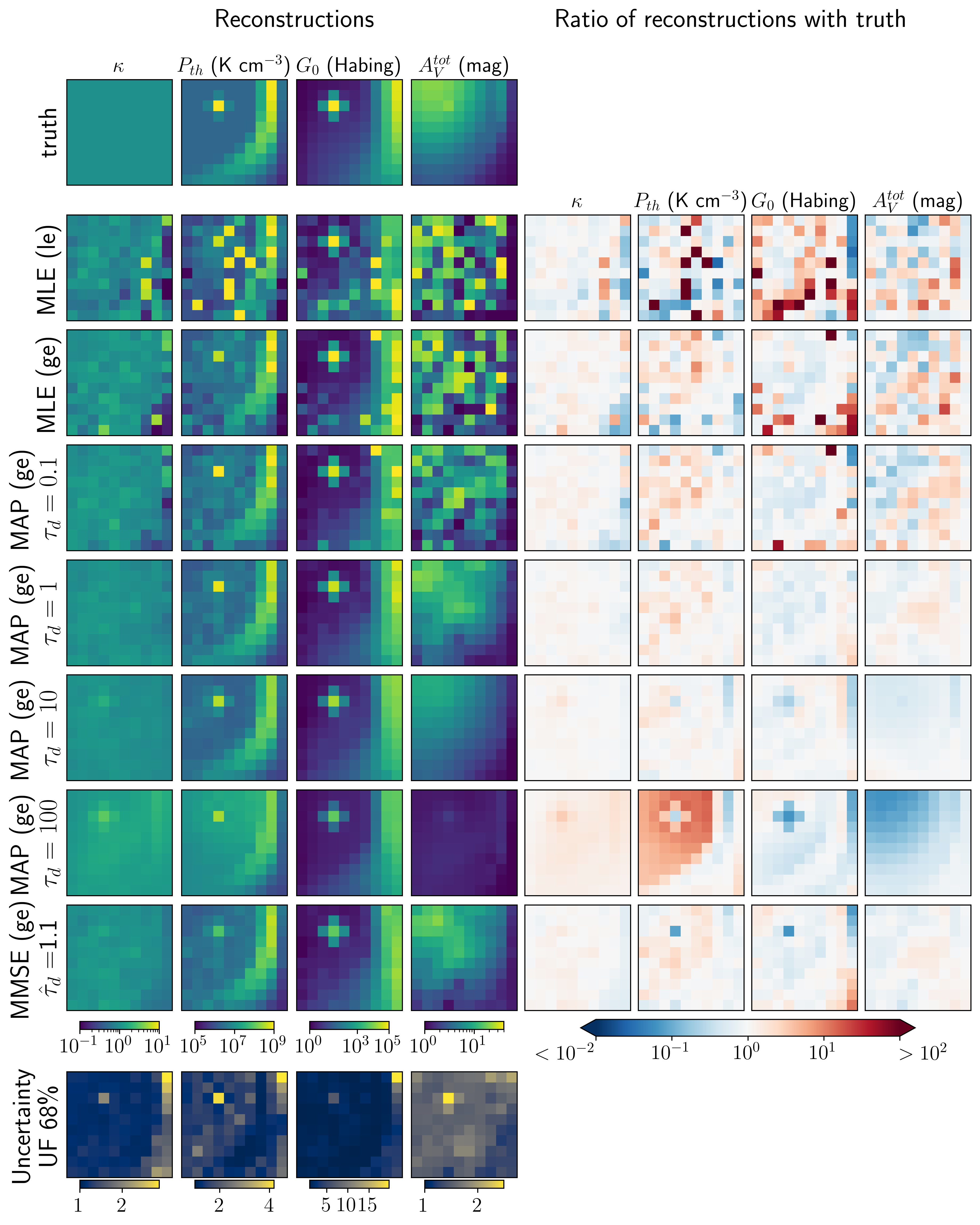}
  \caption{
    Reconstruction results for the $N_\text{side} = 10$ spatial resolution.
    The first row shows the $D=4$ true maps that each estimator tries to reconstruct.
    The obtained reconstructions are shown below the true maps.
    The first estimator is ``MLE (le)'': the MLE evaluated using the PMALA kernel only, that is, only local exploration.
    The second estimator is ``MLE (ge)'': the MLE evaluated using both the PMALA and MTM kernels, that is, a combination of local and global exploration.
    The third to fifth estimators are MAP estimators (i.e., including spatial regularization) with different regularization weight $\tau_d$ values.
    The sixth estimator is the MMSE, obtained with sampling.
    The last row, ``UF 68\%'', quantifies the uncertainties associated with the MMSE by indicating the size of the 68\% CI.
    The maps on the right display the ratios between the estimated and true maps, to better assess the quality of each estimation.
  }%
  \label{fig:synthetic_case_estimation_10}
\end{figure*}

\begin{figure*}[p]
   \centering
  \includegraphics[width=0.9\linewidth]{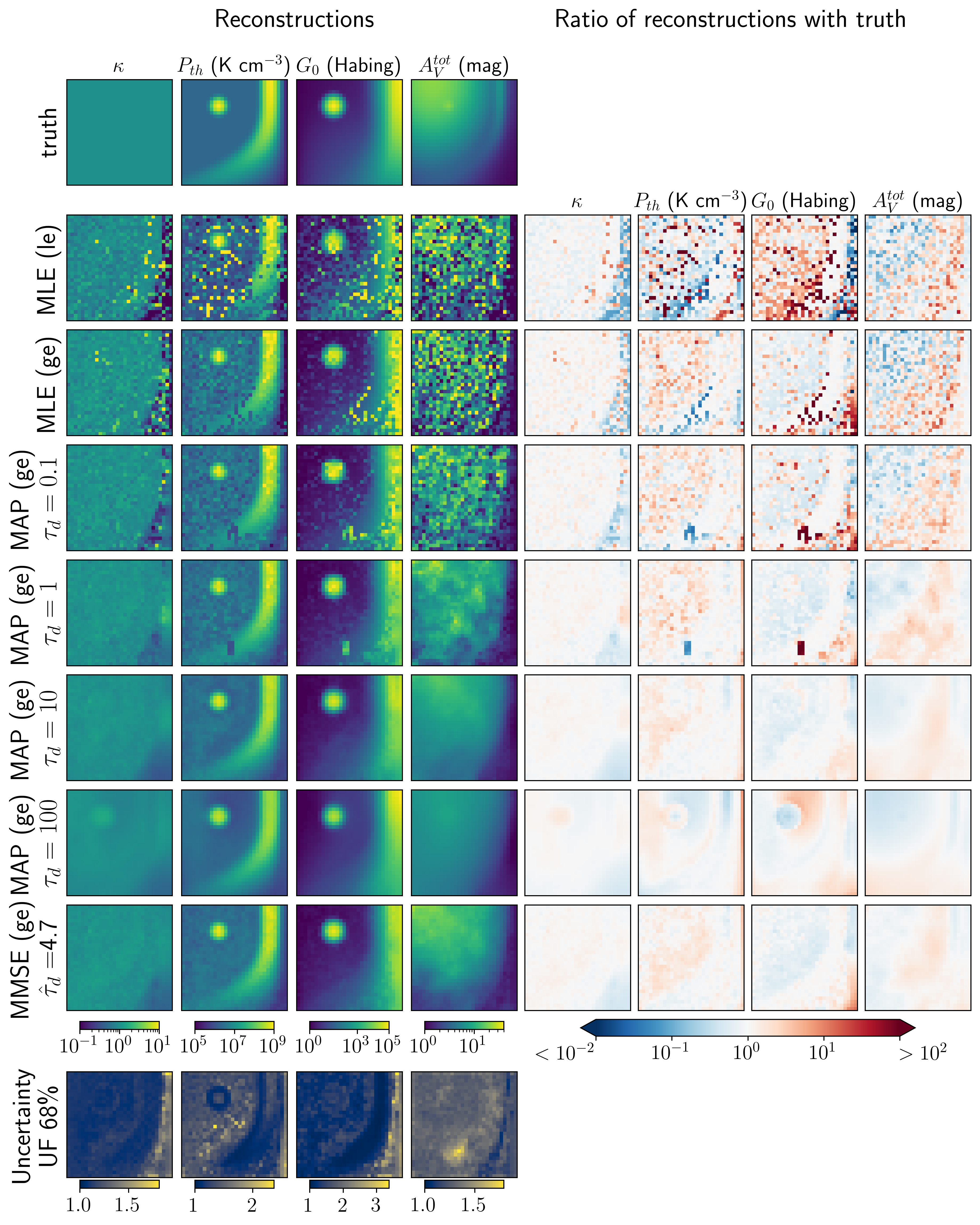}
  \caption{
    Reconstruction results for the $N_\text{side} = 30$ spatial resolution.
    The first row shows the $D=4$ true maps that each estimator tries to reconstruct.
    The obtained reconstructions are shown below the true maps.
    The first estimator is ``MLE (le)'': the MLE evaluated using the PMALA kernel only, that is, only local exploration.
    The second estimator is ``MLE (ge)'': the MLE evaluated using both the PMALA and MTM kernels, that is, a combination of local and global exploration.
    The third to fifth estimators are MAP estimators (i.e., including spatial regularization) with different regularization weight $\tau_d$ values.
    The sixth estimator is the MMSE, obtained with sampling.
    The last row, ``UF 68\%'', quantifies the uncertainties associated with the MMSE by indicating the size of the 68\% CI.
    The maps on the right display the ratios between the estimated and true maps, to better assess the quality of each estimation.
  }%
  \label{fig:synthetic_case_estimation_30}
\end{figure*}

Figures~\ref{fig:synthetic_case_estimation_10} and~\ref{fig:synthetic_case_estimation_30} display the reconstructions for all considered estimators in the $N_\text{side} = 10$ and $30$ use cases, respectively.
Results are commented in Sect.~\ref{sssec:results-comparison}.
Table~\ref{tab:results_synthetic} displays the quantitative results associated with the reconstructions.

\subsection{Optimization: Effect of the regularization weight}%
\label{ssec:effect-of-the-regularization-weight}

\begin{figure}
  \centering
  \includegraphics[width=0.85\linewidth]{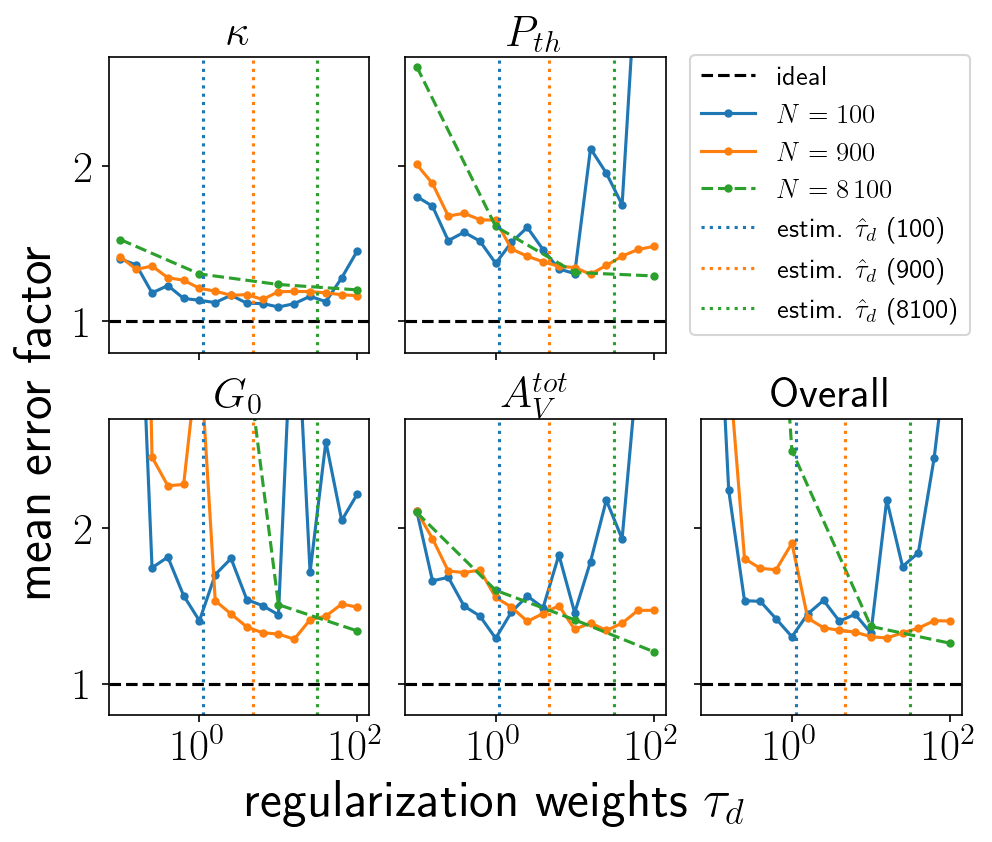}
  \caption{
    Regularization effects: mean EF achieved by the MAP (ge) estimator for different regularization weights $\tau_d$.
  }%
  \label{fig:regularzation-effects}
\end{figure}

Figure~\ref{fig:regularzation-effects} shows how the regularization parameter~$\tau_d$ influences the mean EF~\eqrefp{eq:error-factor} for each physical parameter and their combination.
In all three cases, very low regularization weights do not regularize enough and lead to large errors.
They also complicate the escape from local minima, leading to an artifact in the $N_\text{side}=90$ case, as shown in Fig.~\ref{fig:synthetic_case_estimation_90}.
Conversely, very high regularization weights bias the reconstructions towards over-smoothed maps.
In all three cases, the automatic tuning reaches $\tau_d$ values leading to a good likelihood-prior trade-of.

\subsection{Sampling approach: Model checking results}%
\label{ssec:sampling-approach-model-checking-results}

\begin{figure}
  \centering
  \includegraphics[width=0.8\linewidth]{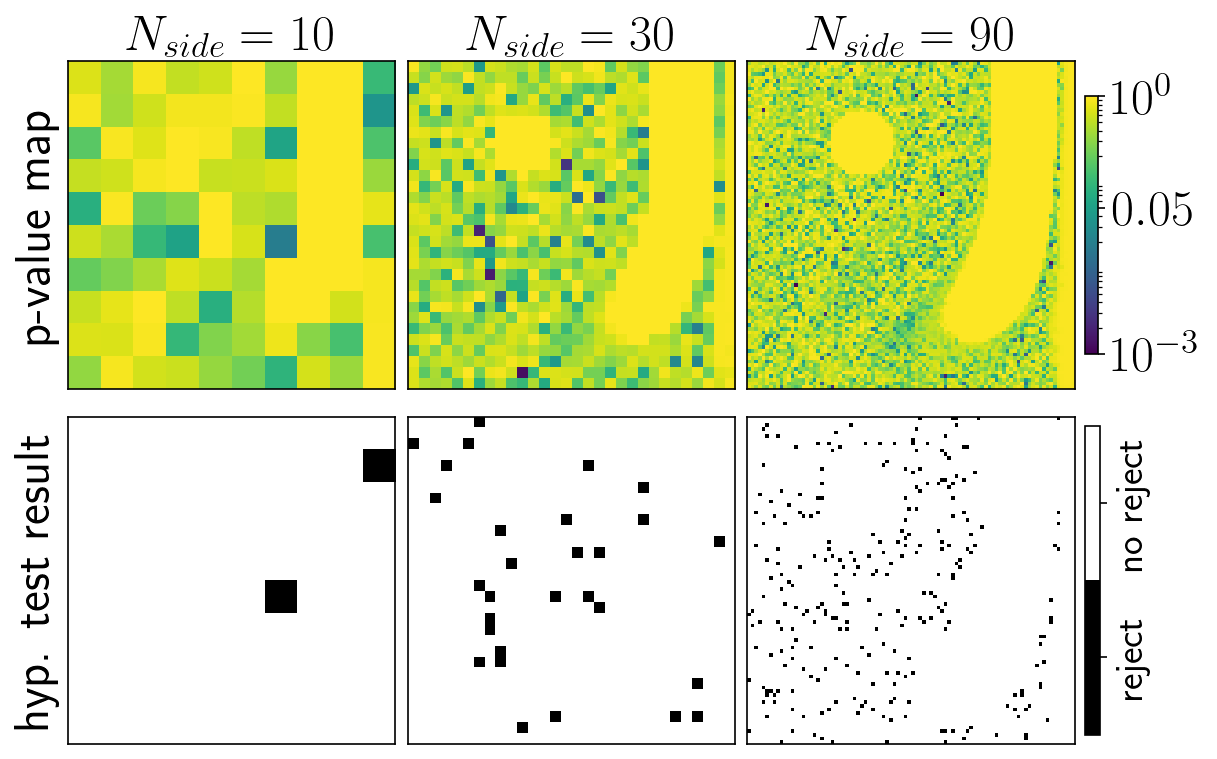}
  \caption{
    Model checking on the three synthetic cases.
  }%
  \label{fig:synthetic-case:model-checking}
\end{figure}

Figure~\ref{fig:synthetic-case:model-checking} shows the results of the model checking performed by \textsc{Beetroots} on the three synthetic cases.
It shows that the null hypothesis, namely ``the astrophysical model $\truef$, noise characteristics $\noiseGeneral$ and prior (i.e., spatial regularization) can reproduce the observations'' is not rejected with the Bayesian hypothesis test for a wide majority of pixels.
For each case, some pixels are rejected with confidence level 5\%:
2 pixels for the $N_\text{side} = 10$ case,
19 pixels for the $N_\text{side} = 30$ case,
and 117 for the $N_\text{side} = 90$ case.
For each case, it means that 2\%, 2.1\% and 1.4\% of the total number of pixels are rejected with confidence level 5\%, respectively.
These pixels are randomly distributed in the map, that is, they do not come from a specific structure in the image.
Besides, if the null hypothesis is true, then the $p$-value has a 5\% chance of being below the confidence level for each pixel.
Therefore, this model checking analysis confirms that the reconstructions obtained with sampling explain the observations.
This means that the reconstructions are meaningful and can be trusted.

In these synthetic cases where the true values are known, the validity of the reconstructions can be assessed directly.
This model checking approach is thus secondary here.
However, true values are unknown for real observations, as well as the best astrophysical model.
Performing model checking is thus critical to assess the relevance of the reconstructions.

\section{More results on OMC-1}%
\label{sec:more-results-on-omc1}

This section provides more details on the OMC-1 application from Sect.~\ref{sec:application-to-real-data}.
First, we display the maps of Gaussian noise STD.
Then, to detail the model checking results, the posterior predictive distribution is shown for the four highlighted positions.

\subsection{Standard deviation maps for the additive Gaussian noise}%
\label{ssec:standard-deviation-maps-of-the-additive-noise}

\begin{figure}[t]
  \centering
  \includegraphics[height=9em]{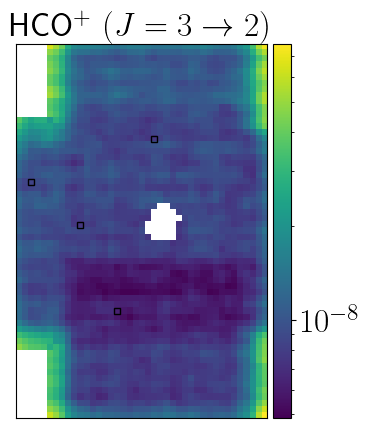}
  \includegraphics[height=9em]{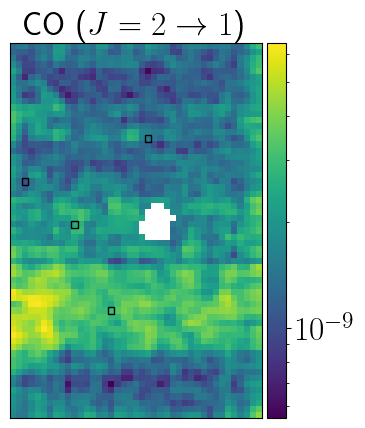}
  \includegraphics[height=9em]{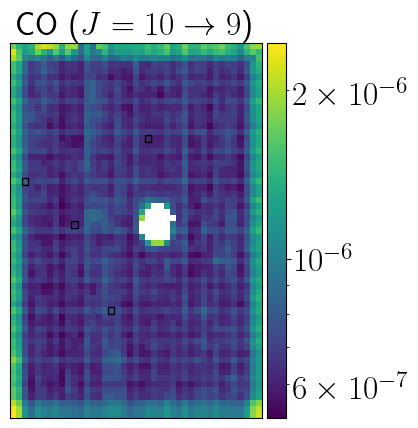} \\
  \includegraphics[height=9em]{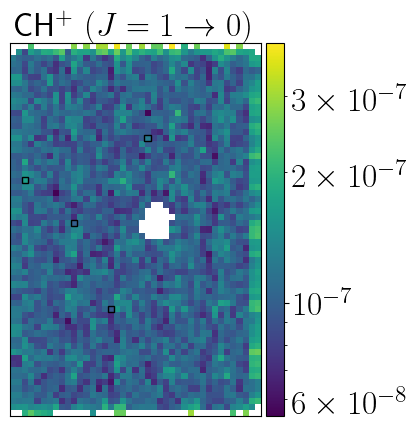}
    \includegraphics[height=9em]{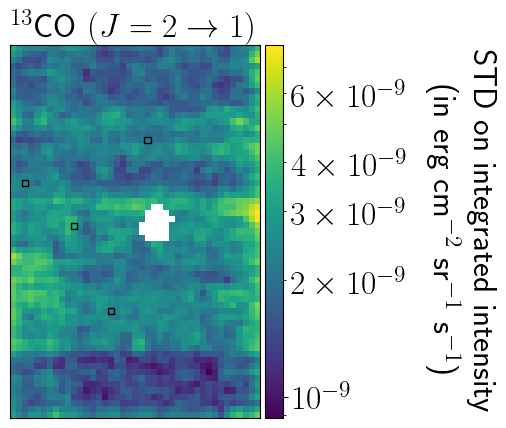}
  \caption{
    Maps of the STD $\sigma_{a,n\ell}$ of the additive Gaussian noise in the observations of OMC-1 used for inversion.
    The white region is the middle of the maps is the BN/KL region, dominated by shocks, and not considered in the inversion.
    %
    %
  }%
  \label{fig:omc1_std}
\end{figure}

Figure~\ref{fig:omc1_std} displays the maps of the STD $\sigma_{a,n\ell}$ of the additive Gaussian noise for the four considered lines in the OMC-1 inversion.
In these maps, the white region is the middle corresponds to the BN/KL region.
This region is dominated by shocks, while the Meudon PDR code does not model shocks.
Therefore, this region is removed from the inversion.
Some other regions are displayed in white on the map borders.
In these regions, the STD is not defined.
In such case, an arbitrarily high value is used so that this line does not affect the reconstruction.
The associated integrated intensities are shown in Fig.~\ref{fig:omc1_observations}.

\subsection{Model checking: Posterior predictive distribution}%
\label{ssec:model-checking-posterior-predictive-distribution}

\begin{figure}
  \centering
  \includegraphics[width=0.4\linewidth]{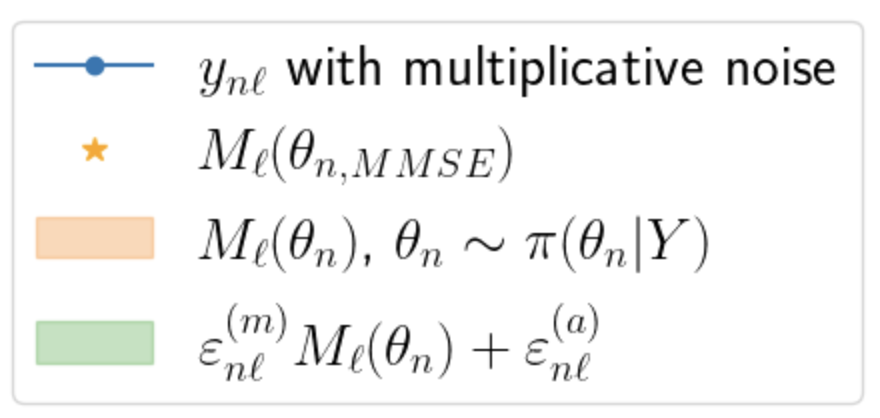}
  \includegraphics[width=0.72\linewidth]{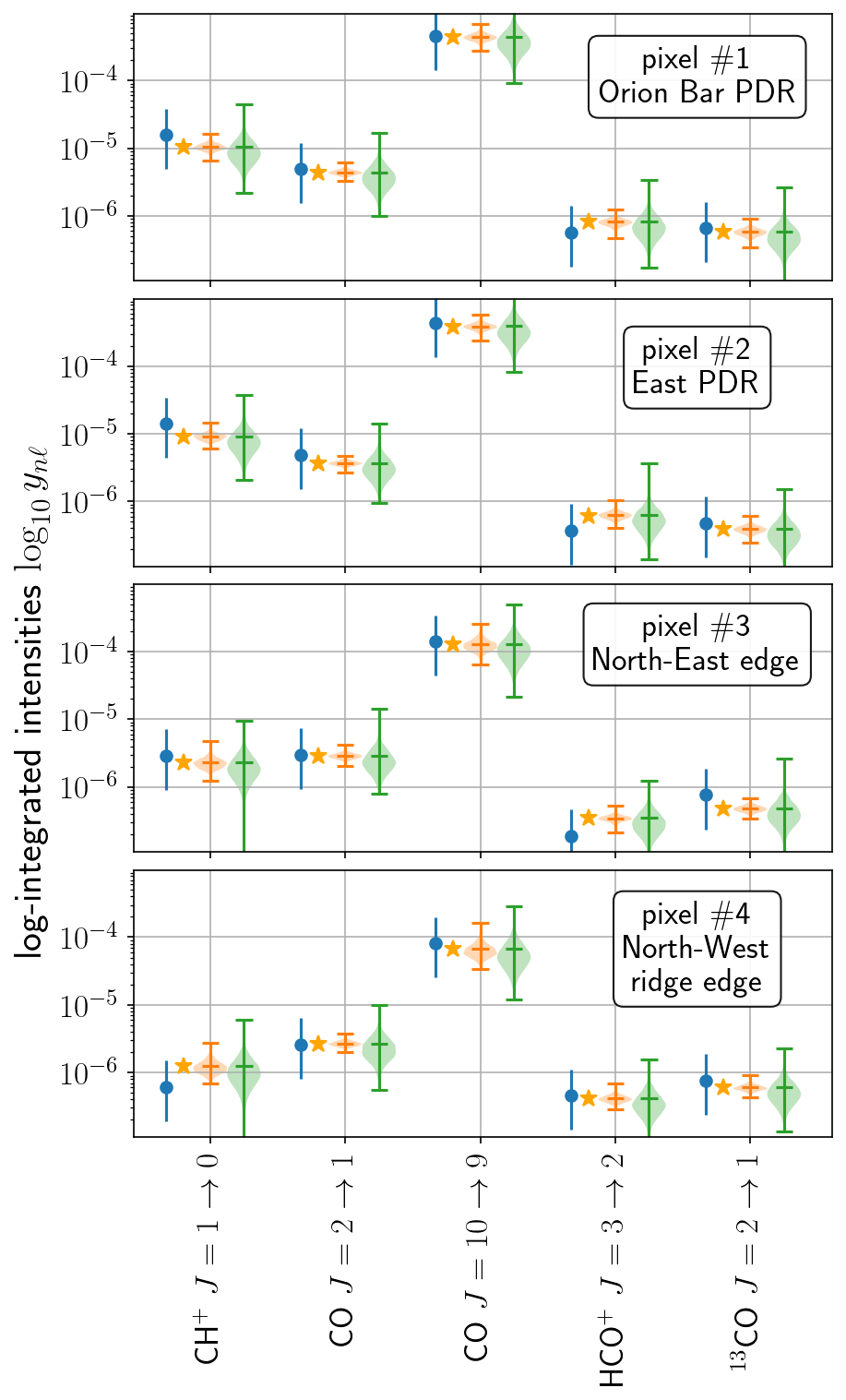}
  \caption{
    Visual diagnostic of the quality of the fit on the lines used for the inversion of OMC-1 for four pixels.
    The blue points are the actual observation with a $1\sigma$ error for the multiplicative noise.
    The orange stars are the prediction of the Meudon PDR code based on the MMSE.
    The orange violin plots show the predictions of the Meudon PDR code on samples of the posterior distribution.
    The green violin plots show the posterior predictive distribution by including addition and multiplicative noises.
  }%
  \label{fig:omc1_posterior_predictive_assess}
\end{figure}

Figure~\ref{fig:omc1_posterior_predictive_assess} compares the posterior predictive distribution (in blue) with the observations $\obselt$ (orange stars).
The predicted integrated intensities show very good agreement with the actual observations.
%
%
The two PDR pixels contain brighter lines than the two other pixels, which is also visible in Fig.~\ref{fig:omc1_observations}.

\end{appendix}

\end{document}
